\documentclass[a4paper,fleqn,usenatbib,useAMS]{mnras}

\usepackage{graphicx}	
\usepackage{amsmath}	
\usepackage{amssymb}	
\usepackage{multicol}        
\usepackage{bm}		
\usepackage{pdflscape}	
\usepackage{grffile}
\usepackage[export]{adjustbox}
\usepackage{titlesec}
\usepackage{float}

\usepackage{epsf}
\usepackage[font=small,labelfont=bf]{caption}
\usepackage{xtab,afterpage}
\usepackage{xtab,booktabs}
\usepackage{mwe}
\usepackage{siunitx,booktabs}
\usepackage[flushleft]{threeparttable}
\usepackage{tabularx}
\newcommand*{\myprime}{^{\prime}\mkern-1.2mu}
\newcommand*{\mydprime}{^{\prime\prime}\mkern-1.2mu}

\usepackage[T1]{fontenc}
\usepackage{ae,aecompl}

\title[]{{Role of Thermal and Non-thermal Processes in the ISM of Magellanic Clouds} \\}

\author[H. Hassani et al. ]{H. Hassani$^{1,2}$\thanks{E-mail: \href{mailto:hamid@ipm.ir}{hamid@ipm.ir}}, F. Tabatabaei$^{2}$, A. Hughes$^{3,4}$, J. Chastenet$^{5,6}$, A. F. McLeod$^{7,8}$,\newauthor E. Schinnerer$^{9}$, S. Nasiri$^{1}$, \\
$^{1}$ Shahid Beheshti University, Faculty of Physics, Department of Astronomy and Astrophysics, Tehran, Iran, 19839 \\
$^{2}$ School of Astronomy, Institute for Research in Fundamental Sciences (IPM), PO Box 19395-5531, Tehran, Iran\\
$^{3}$ CNRS, IRAP, 9 Av. du Colonel Roche, BP 44346, F-31028 Toulouse cedex 4, France\\
$^{4}$ Université de Toulouse, UPS-OMP, IRAP, F-31028 Toulouse cedex 4, France\\
$^{5}$ Center for Astrophysics and Space Sciences, Department of Physics, University of California, San Diego, 9500 Gilman Drive,\\ La Jolla, CA 92093, USA \\
$^{6}$ Sterrenkundig Observatorium, Ghent University, Kri-jgslaan 281-S9, 9000 Gent, Belgium\\
$^{7}$ Department of Astronomy, University of California Berkeley, Berkeley, CA 94720, USA\\
$^{8}$ Centre for Extragalactic Astronomy, Department of Physics, Durham University, South Road, Durham DH1 3LE, UK\\
$^{9}$ Max-Planck-Institut für Astronomie, Königstuhl 17, D-69117, Heidelberg, Germany\\
}

\date{Last updated 2021 October 31}

\pubyear{2021}

\begin{document}
\label{firstpage}
\pagerange{\pageref{firstpage}--\pageref{lastpage}}
\maketitle

\begin{abstract}
The radio continuum emission is a dust-unbiased tracer of both the thermal and non-thermal processes in the interstellar medium. We present new maps of the free-free and synchrotron emission in the Magellanic Clouds (MCs) at 0.166, 1.4, and 4.8\,GHz with no prior assumption about the radio non-thermal spectrum. The maps were constructed using a de-reddened H$\alpha$ map as a template for the thermal radio emission, which we subtract from maps of the total radio continuum emission. To de-redden the H$\alpha$ emission, it is important to know the fraction of dust surface density that attenuates the H$\alpha$ emission along the line-of-sight, $f_\text{d}$. This fraction is obtained by comparing the dust opacity obtained through the infrared emission spectrum and the Balmer decrement method. In star-forming regions, the median $f_\text{d}$ is about 0.1 which is by a factor of 3 lower than that in diffuse regions. We obtain a global thermal radio fraction, $f_\text{th}$, of 30 per cent (35 per cent) in the LMC (SMC) at 1.4\,GHz. Furthermore, we present maps of the equipartition magnetic field strength with average values of $\simeq\,10.1\,\mu$G in the LMC and $\simeq\,5.5\,\mu$G in the SMC. The magnetic field is proportional to the star formation rate to a power of 0.24 and 0.20 for the LMC and SMC, respectively. This study shows that the non-thermal processes control the interstellar medium in the MCs.
\end{abstract}

\begin{keywords}
radio continuum, ISM, Magellanic Clouds, thermal, non-thermal, cosmic rays.
\end{keywords}



\begingroup
\let\clearpage\relax
\endgroup
\newpage

\section{Introduction}

\noindent 
\begin{figure*}
\centering
\includegraphics[width=0.43\textwidth]{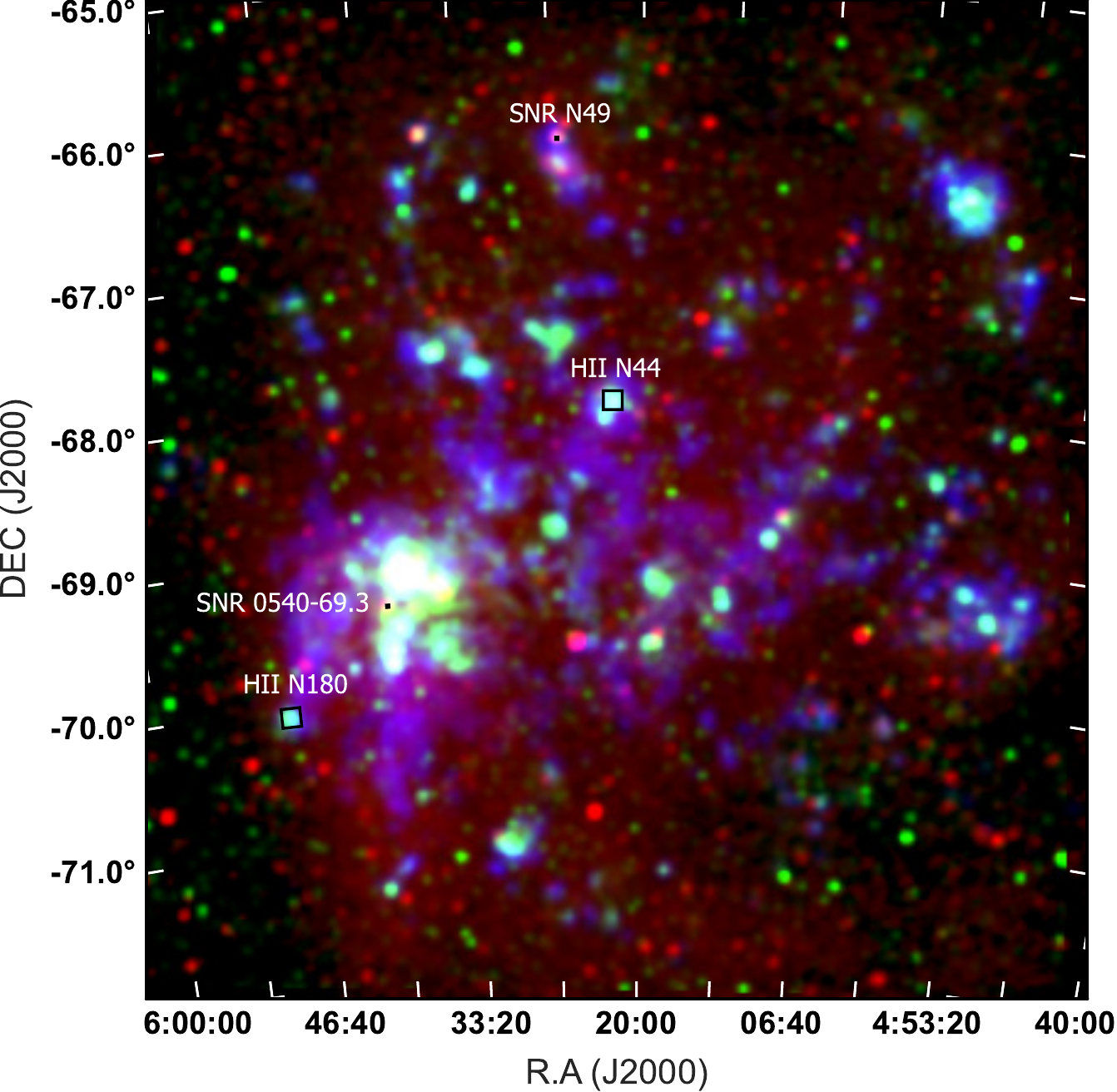}
\includegraphics[width=0.45\textwidth]{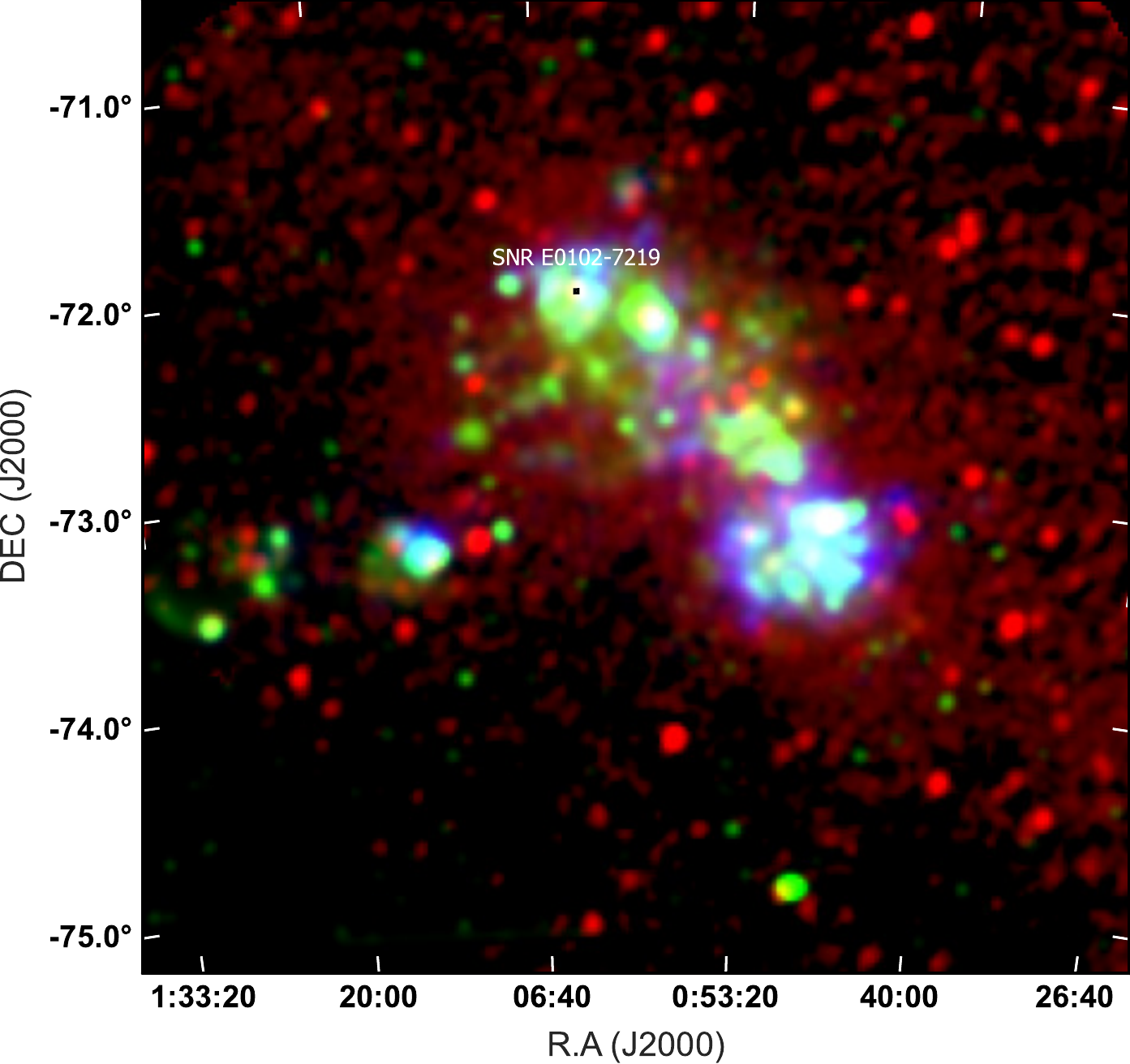}

\caption{Three-colour composite images of the Large Magellanic Cloud ({\it left}) and the Small Magellanic Cloud ({\it right}) showing the radio continuum emission at 0.166\,GHz from the GLEAM survey in log scale ({\it red}), the H$\alpha$ emission from the MCELS survey in linear scale ({\it green}), and the dust mass surface density from \citet{Chastenet_2019} in log scale ({\it blue}). Black squares show the MUSE/VLT observed fields. }

\label{fig:color_map}
\end{figure*} 

The radio continuum (RC) emission observed from astrophysical sources is dominated by the non-thermal (synchrotron) and the thermal (free-free) emission at frequencies $\lambda <10$\,GHz \citep{Condon}. The thermal emission originates from the warm ionized nebular medium (e.g., \ion{H}{ii} regions) with a typical electron temperature of $T_\text{e} \simeq$ 10$^{4}$\,K and has often a flat power-law radio spectral energy distribution $S_{\nu}$ = $\nu ^{-0.1}$ \citep[e.g., ][]{dis_mcs,draine2011}.
 The non-thermal emission is due to cosmic-ray electrons (CREs) accelerated by the magnetic field in the interstellar medium (e.g., supernova remnants) and is described by a power-law spectrum with a variable non-thermal spectral index $\alpha_\text{n}$ ($S_{\nu}$ = $\nu^{\alpha_\text{n}}$). Interaction and energy loss of CREs in different environments (e.g., spiral arms of galaxies or their nuclei) change the non-thermal spectral index $\alpha_\text{n}$ \citep{Longair}. Hence, assuming a constant $\alpha_\text{n}$ in resolved studies can underestimate or overestimate the synchrotron emission depending on the medium. For instance, assuming $\alpha_\text{n} = -1$ results in an underestimation of both the non-thermal emission \citep[by $>30$ per cent ][]{tab_2007} and the magnetic field strength in massive star-forming regions. Avoiding such assumptions, several attempts have been made to map the pure synchrotron emission in nearby galaxies, including M\,33,\,M31, NGC\,6946, and NGC\,1097 \citep{tab_2013b,tab_2013,tab_2018}. Separating the thermal and non-thermal emission is the first step towards understanding the origin of the well-known correlation between the RC and infrared (IR) continuum emission in galaxies \citep{vanderKruit,dejong,helou}. It is also vital in studying the energy balance and structure formation in the interstellar medium (ISM) of galaxies. 
 
The MCs, our nearest neighbour galaxies, provide ideal laboratories for detailed ISM studies in low-mass and low-metallicity systems. The Large Magellanic Cloud (LMC) is at a distance of 49.9\,kpc \citep{Pietrzyski2019} with an almost face-on view that hosts one of the brightest and complex extragalactic \ion{H}{ii} regions, 30~Doradus (hereafter 30~Dor). Being located at about 62.4\,kpc from our Sun \citep{smc_distance_new}, the Small Magellanic Cloud (SMC) is farther away and more inclined (${i}=64\degr$, see e.g., \citealt{smc_i}) than the LMC. These inclinations mean that the distances reported can change up to 10 per cent from one end to other end of each galaxy. General properties of the MCs are summarized in Table~\ref{tab:mc_details}.

They have been extensively studied in RC in both total power \citep{2018for,klein_lmc,smc_l,Haynes91,askap_smc,askap_lmc} and polarization \citep{Gaensler2005,Mao_2008,Mao2012,mcs_bb}. The Herschel space observatory and the Spitzer space telescope have made major breakthroughs in studies of the dust and gas in the MCs \citep{sage_smc,sage_lmc,meixner2013,lacklack}. Moreover, nebular emission lines maps from both the LMC and the SMC are available through optical observations such as the Magellanic Clouds Emission Line Survey {\citep[MCELS; ][]{smith99,smith2005,Points2005,winkler15,Pellegrini_2012}}.
 
Several attempts have been made to study the RC emission components in the MCs, mostly using a fixed $\alpha_\text{n}$ across the galaxies (hereafter classical separation method). \cite{klein_lmc} obtained $\alpha_\text{n} =-0.84$ and found a thermal fraction of $f_\text{th}=55$ per cent at 1.4\,GHz for the LMC. \cite{hughes2006} studied the radio-IR correlation separately for the thermal and non-thermal RC assuming $\alpha_\text{n}=-0.7$ and found a tight correlation in regions with a high thermal fraction.

 \begin{table}
  \begin{threeparttable}
 \caption{General parameters adopted for the MCs.}
 \begin{tabular}{ll}
  \hline
  LMC     \\
  \hline
      \ Centre (J2000)  & $\text{RA}= 05^{\text{h}}\,23^{\text{m}}\,34^{\text{s}}$ $\text{Dec}=-69\degr\,45\myprime\,22\mydprime$ 	  \\
      \ Inclination$^{1}$    &  34.0\degr  \\
      \ Distance$^{2}$ & 49.9\,kpc ($1\arcsec  =  0.24$\,pc)  \\

        \hline
        SMC     \\
  \hline
  
     \  Centre (J2000)  & $\text{RA}= 00^{\text{h}}\,52^{\text{m}}\,44^{\text{s}}$ $\text{Dec}=-72\degr\,49\myprime\,42\mydprime$	  \\
      \ Inclination$^{3}$   &  64.4\degr  \\
      \ Distance$^{4}$  & 62.4\,kpc ($1\arcsec  =  0.30$\,pc)  \\
  \hline
 \end{tabular}
   \label{tab:mc_details}
   
      \begin{tablenotes}
      \small
      \item References:  $^{1}$ \citealt{lmc_ii}. $^{2}$ \citealt{Pietrzyski2019}. $^{3}$ \citealt{smc_i}. $^{4}$ \citealt{smc_distance_new}.
      \end{tablenotes}
  \end{threeparttable}
  
\end{table}

The structure and strength of the magnetic fields in the MCs have been studied by several authors \citep{mcs_bb,Gaensler2005,Mao_2008,Mao2012}. In the LMC, an average magnetic field along the line of sight of $\simeq\,4.3\,\mu$G  was obtained through Faraday rotation studies \citep{Gaensler2005}. Using the classical thermal/non-thermal separation technique, \citep{klein_lmc} reported an equipartition field strength of $\simeq6\,\mu$G in the same galaxy. In the SMC, \cite{Mao_2008} found a galactic-scale mean field strength of $\simeq\,0.2\,\mu$G along the line of sight and a random field strength of $\simeq3\,\mu$G in the sky plane. However, no full map of the magnetic field strength in the MCs is available to study its role in the ISM energy balance and structure formation. The ordered magnetic field is shown to be smaller in dwarf galaxies than in spiral galaxies \citep{dwarf_mag} indicating a large-scale dynamo process may not maintain the production of the magnetic field in dwarf galaxies. As the small-scale dynamo originates from the injection of turbulent energy from  supernova explosions, massive stars may play a primary role in dwarf galaxies such as the MCs. Such a small-scale dynamo can be investigated through its power-law relation with star formation \citep{bsfr}. \cite{dwarf_mag} finds such correlations for dwarf galaxies in global studies. Mapping the magnetic field strength is needed to address those relations in the MCs, taking advantage of our close distance to the MCs and hence inspecting these questions in the physical resolution of less than 80\,pc for the first time.

\begin{table*}
 \caption{Images of the Magellanic Clouds used in this study.}
 \begin{tabular}{llllllllll}
 
  \hline
  Frequency  & Angular Resolution &  RMS  & Ref. & Telescope  \\
		      &	    (arcsec)   	&		&& \\ 
  \hline
  LMC\\
      \ 0.166\,GHz & $221\times221$ &  0.02 (Jy\,beam$^{-1}$) & \cite{2018for} &   MWA \\
  \ 1.4\,GHz & $40\times40$ & 0.30 (mJy\,beam$^{-1}$) & \cite{Huges2007}&   ATCA+Parkes\\
  \ 4.8\,GHz & $33\times33$ & 0.30 (mJy\,beam$^{-1}$)  &  \cite{Dickel2005} &  ATCA+Parkes\\ 
    \ 6563\,{\AA} (H$\alpha$)  & $4.6\times4.6$ &0.12 (10$^{-15}$\,erg\,$\text{s}^{-1}$\,cm$^{2}$) & \cite{smith99} & UM/CTIO \\
      \ 70\,$\mu$m  & $18\times18$ & 0.41 (MJy\,sr$^{-1}$)  & \cite{sage_lmc} & Spitzer/MIPS\\
        \ CO\,(1-0) & $45\times45$ &   0.40 (K\,km\,s$^{-1}$) & \cite{co_2017,co_2011} & Mopra \\
        \ \ion{H}{i} - 21\,cm & $60\times60$ &   15 (mJy\,beam$^{-1}$) & \cite{lmc_HI_merged} & ATCA+Parkes \\
  \ 4800-9300\,{\AA} (IFU)  & $\simeq0.4\times0.4$ & -  &  \cite{muse_2010}  & MUSE/VLT\\

\midrule
    SMC\\
  \ 0.166\,GHz  & $235\times235$ &  0.02 (Jy\,beam$^{-1}$)  &  \cite{2018for} &  MWA \\
  \ 1.4\,GHz &  $98\times98$ & 1.50 (mJy\,beam$^{-1}$) & \cite{wong2011} & ATCA+Parkes\\
    \ 6563\,{\AA} (H$\alpha$)   & $4.6\times4.6$ & 0.01 (10$^{-17}$\,erg\,s$^{-1}$\,cm$^{2}$) & \cite{smith99} & UM/CTIO \\
        \ 70\,$\mu$m  & $18\times18$ & 0.47 (MJy\,sr$^{-1}$) & \cite{sage_smc}  & Spitzer/MIPS\\
  \ 4800-9300\,{\AA} (IFU)  & $\simeq0.4\times0.4$ & - & \cite{muse_2010} & MUSE/VLT\\

  \hline
 \end{tabular}
 \label{tab:mcs_data}
\end{table*}

 \begin{table*}
   \begin{threeparttable}
 \caption{ MUSE/VLT observed fields in the MCs.}
 \begin{tabular}{llllll}
  \hline
  Name   & R.A.& DEC & Area &Program ID  & Ref.   \\
               & (J2000) & (J2000) & &&\\
  \hline
        \ LMC/N44 		 & 80.534& -67.936& $8\arcmin \times 8\arcmin$ & 096.C-0137(A) &   \cite{McLeod_2018} \\
      \ LMC/N49                 & 81.505& -66.083 &$60\arcsec \times 60\arcsec$&0100.D-0037(A) &  Van Loon, J.T$^{1}$ \\
       \ LMC/SNR 0540-69.3  &85.046 &-69.332&$60\arcsec \times 60\arcsec$ & 0102.D-0769(A) & Lyman, Joseph$^{1}$ \\
      \ LMC/N180			& 87.206& -70.055& $8\arcmin \times 8\arcmin$ &096.C-0137(A) & \cite{McLeod_2018} \\
      \ SMC/1E0102–7219     &16.012 & -72.031&$60\arcsec \times 60\arcsec$& 297.D-5058(A) & \cite{snr102}$^{1}$ \\
  \hline
 \end{tabular}
   \label{tab:muse_data}        
         \begin{tablenotes}
      \small
      \item $^{1}$ Data retrieved from the ESO Phase 3 Archive.\end{tablenotes}
  \end{threeparttable}
\end{table*}

The paper aims to study the thermal and non-thermal properties of the ISM, map the total magnetic field strength, and investigate its correlation with SFR in the MCs. Using recent Murchison Widefield Array \citep{mwa_design,mwa_1,Tingay2013} low-frequency observations as well as archival ATCA and Parkes data \citep{atca,lmc_HI_merged,mc_parks,filp,Filipovic96,Filipovic97}, we present full maps of the thermal and non-thermal emission for both the LMC and the SMC.

Unlike the classical separation method, we do not assume a fixed $\alpha_\text{n}$ to obtain the thermal and non-thermal RC maps. We use the Thermal Radio Template (TRT) technique developed for NGC\,6946 \citep{tab_2013}. This method uses a recombination line such as H$\alpha$ emission to trace the thermal free-free emission. In this work, to de-redden the H$\alpha$ emission, no assumption for the fraction of dust attenuating the emission ($f_\text{d}$) is applied unlike our previous studies. We determine $f_\text{d}$ for the MCs by comparing the dust mass obtained using the Balmer-line-decrement ratio method \citep[e.g.,][]{Cardelli,Calzetti} with that extracted from the dust emission Spectral Energy Distribution (SED) studies. Hence, this paper presents distributions of the thermal and non-thermal RC emission across the MCs more precisely than before.

The data is described in Section~\ref{sec:data}. In Section~\ref{sec:extinction}, we determine $f_\text{d}$ for a sample of \ion{H}{ii} regions as well as for more diffuse regions and present a general $f_\text{d}$ calibration relation using its correlation with the neutral gas. In Section~\ref{sec:trt}, we present the thermal and non-thermal maps at different frequencies at an angular resolution of 221$\arcsec$ ($\sim 53$\,pc) for the LMC and 235$\arcsec$ ($\sim71$\,pc) for the SMC. The non-thermal spectral index maps are presented in Section~\ref{sec:nt}. After mapping the magnetic field strength, we investigate its correlation with the star formation rate (SFR) at different spatial resolutions, compare the thermal and non-thermal energy densities of the ISM, and discuss the propagation of CREs in Section~\ref{sec:dis}. 

\section{Data} \label{sec:data}
The data used in this study are summarised in Table~\ref{tab:mcs_data}. The MCs were observed with the Murchison Widefield Array telescope \citep{mwa_design,Tingay2013} as part of the GaLactic Extragalactic All-sky MWA (GLEAM) survey in the frequency range of 0.074-0.231\,GHz \citep{Wayth2015}. The data reduction is explained in detail by \cite{Hurley-Walker2016}. Due to slight changes in ionospheric conditions, a point spread function (PSF) map is created for each mosaic with a variation of 15-20 per cent across each field. We used radio continuum at 0.166\,GHz with a robust 0 weighting from \citet{2018for}. Calibration uncertainty in the flux density is 8.5 per cent for the LMC and 13 per cent for the SMC \citep{Hurley-Walker2016}.

At higher frequencies, the LMC and the SMC have been observed with the Australia Telescope Compact Array (ATCA) and Parkes telescope at 1.4\,GHz and 4.8\,GHz by several authors \citep{Huges2007,Dickel2005,dickel_2010,wong2011,Crawford2011}. For the LMC, we used the merged ATCA+Parkes data at 1.4\,GHz and 4.8\,GHz as presented by \cite{Huges2007}  and \cite{Dickel2005}, respectively. We assume a calibration uncertainty of 6.5 per cent at 1.4\,GHz and 8.5 per cent at 4.8\,GHz for the LMC data.
For the SMC, we used these observations only at 1.4\,GHz assuming a 5 per cent calibration uncertainty  \citep{wong2011}. The quality and signal-to-noise ratio of the available RC data of the SMC are much poorer at 4.8\,GHz \citep{dickel_2010,Crawford2011} than at 1.4\,GHz. The RMS noise value of the RC map is about 2.6 times higher at 4.8\,GHz than at 1.4\,GHz. We note that, in the SMC, the signal-to-noise ratio is by a factor of 2 lower that that in the LMC at 4.8\,GHz. These data are also severely affected by artefacts around bright sources. These prevent studying low-surface brightness regions of the SMC that is the main interest of this study. Hence, we opt not to use these data at 4.8\,GHz for the SMC.

The H$\alpha$ map of the MCs was taken through the MCELS survey \citep{winkler15,smith99}. This survey covers the central $8 \degr \times 8 \degr$ of the LMC and $4.5\degr \times 3.5\degr$ of the SMC. It used a Curtis Schmidt telescope with a pixel size of 2.3$\arcsec$ \citep{smith99}. The H$\alpha$ maps were continuum subtracted for the entire mosaic, rather than field-by-field using two continuum band filters as detailed in \cite{paredes15}. We masked several bright background point sources with H$\alpha$ flux of about 100 times higher than in \ion{H}{ii} regions as they seem to be suffering from saturation in the reduction process. We use a calibration uncertainty of 10 per cent for the H$\alpha$ maps provided by the MCELS Team (private communication).
 
Absolute photometry of H$\alpha$ and H$\beta$ for 74 (27) \ion{H}{ii} regions in the LMC (SMC) has been reported by \citet{Caplan85,Caplan96} (hereafter Caplan catalog). Photometry measurements were taken in circular apertures of 4.89$\arcmin$, 2.00$\arcmin$, and 1.06$\arcmin$ diameter. These aperture sizes are adopted to ensure integration over the whole \ion{H}{ii} regions and for comparison with Parkes radio continuum at 4.8\,GHz data. The LMC H$\alpha$/H$\beta$ flux ratio and their emission lines suffer from 4 per cent uncertainty in random errors, imprecise atmospheric attenuation, and other systematic errors. However, the uncertainty in the H$\alpha$/H$\beta$ flux ratio is estimated to be about 7 per cent in the SMC \citep{Caplan85,Caplan96}.
 
The MCs were observed with Herschel and Spitzer in different bands from 3.6\,$\mu$m to 500\,$\mu$m by the HERITAGE and SAGE projects \citep{sage_lmc,mx2010,meixner2013,meixner2015,sage_smc}. \citet{Chastenet_2019} modelled the FIR dust emission using the \cite{draine2007} dust model. We use their total dust mass surface density maps with an angular resolution of 36$\arcsec$.

The 21-cm \ion{H}{i} line emission was observed with ATCA and Parkes \citep{lmc_HI_merged} and the CO\,(1-0) line emission is provided by the Magellanic Mopra Assessment (MAGMA) survey \citep{co_2011,co_2017} for the LMC. 

We use optical IFU MUSE/VLT observations toward several fields in the MCs (see Fig.~\ref{fig:color_map}), including \ion{H}{ii} regions and Supernova remnants (SNRs) to investigate the extinction properties of the diffuse gas in SNRs vs star-forming regions. These data were taken by Wide Field Mode (WFM) with a FoV$\approx 1\arcmin$ and are summarized in Table~\ref{tab:muse_data}. The two \ion{H}{ii} regions of the LMC are reduced and mosaiced by \cite{McLeod_2018}. Our sample of MUSE SNRs fields was taken from the Phase 3 ESO Science Archive and reduced using ESO’s automatic pipeline \citep{muse_pipeline}. We note that this automatic pipeline may over-subtract emission lines and produce artificial absorption lines in the spectrum. Comparing MUSE SNRs H$\alpha$ flux with MCELS survey data in 1$\arcmin$ rectangular aperture shows an agreement within <18 per cent of total flux. We take extra 20 per cent uncertainty in flux of emission line ratio ($F_{\text{H}\alpha}$/$F_{\text{H}\beta}$) for the sky over-subtraction effect in the LMC SNRs. We did not find any noticeable absorption lines in the integrated spectra or each pixel of the SNR 1E0102–7219.  

All maps were convolved to the lowest common resolution of our datasets, $221 \arcsec \times 221\arcsec$ for the LMC and $235\arcsec \times 235 \arcsec$ for the SMC using Gaussian kernels. The smoothed maps were then regridded to a common astrometric grid. 
We considered not only the RMS noise ($\sigma_{\rm rms}$) of the observed maps, but also the calibration uncertainty ($\sigma_{\rm cal}$) of the instruments to estimate uncertainties in fluxes (following $\sigma=\sqrt{\sigma_{\rm rms}^2 + \sigma_{\rm cal}^2}$). These errors were then propagated to obtain uncertainties in other parameters reported throughout the paper.

\section{Extinction in the Magellanic Clouds} \label{sec:extinction}
The Herschel and Spitzer space telescopes have made a major breakthrough in mapping the dust content of galaxies. These observations can be used to map extinction and de-redden the optical H$\alpha$ emission provided that the relative distribution of emitting sources and dust particles is known along the line of sight \citep{tab_2013}. A uniform mix of dust and ionized gas emitting H$\alpha$ requires half of the total dust mass (or optical depth) to be considered in de-reddening. A smaller fraction must be used in the more realistic case of non-uniformity, such as a smaller z-distribution of dust than ionized gas \citep[][]{di2013} or due to clumpiness. As follows, the total dust opacity is first obtained in the MCs. Then we derive the fraction of it that is linked to reddening of the H$\alpha$ emission ($f_\text{d}$) by comparing the total dust optical depth with true extinction obtained using the H$\alpha$-to-H$\beta$ ratio method. This is done for \ion{H}{ii} regions and a few MUSE fields including more diffuse ISM. Moreover, a calibration relation for $f_\text{d}$ is obtained by investigating its correlation with neutral gas surface density across the LMC.

\subsection{Dust opacity}\label{ssec:tau_dust}

Using the Herschel and Spitzer data, \cite{Chastenet_2019} mapped different dust properties including dust mass surface density in the MCs based on the DL07 model \citep{draine2007}. A composition of carbonaceous and amorphous silicates were assumed for dust grains which are heated by a variable radiation field U with a delta-function distribution. Taking a fixed $U_\text{max}=10^{7}$\,$U_{\sun}$, the minimum interstellar radiation field intensity spans a range of 0.1\,$U_{\sun}$ $\leq$ $U_\text{min}$ < 50\,$U_{\sun}$. Considering the Galactic $R_\text{V}=3.1$ \citep{Cardelli}, dust SEDs were fitted based on this model from 3.6 to 500\,$\mu$m. This results in 36$\arcsec$ resolution maps of dust mass surface density $\Sigma_\text{d}$ with a pixel size of 14$\arcsec$ for the MCs. This leads to a dust optical depth following $\tau_\text{dust}$ = $ \kappa_{\Sigma}\,\Sigma_\text{d}$, where $\kappa_{\Sigma}$ is the dust opacity. At the H$\alpha$ wavelength, $\kappa_{\Sigma}=1.4701\times10^{4}\,\text{cm}^{2}\,\text{g}^{-1}$ for the LMC and $\kappa_{\Sigma}=1.2026\times10^{4}\,\text{cm}^{2}\,\text{g}^{-1}$ for the SMC \citep{Weingartner,Gordon_2003} taking into account both absorption and scattering by dust grains.

Examining the $\tau_\text{dust}$ map obtained reveals a mean $\tau_\text{dust}=1.08 \pm 0.04$ within 5$\degr$ radius centred on (05$^{\text{h}}$\,23$^{\text{m}}$,\,-69$\degr$\,45$\myprime$)$_\text{J2000}$ for the LMC. It is $\tau_\text{dust}=0.48 \pm 0.02 $ within 3$\degr$ radius centred on (00$^{\text{h}}$\,52$^{\text{m}}$,\,-72$\degr$\,49$\myprime$)$_\text{J2000}$ in the SMC. In other words, the SMC’s ISM is more transparent to the H$\alpha$ emission than the LMC, since it has a relatively lower dust mass surface density $\Sigma_\text{d} <0.1$ and metallicity \citep[and references there in]{all_gas_mcs}.

\subsection{Effective extinction} \label{ssec:eff_tau_dust}

The dust optical depth obtained in Section~\ref{ssec:tau_dust} can be used to de-redden the observed H$\alpha$ emission if emitting sources (here ionized gas) are all behind a slab of dust with that optical depth ($\tau_\text{dust}$). However, in reality, ionized gas and dust are mixed and hence only a fraction of $\tau_\text{dust}$ must be considered. This fraction depends on the relative distributions of ionized gas and dust along the line of sight that is often unknown.
Following \cite{di2013} and \cite{tab_2013} we define an effective dust optical depth that should be considered to obtain the true extinction. It is given by multiplication of $\tau_\text{dust}$ by the factor $f_\text{d}$ that is the fraction of dust attenuating the H$\alpha$ emission along the line of sight: 

\begin{equation}
\label{eqn:f_d_for}
\tau_\text{eff}\equiv f_\text{d} \times  \tau_\text{dust}
\end{equation}

\noindent 
with $\tau_\text{eff}$ as the effective dust optical depth obtained at the H$\alpha$ wavelength. We note that $\tau_\text{eff}$ in terms of visual extinction, $A_\text{V}$, is given by:

\begin{equation}
\label{eqn:tau_eff_for}
\tau_\text{eff} = \frac {A_\text{V} \kappa_{H{\alpha}}  }  {1.086  R_\text{V}}
\end{equation}

\begin{figure}
\includegraphics[width=0.8\columnwidth,center]{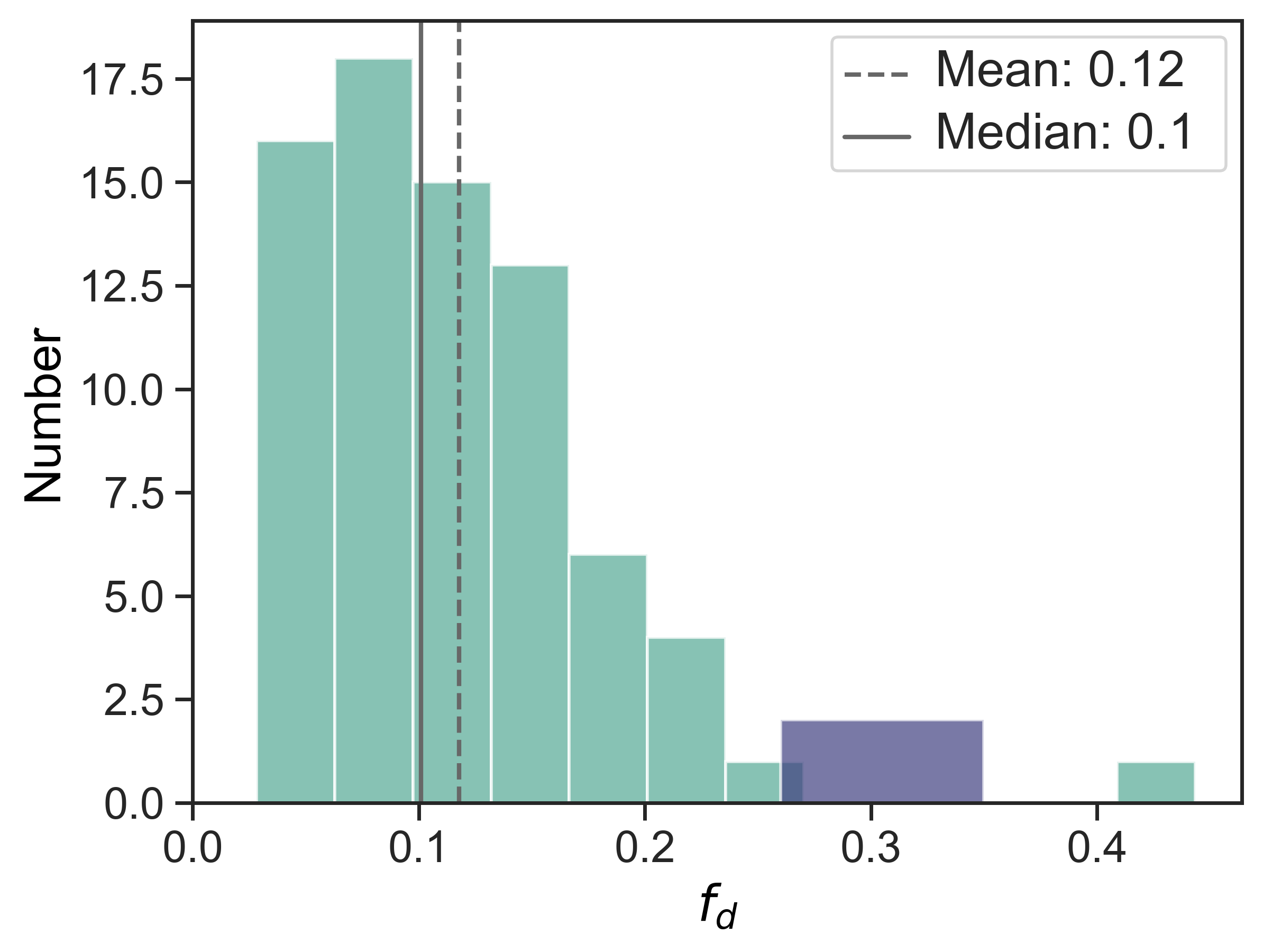}
\includegraphics[width=0.8\columnwidth,center]{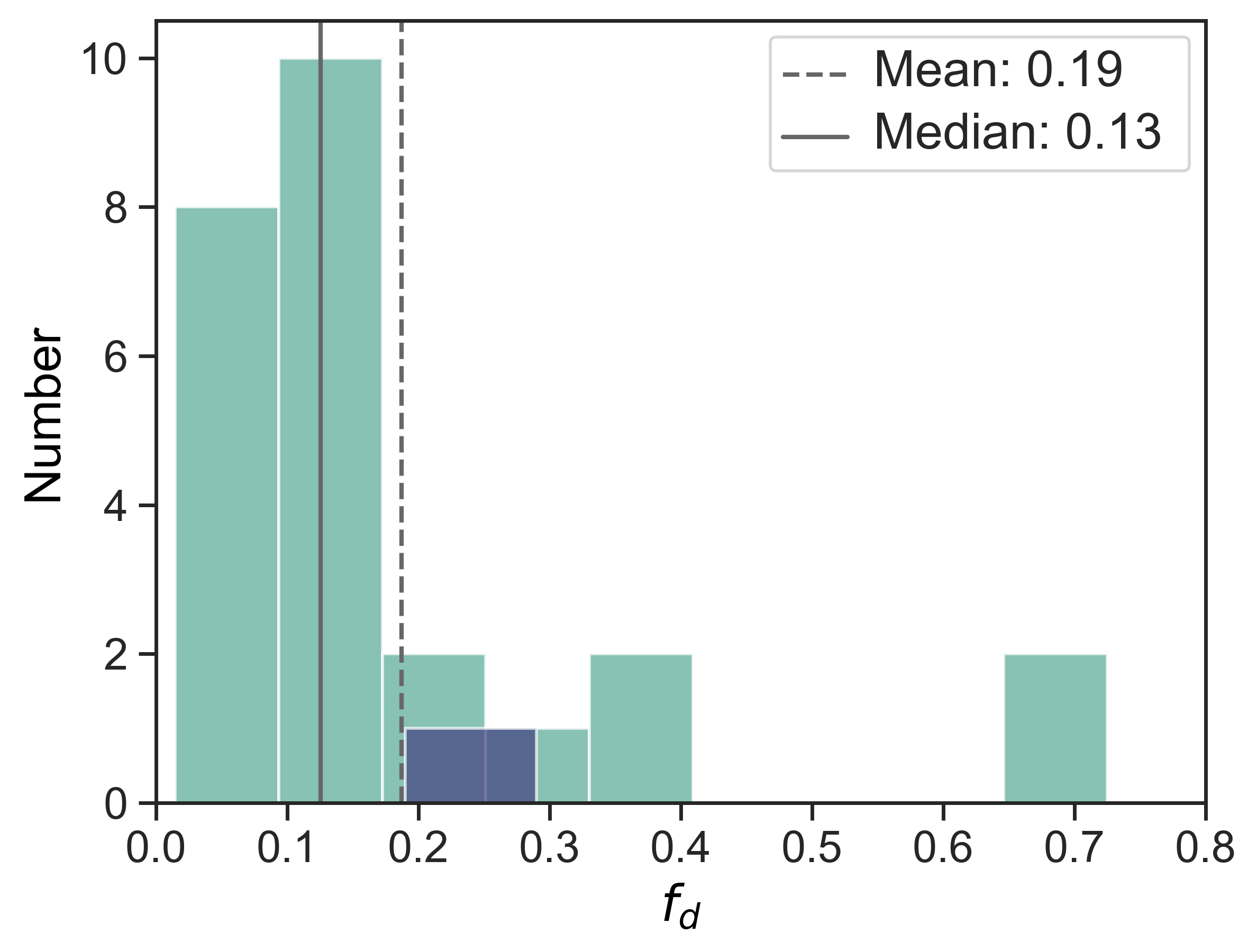}
 \caption{ A histogram of $f_\text{d}$ in \ion{H}{ii} regions ({\it green}) and diffuse regions ({\it purple}) for the LMC ({\it top}) and the SMC ({\it bottom}). }
 \label{fig:fd_histogram}
\end{figure} 

\begin{figure*}
\centering
\includegraphics[width=0.45\textwidth]{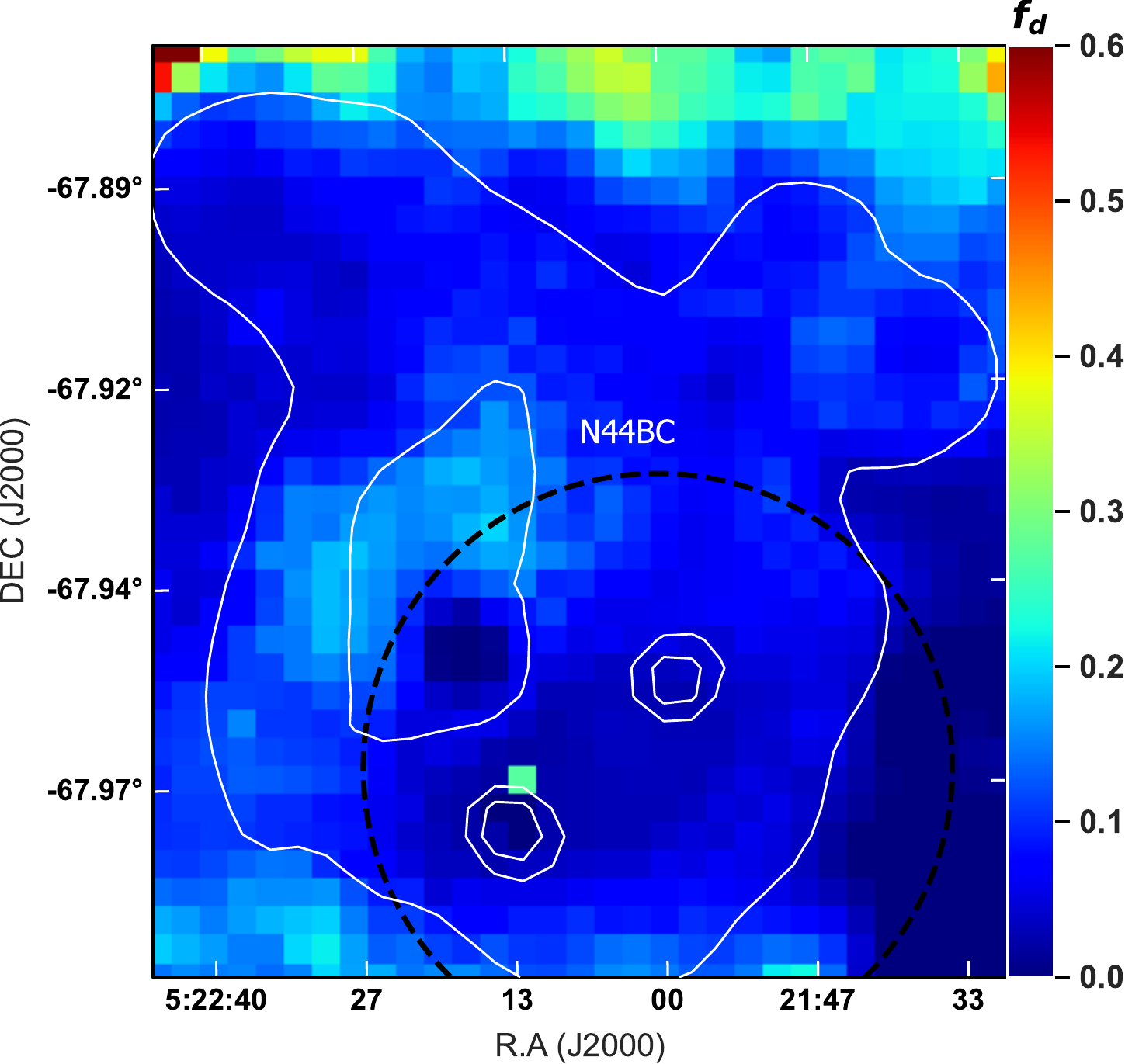}
\includegraphics[width=0.45\textwidth]{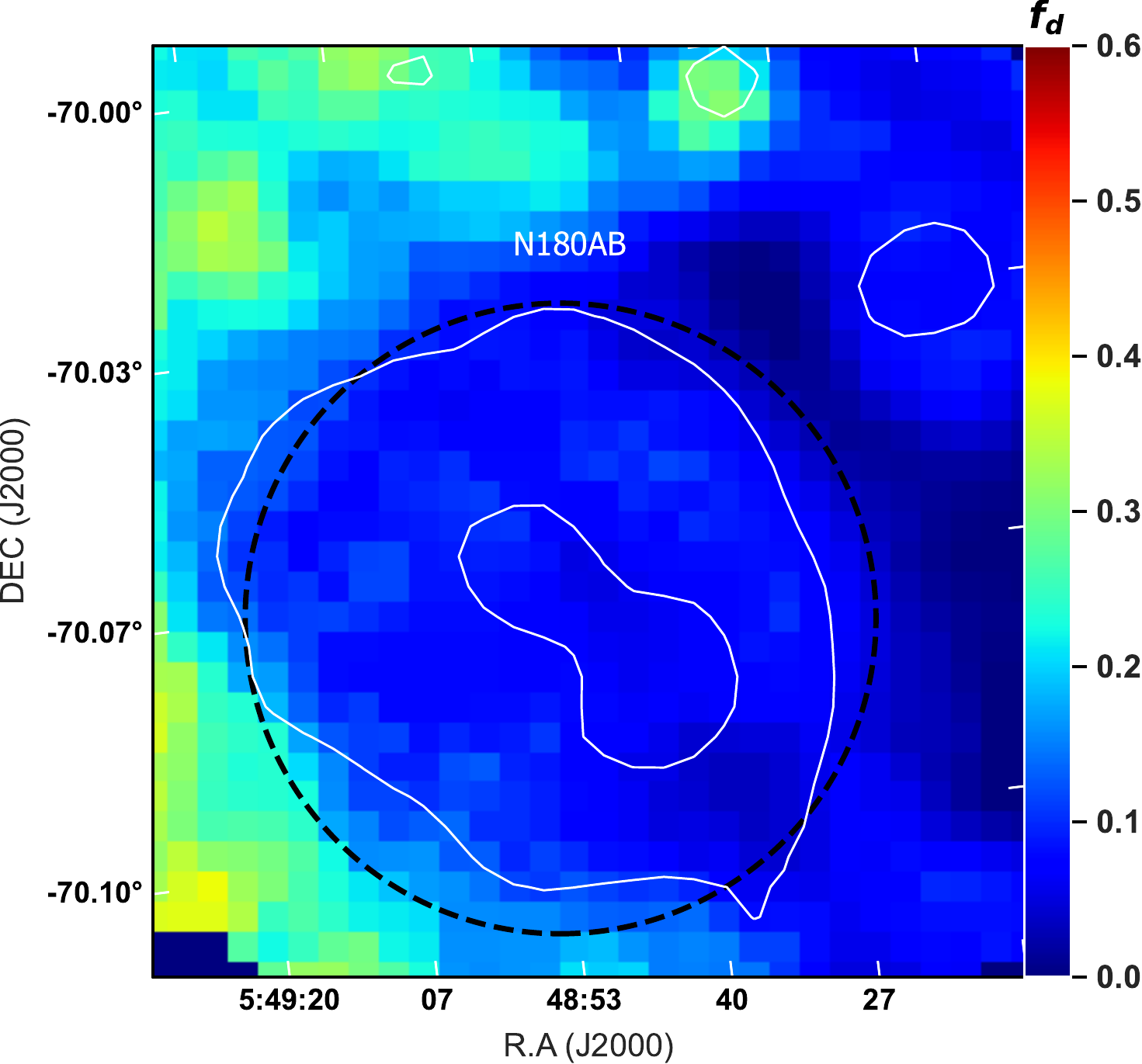}

 \caption{ Map of the fraction of dust attenuating H$\alpha$ emission $f_\text{d}$ in the \ion{H}{ii} regions N44 (left) and N180 (right), overlaid with the H$\alpha$ emission contours (white). Contours levels are 1, 6, and 8 $\times$\,10$^{-16}$\,erg\,s$^{-1}$\,cm$^{2}$ (left) and 1, 3, and 6 $\times$\,10$^{-16}$\,erg\,s$^{-1}$\,cm$^{2}$ (right). Black dashed circles are the apertures reported by \protect\cite{Caplan85}.  }

 \label{fig:fd_HIIregions_muse}
\end{figure*} 

\noindent
with a total to selective ratio $R_\text{V} = 3.41$ for the LMC and 2.74 for the SMC \citep{Gordon_2003}. \cite{di2013} found that $f_\text{d} = 0.33$ can best reproduce the Galactic plane's observed data, interpreting it as only one-third of dust along the line of sight is responsible for reddening. As the distribution of the ISM in dwarf systems can be different from normal-mass spiral galaxies, we expect that $f_\text{d}$ also differs. Hence, we try to estimate $f_\text{d}$ using observations of the Balmer decrements.

We use the H$\alpha$ and H$\beta$ emission data to measure the extinction using the following relations \citep[e.g.,][]{Cardelli,Calzetti}:

\begin{equation}
\label{eqn:redd}
E(B-V)_\text{Balmer} = \frac{2.5}{\kappa(H\beta) - \kappa(H\alpha)} log_{10}  \left( \frac{{F}_{\text{H}\alpha} / {F}_{\text{H}\beta} } {2.86 } \right)
\end{equation}

\begin{equation}
\label{eqn:av_for}
A_\text{V} = R_\text{V} \: E(B-V)_\text{Balmer}
\end{equation}

\noindent
with $E(B-V)$ the reddening and $F_{\text{H}\alpha}$, $F_{\text{H}\beta}$ the intrinsic  H$\alpha$ and H$\beta$ fluxes. We adopted the theoretical $F_{\text{H}\alpha}/F_{\text{H}\beta}=2.86$ \citep{Brocklehurst,Osterbrock}.

Using the \cite{Calzetti} attenuation law, we obtain $\kappa$(H$\alpha$) and $\kappa$(H$\beta$) as 2.72 and 4.40 for the LMC and 2.19 and 3.76 for the SMC. Considering that $A_\text{V}$ from equation~(\ref{eqn:tau_eff_for}) must be equal to that given by equation~(\ref{eqn:av_for}), $f_\text{d}$ is obtained separately for \ion{H}{ii} regions and diffuse ISM as follows.

\subsubsection{Calibrating $f_\text{d}$ in \ion{H}{ii} regions} \label{sssec:num1}

The $F_{\text{H}\alpha}$ and $F_{\text{H}\beta}$ fluxes were extracted from the Caplan catalog for a sample of 74 \ion{H}{ii} regions in the LMC and 25 \ion{H}{ii} regions in the SMC. We first derive their visual extinction $A_\text{V}$ using equations~(\ref{eqn:redd}) and~(\ref{eqn:av_for}). Extracting $\tau_\text{dust}$ for the apertures in the Caplan catalog, we then calibrate and obtain $f_\text{d}$ using equation~(\ref{eqn:tau_eff_for}). The $A_\text{V}$ changes between 0.09 and 1.03 for the LMC with the median value of $A_\text{V}=0.35$, which is larger by 34 per cent than in the SMC. Uncertainties in the observed fluxes affect $A_\text{V}$ by less than 0.09 magnitudes in the LMC and 0.14 magnitudes in the SMC. The dust filling factor $f_\text{d}$ obtained using equation~(\ref{eqn:tau_eff_for}) varies between 0.02 and 0.46 for the LMC and it changes between 0.01 and 0.72 for the SMC with a median value of $f_\text{d}$ $\simeq$ 0.1 in both the LMC and the SMC \ion{H}{ii} regions (green bars in Fig.~\ref{fig:fd_histogram}). Tables~\ref{tab:lmc_hII} and~\ref{tab:smc_hII} list the resulting visual extinction $A_\text{V}$, dust mass surface density $\Sigma_\text{d}$ and $f_\text{d}$. 

Using the Caplan catalog we can only derive an average $f_\text{d}$ in \ion{H}{ii} regions, but taking advantage of the MUSE observations we can map $f_\text{d}$ in the two \ion{H}{ii} regions N44 and N180 \citep{McLeod_2018}. We take 32\,{\AA} wide (H$\alpha$ + \ion{N}{ii}) line based on MUSE observations covering a 8$\arcmin$ $\times$ 8$\arcmin$ area centred at (5$^{\text{h}}$\,48$^{\text{m}}$\,49.46$^{\text{s}}$,\,-70$\degr$\,03$\myprime$\,19.53$\mydprime$)$_\text{J2000}$ and (5$^{\text{h}}$\,22$^{\text{m}}$\,08.13$^{\text{s}}$,\,-67$\degr$\,56$\myprime$\,08.26$\myprime$)$_\text{J2000}$ for N180 and N44, respectively. Details of spectral extraction are found in \cite{McLeod_2018}. We performed aperture photometry on the H$\alpha$ and H$\beta$ emission lines in N44 and N180 for each MUSE pixel resulting in a map of $f_\text{d}$ (Fig.~\ref{fig:fd_HIIregions_muse}), indicating a good agreement with our integrated measurement within a 4.89$\arcmin$ diameter circular aperture of the Caplan catalog, $f_\text{d}=0.06$ and $f_\text{d}= 0.11$ in N44BC and N180AB. We note that the integrated H$\alpha$ flux increases by about 20 per cent adopting the (H$\alpha$ + \ion{N}{ii}) bandwidth in comparison with the slab of narrow 6{\AA} H$\alpha$ line. Inspecting the $f_\text{d}$ maps, it is inferred that clumps of dust corresponding to dense regions of molecular gas have lower $f_\text{d}$ than more diffuse regions (Fig.~\ref{fig:extinction_map}, top left).

\subsubsection{Calibrating $f_\text{d}$ in the diffuse ISM} \label{sssec:2}
Determining $f_\text{d}$ and $\tau_\text{eff}$ should not be limited to only dense ionized gas in the \ion{H}{ii} regions because there is a considerable amount of H$\alpha$ emission from  lower density regions in the MCs. In this section, we use the MUSE observations of few SNRs (see Table~\ref{tab:muse_data}) as they represent more diffuse regions than the \ion{H}{ii} regions. Using a Gaussian fit, we first derive the centre of each specific emission line (listed in Table~\ref{tab:muse_res}), and then we obtain zero moments of the H$\alpha$ line with a 32\,{\AA} width and H$\beta$ line using a 6\,{\AA} width slab on each central line. The resulting $f_\text{d}$ is higher in the SNRs by a factor of $\sim$~3 compared to the \ion{H}{ii} regions, indicating that dust is better mixed with the ionized gas in the SNRs than in the \ion{H}{ii} regions (purple bars in Fig.~\ref{fig:fd_histogram}). In other words, as dust is clumpier in the \ion{H}{ii} regions than in the SNRs, the dust filling factor is smaller in the \ion{H}{ii} regions. Table~\ref{tab:muse_res} shows the resulting visual extinction $A_\text{V}$, dust mass surface density $\Sigma_\text{d}$, and $f_\text{d}$ for the MUSE/VLT fields.

\noindent 
\begin{figure*}
\centering
\includegraphics[width=0.43\textwidth]{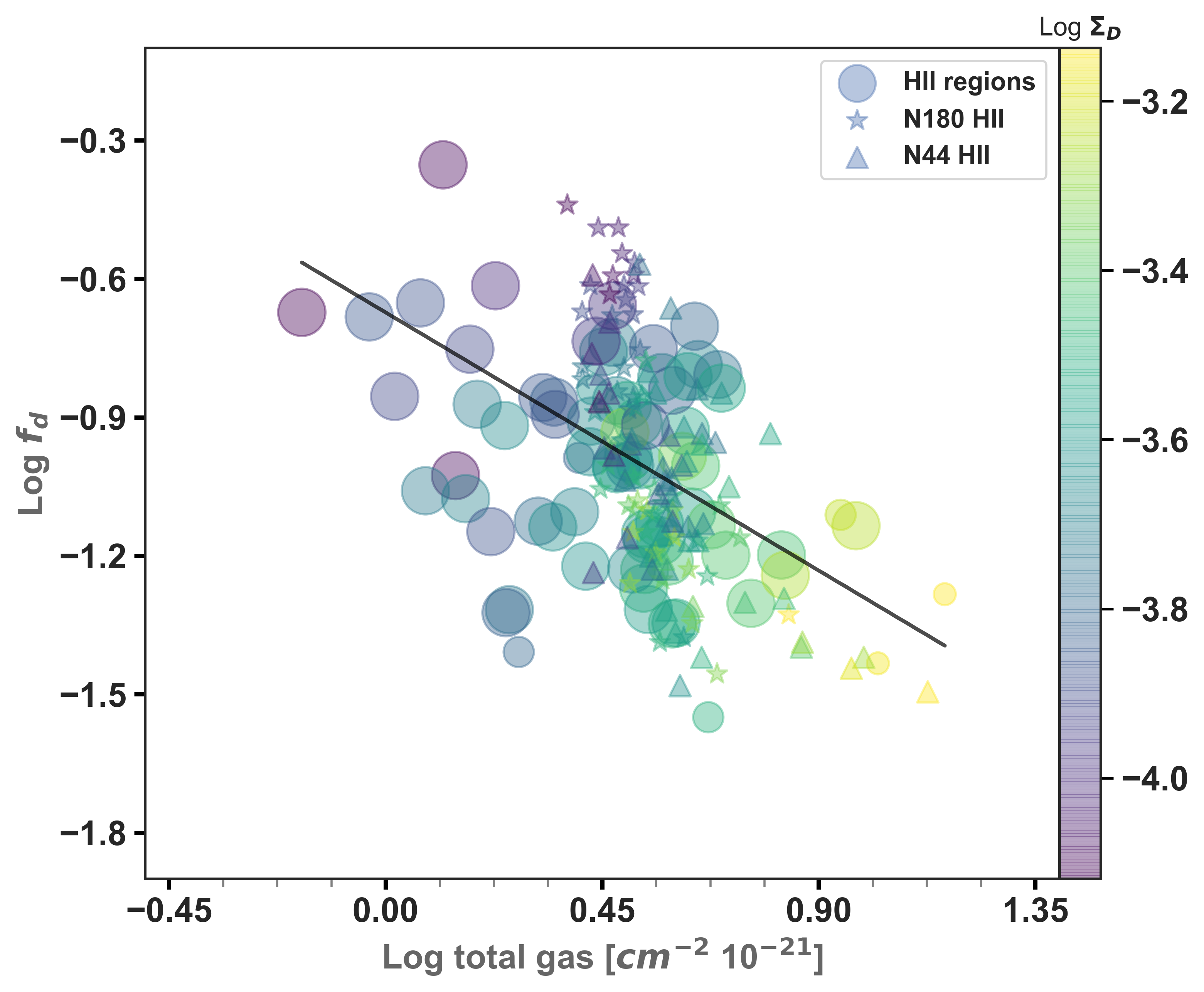}
\includegraphics[width=0.41\textwidth]{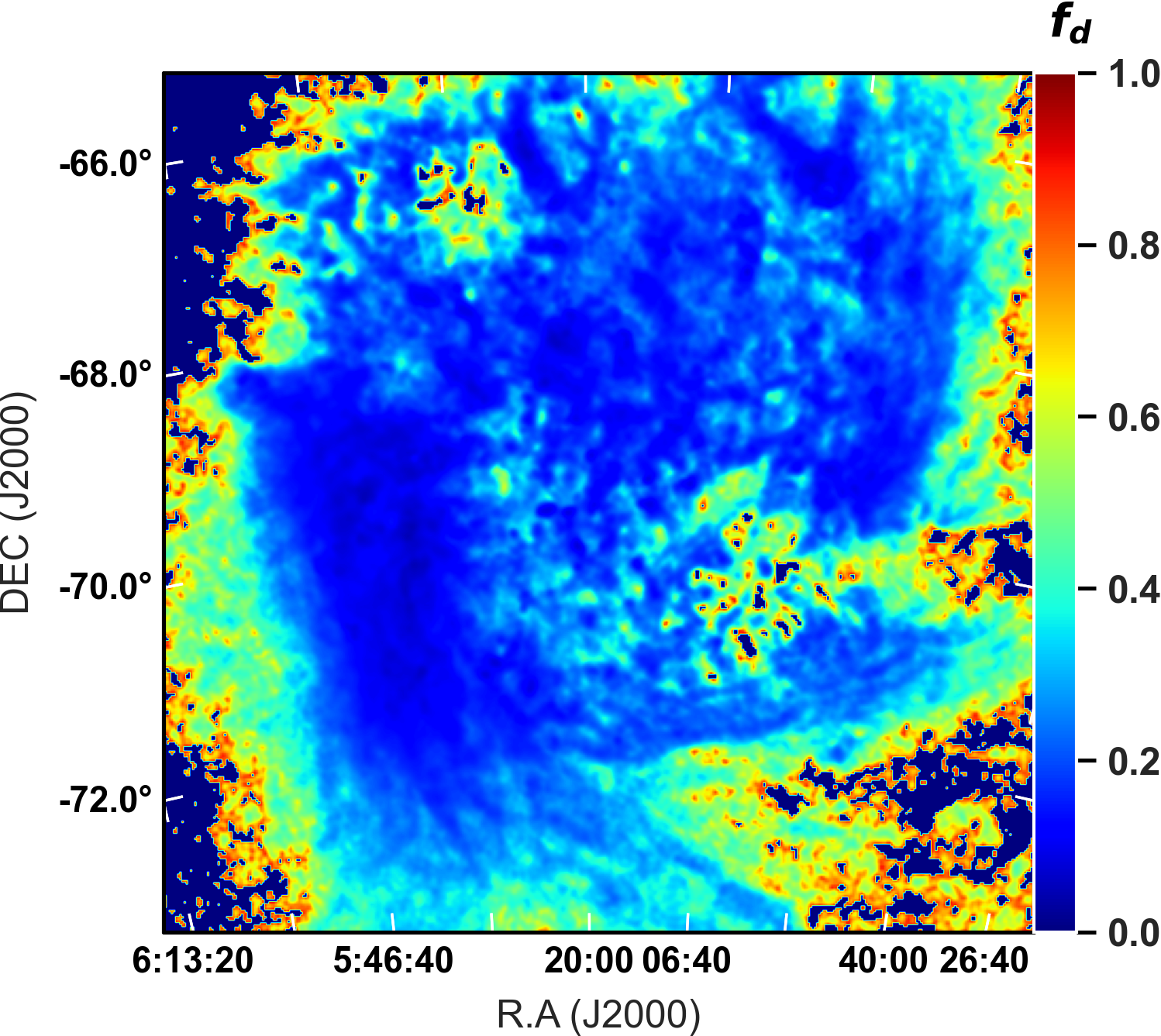}

\includegraphics[width=0.43\textwidth]{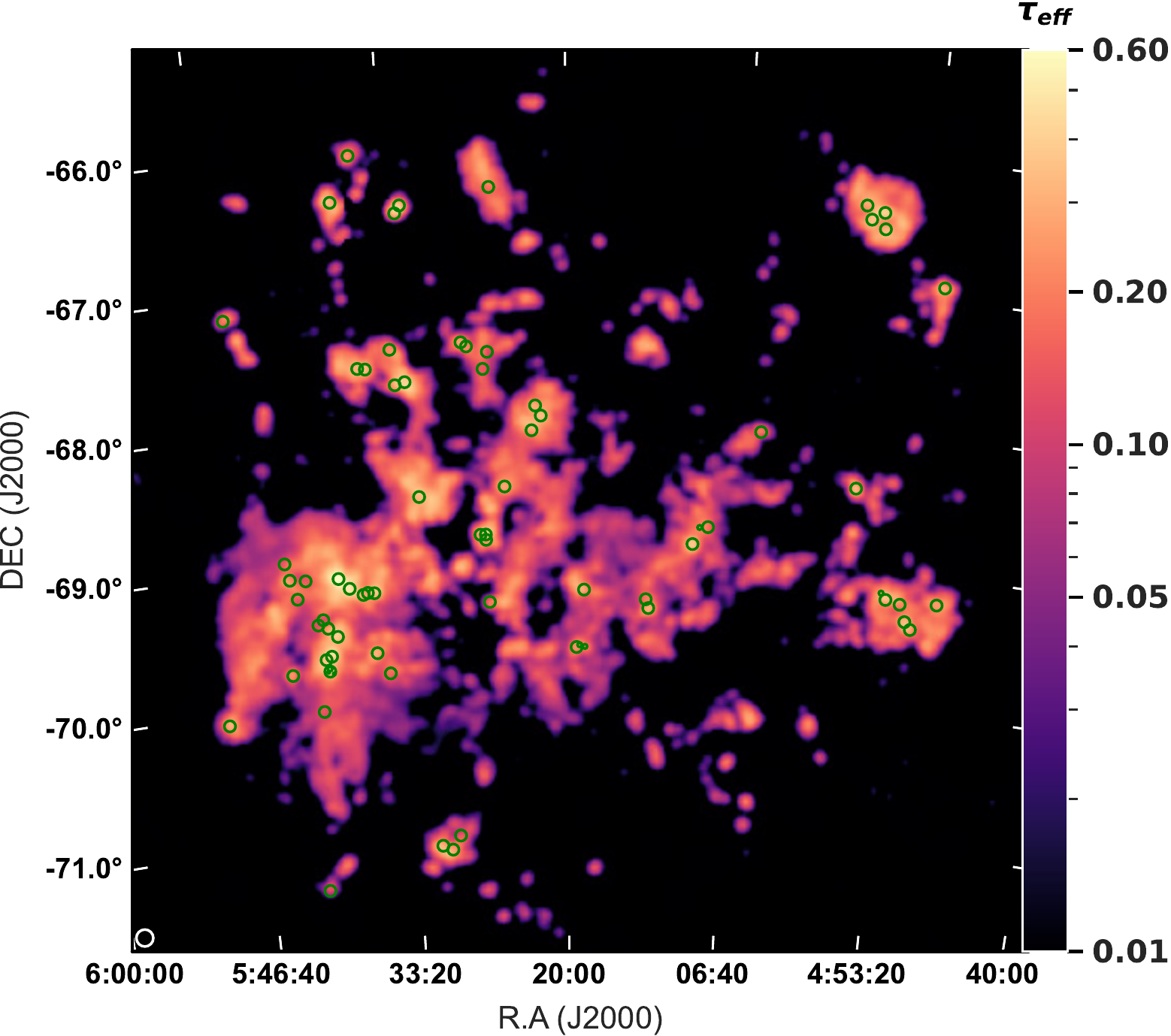}
\includegraphics[width=0.43\textwidth]{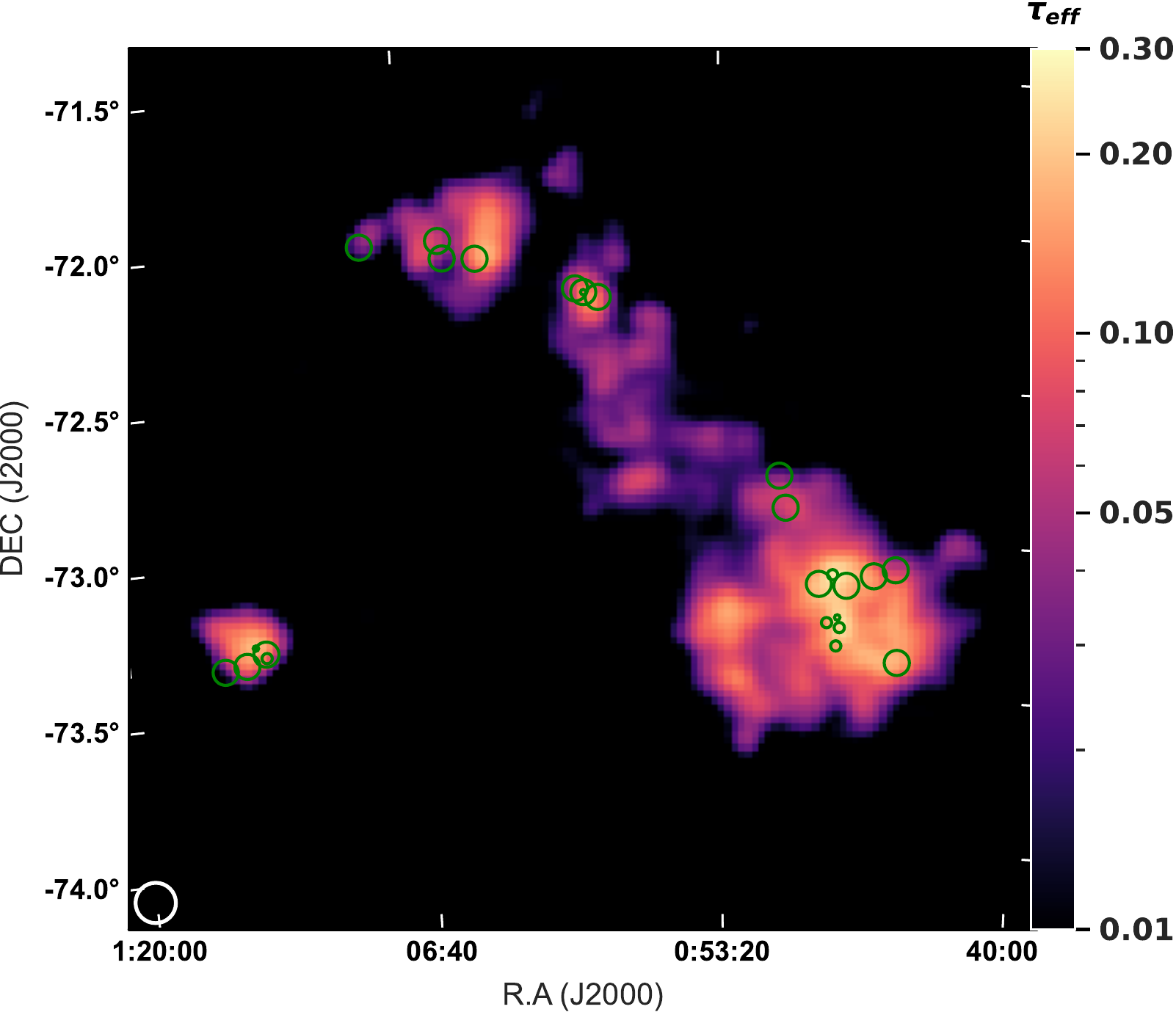}

 \caption{Top: the fraction of dust attenuating H$\alpha$ emission $f_\text{d}$ vs the total gas density $\Sigma_\text{gas}$ (left) in separate fields and the map of $f_\text{d}$ (right) in the LMC. Solid line shows the Ordinary Least Squares (OLS) bisector fit in log-log space used as an independent calibration of $f_\text{d}$. Bottom: effective dust optical depth ($\tau_\text{eff}$) at H$\alpha$ wavelength for the LMC (left) and the SMC (right) shown in log scale. Green circles indicate the position of \ion{H}{ii} regions inside the MCs. The beam sizes of 221\arcsec (for the LMC) and 235\arcsec (for the SMC) are shown in the lower left corner of the maps.}

\label{fig:extinction_map}
\end{figure*}

Maps of $f_\text{d}$ could only be derived for a few \ion{H}{ii} regions and SNRs using MUSE observations. To map the variation of $f_\text{d}$ for the entire galaxy, its correlation with the total neutral gas is first investigated for the given fields and sources (Fig.~\ref{fig:extinction_map}, top left). Then a calibration relation is built to map $f_\text{d}$ over the entire galaxy. We measured the total gas surface density as $\Sigma_\text{gas} = \Sigma_\text{\ion{H}{i}} + \Sigma _{\text{H}_\text{mol}}$ where $\Sigma _{\text{H}_\text{mol}}= 2\,N(H_{2}) = 2\,X_\text{CO}$ $\times$\, $I_\text{CO}$ and adopting a conversion factor $X_\text{CO} = 2$ $\times$\,10$^{20}$\,cm$^{-2}$ (K\,km\,s$^{-1}$)$^{-1}$ \citep{x_co}. The similar conversation factor $X_\text{CO} \approx 2.76$ $\times$\,10$^{20}$\,cm$^{-2}$ (K\,km\,s$^{-1}$)$^{-1}$ was reported for the LMC \citep{Leroy_CO}. Excluding pixels weaker then 3$\sigma$ RMS ($\sigma = 3.3$ $\times$\,10$^{19}$ H\,cm$^{-2}$) in total gas surface density, we found a relation between $f_\text{d}$ and $\Sigma_\text{gas}$ in the LMC:

 \begin{equation}
 \label{eqn:fd_cal}
 \text{Log} \: f_\text{d} \: \text{(LMC)} =  (-0.62 \pm  0.04) \: \text{Log} \: \Sigma_\text{gas}  - (0.67 \pm  0.03) 
 \end{equation}
  
 \noindent
We then construct a synthesized map for $f_\text{d}$ using equation~(\ref{eqn:fd_cal}) in the LMC (Fig.~\ref{fig:extinction_map}, top right). In the \ion{H}{ii} regions, the median value of $f_\text{d}$ is about 0.1 that agrees with the more direct estimate outlined in Section~\ref{sssec:num1} with a mean $\Sigma_\text{gas} = 3 \times 10^{21}\,$ H\,cm$^{-2}$. This factor is higher in regions of lower density gas ($\Sigma_\text{gas} <8 \times 10^{20}$\,H\,cm$^{-2}$) as expected. We found a mean value $f_\text{d} = 0.29 \pm 0.04$ in the LMC within 5$\degr$ of the central point (05$^{\text{h}}$\,23$^{\text{m}}$\,34$^{\text{s}}$,\,-69$\degr$\,45$\myprime$22$\mydprime$)$_\text{J2000}$, which is in a good agreement with Galactic value \citep{di2013}. However, we should note that we found a variation of $0.2-0.3$ in different ISM gas regimes. 

We note that no correlation is found between $f_\text{d}$ and neutral gas in the SMC, and hence no calibration relation can be extracted, perhaps due to the lack of a deep and full-coverage CO map. Thus, we take the median value of $f_\text{d}=0.1$ to map the extinction in the SMC. In diffuse regions where it is expected to have deviations from $f_\text{d}=0.1$, changing $f_\text{d}$ from 0.1 to 0.3 affects (increases) the intrinsic H${\alpha}$ and hence the thermal fraction by less than 22 per cent as in these regions $\tau_\text{eff}<1$.

\setcounter{table}{3}
\begin{table*}
  \renewcommand{\arraystretch}{1.35}
 \caption{Dust surface density ($\Sigma_\text{d}$), visual extinction ($A_\text{V}$), the fraction of dust attenuating H$\alpha$ emission ($f_\text{d}$), and thermal fractions at 0.166\,GHz ($f^\text{0.166\,GHz}_\text{th}$) and 1.4\,GHz ($f^\text{1.4\,GHz}_\text{th}$) obtained for \ion{H}{ii} regions in the LMC. The Balmer-line-decrement ratio ($F_{\text{H}\alpha}$/$F_{\text{H}\beta}$), and the size of the sources  are taken from \protect\cite{Caplan85}. }
     \label{tab:lmc_hII}
 \begin{tabular}{llllllllll}
  \hline
  Source & R.A (J2000) & Dec (J2000)  & Radius & $F_{\text{H}\alpha}$/$F_{\text{H}\beta}$   &  $\Sigma_\text{d}$ &$A_\text{V}$ & $f_\text{d}$  & $f^\text{0.166\,GHz}_\text{th}$ & $f^\text{1.4\,GHz}_\text{th}$  \\
    &  (deg) & (deg) &  (arcmin)  &     &  ($10^{-5}$\,$\text{g}$\,$\text{cm}^{-2}$) & &   & (\%)  & (\%)     \\
  \hline

\ N77E & 72.444 & -69.199 & 4.89 & 3.52 & 16.73 $\pm$ 0.69 & 0.46 & 0.14 $\pm$ 0.03 & 20.4 $\pm$  3.5 & 80.2 $\pm$  10.1 \\
\ N4AB & 72.977 & -66.907 & 4.89 & 3.24 & 19.62 $\pm$ 0.61 & 0.28 & 0.07 $\pm$ 0.02 & 46.1 $\pm$  8.7 & 69.6 $\pm$  9 \\
\ N79AB & 72.931 & -69.398 & 4.89 & 3.78 & 17.63 $\pm$ 0.52 & 0.61 & 0.17 $\pm$ 0.03 & 28.2 $\pm$  4.2 & 64.3 $\pm$  8.1 \\
\ N79DE & 73.061 & -69.345 & 4.89 & 3.50 & 22.91 $\pm$ 0.49 & 0.45 & 0.10 $\pm$ 0.02 & 36.3 $\pm$  5.5 & 94.8 $\pm$  12.1 \\
\ N81 & 73.195 & -69.224 & 4.89 & 3.65 & 14.72 $\pm$ 0.66 & 0.54 & 0.18 $\pm$ 0.03 & - & 127.7 $\pm$  16.8 \\
\ N83A & 73.494 & -69.200 & 4.89 & 3.77 & 24.77 $\pm$ 0.79 & 0.61 & 0.12 $\pm$ 0.02 & 55.8 $\pm$  8.4 & 79.3 $\pm$  10.2 \\
\ N83north & 73.596 & -69.154 & 2.00 & 3.68 & 19.24 $\pm$ 0.55 & 0.56 & 0.14 $\pm$ 0.02 & 98.2 $\pm$  16 & 83.8 $\pm$  10.6 \\
\ N11F & 74.164 & -66.522 & 4.89 & 3.17 & 18.99 $\pm$ 0.53 & 0.23 & 0.06 $\pm$ 0.02 & 126.9 $\pm$  20.4 & 117.1 $\pm$  15 \\
\ N11B & 74.201 & -66.404 & 4.89 & 3.36 & 30.22 $\pm$ 0.89 & 0.35 & 0.06 $\pm$ 0.01 & 137.2 $\pm$  20.3 & 101 $\pm$  13.2 \\
\ N91 & 74.291 & -68.415 & 4.89 & 3.42 & 14.70 $\pm$ 0.53 & 0.39 & 0.13 $\pm$ 0.03 & 17.1 $\pm$ 2.7 & 111.1 $\pm$  14.1 \\
\ N11CD & 74.428 & -66.461 & 4.89 & 3.21 & 28.33 $\pm$ 0.77 & 0.25 & 0.04 $\pm$ 0.02 & 130.5 $\pm$  19.3 & 103.8 $\pm$  13.4 \\
\ N11E & 74.535 & -66.362 & 4.89 & 3.36 & 22.68 $\pm$ 0.55 & 0.35 & 0.08 $\pm$ 0.02 & 87.4 $\pm$  14.9 & 82.8 $\pm$  10.5 \\
\ N23A & 76.222 & -68.056 & 4.89 & 3.28 & 8.56 $\pm$ 0.42 & 0.30 & 0.18 $\pm$ 0.05 & - & 89.6 $\pm$  11.1 \\
\ N103B & 77.178 & -68.764 & 4.89 & 3.24 & 9.92 $\pm$ 0.49 & 0.28 & 0.14 $\pm$ 0.04 & 11.4 $\pm$  1.6 & 44.5 $\pm$  5.5 \\
\ N103A & 77.344 & -68.768 & 2.00 & 3.17 & 11.06 $\pm$ 0.68 & 0.23 & 0.10 $\pm$ 0.04 & -  & 48.4 $\pm$  6.1 \\
\ N105A & 77.477 & -68.891 & 4.89 & 3.28 & 28.19 $\pm$ 0.78 & 0.30 & 0.05 $\pm$ 0.02 & 77.6 $\pm$  11.2 & 74.9 $\pm$  9.6 \\
\ N113south & 78.340 & -69.365 & 4.89 & 3.16 & 24.44 $\pm$ 0.58 & 0.22 & 0.04 $\pm$ 0.02 & 56.8 $\pm$  8.2 & 90.4 $\pm$  11.3 \\
\ N113north & 78.397 & -69.301 & 4.89 & 3.27 & 10.89 $\pm$ 0.50 & 0.29 & 0.14 $\pm$ 0.04 & 65.2 $\pm$  9.5 & 92.2 $\pm$  11.4 \\
\ N119 & 79.663 & -69.236 & 4.89 & 3.04 & 13.93 $\pm$ 0.34 & 0.13 & 0.05 $\pm$ 0.03 & 100.5 $\pm$  14.5 & 118.9 $\pm$  15 \\
\ N120(SNR) & 79.643 & -69.648 & 2.00 & 2.98 & 11.68 $\pm$ 0.37 & 0.09 & 0.04 $\pm$ 0.04 & - & 54.2 $\pm$  6.7 \\
\ N120AB & 79.751 & -69.637 & 2.00 & 3.06 & 26.49 $\pm$ 0.31 & 0.15 & 0.03 $\pm$ 0.02 & - & 71.3 $\pm$  8.8 \\
\ N120ABC & 79.817 & -69.653 & 4.89 & 3.14 & 17.48 $\pm$ 0.48 & 0.21 & 0.06 $\pm$ 0.03 & 40.6 $\pm$  5.7 & 75.6 $\pm$  9.3 \\
\ N44BC & 80.503 & -67.970 & 4.89 & 3.33 & 33.73 $\pm$ 0.73 & 0.34 & 0.05 $\pm$ 0.01 & 62.6 $\pm$  8.9 & 70.5 $\pm$  8.9 \\
\ N44I & 80.609 & -67.896 & 4.89 & 3.14 & 21.40 $\pm$ 0.56 & 0.21 & 0.05 $\pm$ 0.02 & 42.9 $\pm$  6.1 & 59 $\pm$  7.5 \\
\ N44D & 80.686 & -68.076 & 4.89 & 3.44 & 32.21 $\pm$ 0.79 & 0.41 & 0.06 $\pm$ 0.01 & - & 65.9 $\pm$  8.3 \\
\ N138A & 81.238 & -68.480 & 4.89 & 3.44 & 19.78 $\pm$ 0.60 & 0.41 & 0.10 $\pm$ 0.02 & - & 69.7 $\pm$  8.8 \\
\ N48B & 81.396 & -66.304 & 4.89 & 3.64 & 16.24 $\pm$ 0.67 & 0.53 & 0.16 $\pm$ 0.03 & 8.5 $\pm$  1.2 & 20.3 $\pm$  2.5 \\
\ N51D & 81.508 & -67.499 & 4.89 & 2.99 & 5.22 $\pm$ 0.39 & 0.10 & 0.09 $\pm$ 0.09 & 93.3 $\pm$  13.8 & 104.8 $\pm$  12.8 \\
\ N51E & 81.600 & -67.622 & 4.89 & 3.31 & 6.65 $\pm$ 0.50 & 0.32 & 0.24 $\pm$ 0.07 & 26.5 $\pm$  4.2 & 88.8 $\pm$  10.9 \\
\ N143 & 81.598 & -69.315 & 4.89 & 3.14 & 4.86 $\pm$ 0.35 & 0.21 & 0.21 $\pm$ 0.09 & 13.6 $\pm$  2.1 & 83.8 $\pm$  10.5 \\
\ N144AB & 81.637 & -68.825 & 4.89 & 3.47 & 17.67 $\pm$ 0.56 & 0.43 & 0.12 $\pm$ 0.03 & 114 $\pm$  16.9 & 128.3 $\pm$  16.6 \\
\ N144 & 81.635 & -68.863 & 4.89 & 3.46 & 9.43 $\pm$ 0.50 & 0.42 & 0.22 $\pm$ 0.05 & 94.2 $\pm$  14.1 & 123 $\pm$  15.6 \\
\ N144 & 81.745 & -68.825 & 4.89 & 3.45 & 9.97 $\pm$ 0.51 & 0.41 & 0.21 $\pm$ 0.05 & 102.9 $\pm$  15.3 & 130 $\pm$  16.6 \\
\ N51C & 81.901 & -67.455 & 4.89 & 3.11 & 12.37 $\pm$ 0.54 & 0.18 & 0.07 $\pm$ 0.04 & 85.4 $\pm$  12.8 & 110.8 $\pm$  13.8 \\
\ N51A & 82.002 & -67.424 & 4.89 & 3.21 & 16.21 $\pm$ 0.52 & 0.25 & 0.08 $\pm$ 0.03 & 49.7 $\pm$  7.4 & 104.6 $\pm$  13.2 \\
\ N206 & 82.400 & -71.001 & 4.89 & 3.51 & 5.10 $\pm$ 0.30 & 0.45 & 0.44 $\pm$ 0.09 & - & 87.9 $\pm$  10.9 \\
\ N206 & 82.587 & -71.101 & 4.89 & 3.46 & 17.00 $\pm$ 0.60 & 0.42 & 0.12 $\pm$ 0.03 & 68.9 $\pm$  10.3 & 81.7 $\pm$  10.3 \\
\ N206A & 82.805 & -71.069 & 4.89 & 3.46 & 21.29 $\pm$ 0.59 & 0.42 & 0.10 $\pm$ 0.02 & 81.5 $\pm$  11.7 & 83.5 $\pm$  10.5 \\
\ N148 & 82.937 & -68.532 & 4.89 & 3.34 & 25.23 $\pm$ 0.63 & 0.34 & 0.07 $\pm$ 0.02 & - & 78.5 $\pm$  10.2 \\
\ N55north & 83.024 & -66.411 & 4.89 & 3.23 & 16.02 $\pm$ 0.53 & 0.27 & 0.08 $\pm$ 0.03 & 119.4 $\pm$  22.7 & 128.4 $\pm$  16.8 \\
\ N55A & 83.118 & -66.467 & 4.89 & 3.21 & 14.59 $\pm$ 0.41 & 0.25 & 0.09 $\pm$ 0.03 & 147.6 $\pm$  25.7 & 133.5 $\pm$  17.3 \\
\ N57A & 83.099 & -67.694 & 4.89 & 3.41 & 31.55 $\pm$ 0.58 & 0.39 & 0.06 $\pm$ 0.01 & 65.2 $\pm$  9.8 & 124.4 $\pm$  16.4 \\
\ N57C & 83.290 & -67.713 & 4.89 & 3.42 & 18.64 $\pm$ 0.48 & 0.39 & 0.11 $\pm$ 0.02 & - & 88.6 $\pm$  11.4 \\
\ Filaments & 83.353 & -67.453 & 4.89 & 3.00 & 11.08 $\pm$ 0.49 & 0.10 & 0.05 $\pm$ 0.04 & 37.4 $\pm$  5.6 & 69.7 $\pm$  8.7 \\

  \hline
 \end{tabular}
  \vspace{1ex}
     
\end{table*}

\setcounter{table}{3}
\begin{table*}
  \renewcommand{\arraystretch}{1.35}
 \caption{Continued.}
     \label{tab:lmc_hII}
 \begin{tabular}{llllllllll}
  \hline
  Source & R.A (J2000) & Dec (J2000)  & Radius & $F_{H\alpha}$/$F_{H\beta}$   &  $\Sigma_\text{d}$ &$A_\text{V}$ & $f_\text{d}$  & $f^\text{0.166\,GHz}_\text{th}$ & $f^\text{1.4\,GHz}_\text{th}$  \\
    &  (deg) & (deg) &  (arcmin)  &     &  ($10^{-5}$\,$\text{g}$\,$\text{cm}$$^{-2}$) & &   & (\%)  & (\%)     \\
  \hline
\ N154south & 83.725 & -69.798 & 4.89 & 3.20 & 9.70 $\pm$ 0.48 & 0.25 & 0.13 $\pm$ 0.05 & 37 $\pm$  5.2 & 66.7 $\pm$  8.2 \\
\ N63A & 83.892 & -66.031 & 4.89 & 3.19 & 8.61 $\pm$ 0.44 & 0.24 & 0.14 $\pm$ 0.05 & - & 18.5 $\pm$  2.3 \\
\ N59A & 83.840 & -67.585 & 4.89 & 3.84 & 24.33 $\pm$ 0.50 & 0.65 & 0.13 $\pm$ 0.02 & 97.9 $\pm$  14.2 & 75.5 $\pm$  9.7 \\
\ N59B & 83.986 & -67.577 & 4.89 & 3.48 & 21.81 $\pm$ 0.59 & 0.43 & 0.10 $\pm$ 0.02 & 66.8 $\pm$  9.7 & 75.6 $\pm$  9.7 \\
\ N157 & 83.952 & -69.210 & 4.89 & 3.83 & 32.62 $\pm$ 0.78 & 0.64 & 0.10 $\pm$ 0.01 & 29.2 $\pm$  4.2 & 52.5 $\pm$  6.8 \\
\ N157 & 84.086 & -69.205 & 4.89 & 3.70 & 18.51 $\pm$ 0.45 & 0.57 & 0.15 $\pm$ 0.02 & 29.3 $\pm$  4.1 & 54.2 $\pm$  6.9 \\
\ N154A & 83.973 & -69.646 & 4.89 & 3.35 & 23.51 $\pm$ 0.68 & 0.35 & 0.07 $\pm$ 0.02 & 62.6 $\pm$  8.9 & 86.6 $\pm$  11 \\
\ N157 & 84.177 & -69.216 & 4.89 & 3.69 & 23.89 $\pm$ 0.73 & 0.56 & 0.12 $\pm$ 0.02 & 31.9 $\pm$  4.5 & 58.3 $\pm$  7.4 \\
\ N64AB & 84.274 & -66.361 & 4.89 & 3.23 & 18.44 $\pm$ 0.58 & 0.27 & 0.07 $\pm$ 0.02 & 64.6 $\pm$  11.5 & 112.6 $\pm$  14.6 \\
\ N157B & 84.449 & -69.165 & 4.89 & 4.22 & 41.44 $\pm$ 0.72 & 0.86 & 0.10 $\pm$ 0.01 & 43.1 $\pm$  6.8 & 47.6 $\pm$  6.6 \\
\ N157A-30dor & 84.660 & -69.088 & 4.89 & 4.57 & 44.22 $\pm$ 3.32 & 1.03 & 0.12 $\pm$ 0.01 & 110.4 $\pm$  21.9 & 65.6 $\pm$  9.8 \\
\ N158C & 84.769 & -69.506 & 4.89 & 3.62 & 35.33 $\pm$ 0.39 & 0.52 & 0.07 $\pm$ 0.01 & 73.2 $\pm$  10.7 & 87 $\pm$  11.4 \\
\ N159A & 84.908 & -69.771 & 1.06 & 4.01 & 71.59 $\pm$ 1.08 & 0.74 & 0.05 $\pm$ 0.01 & 47.2 $\pm$  6.6 & 37.7 $\pm$  4.8 \\
\ N16OAD & 84.926 & -69.646 & 4.89 & 3.73 & 51.23 $\pm$ 0.95 & 0.58 & 0.06 $\pm$ 0.01 & 6.6 $\pm$  1 & 65.4 $\pm$  8.7 \\
\ N158 & 84.955 & -69.441 & 4.89 & 3.34 & 17.32 $\pm$ 0.40 & 0.34 & 0.10 $\pm$ 0.03 & 43.4 $\pm$  6.1 & 70.2 $\pm$  8.8 \\
\ N159 & 84.988 & -69.752 & 4.89 & 4.09 & 53.88 $\pm$ 1.05 & 0.79 & 0.07 $\pm$ 0.01 & 28.9 $\pm$  4.1 & 42.5 $\pm$  5.5 \\
\ N159BD & 84.989 & -69.733 & 2.00 & 4.24 & 56.16 $\pm$ 1.25 & 0.87 & 0.08 $\pm$ 0.01 & - & 46.5 $\pm$  6.1 \\
\ N159C & 85.025 & -69.756 & 1.06 & 3.59 & 68.06 $\pm$ 0.14 & 0.50 & 0.04 $\pm$ 0.01 & 50.3 $\pm$  7.2 & 41.4 $\pm$  5.4 \\
\ N158A & 85.040 & -69.376 & 4.89 & 3.30 & 7.19 $\pm$ 0.45 & 0.32 & 0.22 $\pm$ 0.06 & 17.8 $\pm$  2.4 & 36.2 $\pm$  4.4 \\
\ N16OBCE & 85.046 & -69.667 & 4.89 & 3.47 & 33.74 $\pm$ 0.54 & 0.43 & 0.06 $\pm$ 0.01 & - & 89.7 $\pm$  11.7 \\
\ N158 & 85.151 & -69.412 & 4.89 & 3.25 & 7.68 $\pm$ 0.41 & 0.28 & 0.18 $\pm$ 0.06 & 21.3 $\pm$  2.9 & 41.6 $\pm$  5 \\
\ N175 & 85.180 & -70.041 & 4.89 & 3.52 & 11.55 $\pm$ 0.45 & 0.46 & 0.20 $\pm$ 0.04 & 12.4 $\pm$  1.8 & 28.8 $\pm$  3.6 \\
\ Filaments & 85.331 & -69.084 & 4.89 & 3.37 & 10.20 $\pm$ 0.57 & 0.36 & 0.18 $\pm$ 0.04 & 20.5 $\pm$  2.8 & 48.8 $\pm$  6 \\
\ N214C & 85.410 & -71.337 & 4.89 & 3.29 & 10.72 $\pm$ 0.43 & 0.31 & 0.14 $\pm$ 0.04 & 10.2 $\pm$ 1.5 & 75.3 $\pm$  9.1 \\
\ NGC2100 & 85.526 & -69.211 & 4.89 & 3.19 & 9.96 $\pm$ 0.64 & 0.24 & 0.12 $\pm$ 0.04 & 17.5 $\pm$  2.4 & 47.5 $\pm$  5.8 \\
\ N164 & 85.648 & -69.068 & 4.89 & 3.83 & 20.90 $\pm$ 0.38 & 0.64 & 0.15 $\pm$ 0.02 & 37.1 $\pm$  5.2 & 62.7 $\pm$  7.8 \\
\ N165 & 85.723 & -68.947 & 4.89 & 3.43 & 12.88 $\pm$ 0.55 & 0.40 & 0.16 $\pm$ 0.04 & - & 37.1 $\pm$  4.6 \\
\ N163 & 85.770 & -69.759 & 4.89 & 3.89 & 23.30 $\pm$ 0.45 & 0.68 & 0.15 $\pm$ 0.02 & 10.2 $\pm$  1.4 & 49 $\pm$  6.1 \\
\ N74A & 86.427 & -67.149 & 4.89 & 3.04 & 9.44 $\pm$ 0.53 & 0.13 & 0.07 $\pm$ 0.05 & - & 126.8 $\pm$  23 \\
\ N180AB & 87.215 & -70.071 & 4.89 & 3.30 & 23.07 $\pm$ 0.56 & 0.32 & 0.07 $\pm$ 0.02 & 63.9 $\pm$  9.3 & 106.1 $\pm$  13.4 \\

  \hline
 \end{tabular}
  \vspace{1ex}
     
\end{table*}
\bigskip

\begin{table*}
  \renewcommand{\arraystretch}{1.35}
 \caption{Dust surface density ($\Sigma_\text{d}$), visual extinction ($A_\text{V}$), the fraction of dust attenuating H$\alpha$ emission ($f_\text{d}$), and thermal fractions at 0.166\,GHz ($f^\text{0.166\,GHz}_\text{th}$) and 1.4\,GHz ($f^\text{1.4\,GHz}_\text{th}$) obtained for \ion{H}{ii} regions in the SMC. The Balmer-line-decrement ratio ($F_{\text{H}\alpha}$/$F_{\text{H}\beta}$), and the size of the sources are taken from \protect\cite{Caplan85}.}
     \label{tab:smc_hII}
 \begin{tabular}{llllllllll}
  \hline
  Source & R.A (J2000) & Dec (J2000)  & Radius & $F_{\text{H}\alpha}$/$F_{\text{H}\beta}$   &  $\Sigma_\text{d}$ &$A_\text{V}$ & $f_\text{d}$  & $f^\text{0.166\,GHz}_\text{th}$ & $f^\text{1.4\,GHz}_\text{th}$  \\
    &  (deg) & (deg) &  (arcmin)  &     &  ($10^{-5}$\,$\text{g}$\,$\text{cm}^{-2}$) & &   & (\%)  & (\%)     \\

  \hline
\ N13AB & 11.347 & -73.380 & 4.89 & 3.51 & 15.60 $\pm$ 0.30 & 0.39 & 0.15 $\pm$ 0.05 & 27.5 $\pm$  5.8 & 59.3 $\pm$  6.3 \\
\ N12Ba & 11.387 & -73.080 & 4.89 & 3.42 & 5.63 $\pm$ 0.48 & 0.34 & 0.37 $\pm$ 0.15 & 39.1 $\pm$  8.2 & 87.6 $\pm$  10.1 \\
\ N12-Ab & 11.629 & -73.101 & 4.89 & 3.34 & 11.86 $\pm$ 0.39 & 0.29 & 0.15 $\pm$ 0.07 & - & 71.5 $\pm$  7.4 \\
\ N19 & 11.932 & -73.134 & 4.89 & 3.46 & 15.37 $\pm$ 0.14 & 0.36 & 0.14 $\pm$ 0.05 & - & 56.1 $\pm$  5.3 \\
\ N22 & 12.004 & -73.270 & 2.00 & 3.46 & 24.31 $\pm$ 0.17 & 0.36 & 0.09 $\pm$ 0.03 & 53 $\pm$  9.6 & 75.4 $\pm$  7.1 \\
\ N25-N26 & 12.029 & -73.237 & 1.06 & 3.78 & 31.18 $\pm$ 0.44 & 0.53 & 0.10 $\pm$ 0.03 & 51.5 $\pm$  9.6 & 79.8 $\pm$  7.6 \\
\ N24 & 12.040 & -73.329 & 2.00 & 3.15 & 15.57 $\pm$ 0.27 & 0.18 & 0.07 $\pm$ 0.05 & - & 51 $\pm$  5.1 \\
\ N27 & 12.088 & -73.099 & 2.00 & 3.43 & 37.23 $\pm$ 0.36 & 0.34 & 0.06 $\pm$ 0.02 & - & 84.7 $\pm$  7.8 \\
\ N28 & 12.148 & -73.254 & 2.00 & 3.29 & 10.18 $\pm$ 0.20 & 0.26 & 0.16 $\pm$ 0.08 & 40.3 $\pm$  7.3 & 72.6 $\pm$  7.2 \\
\ N30 & 12.240 & -73.129 & 4.89 & 3.45 & 17.92 $\pm$ 0.07 & 0.35 & 0.12 $\pm$ 0.04 & 31.4 $\pm$  5.6 & 79.7 $\pm$  7.7 \\
\ N36cg & 12.616 & -72.884 & 4.89 & 3.16 & 6.26 $\pm$ 0.45 & 0.19 & 0.18 $\pm$ 0.13 & 44.5 $\pm$  8.3 & 92.7 $\pm$  10 \\
\ N37dg & 12.687 & -72.780 & 4.89 & 3.40 & 2.88 $\pm$ 0.28 & 0.33 & 0.69 $\pm$ 0.29 & 23.6 $\pm$  4.4 & 81.5 $\pm$  9.2 \\
\ N66i & 14.619 & -72.195 & 4.89 & 3.08 & 9.97 $\pm$ 0.22 & 0.14 & 0.09 $\pm$ 0.08 & - & 91.6 $\pm$  9.5 \\
\ N66ii & 14.771 & -72.178 & 4.89 & 3.08 & 11.84 $\pm$ 0.28 & 0.14 & 0.07 $\pm$ 0.07 & 20.5 $\pm$  3.5 & 78.2 $\pm$  8 \\
\ N66iii & 14.771 & -72.178 & 1.06 & 2.93 & 19.13 $\pm$ 0.07 & 0.05 & 0.01 $\pm$ 0.04 & - & 76.6 $\pm$  7.7 \\
\ N66iv & 14.858 & -72.165 & 4.89 & 3.13 & 8.09 $\pm$ 0.30 & 0.17 & 0.13 $\pm$ 0.10 & 32.7 $\pm$  5.6 & 68.2 $\pm$  7.1 \\
\ N76 & 15.902 & -72.057 & 4.89 & 3.04 & 12.01 $\pm$ 0.27 & 0.12 & 0.06 $\pm$ 0.07 & 29.5 $\pm$  5.2 & 65.3 $\pm$  6.6 \\
\ N78ii & 16.285 & -71.993 & 4.89 & 3.35 & 6.10 $\pm$ 0.47 & 0.30 & 0.30 $\pm$ 0.13 & - & 66.1 $\pm$  7.2 \\
\ N80 & 17.106 & -71.999 & 4.89 & 2.96 & 3.42 $\pm$ 0.45 & 0.07 & 0.12 $\pm$ 0.24 & 30.3 $\pm$  7 & 79.8 $\pm$  9.7 \\
\ N81 & 17.284 & -73.202 & 4.89 & 2.97 & 1.07 $\pm$ 0.13 & 0.07 & 0.41 $\pm$ 0.76 & - & 129.8 $\pm$  24.6 \\
\ N83A & 18.451 & -73.301 & 2.00 & 3.04 & 13.29 $\pm$ 0.26 & 0.12 & 0.05 $\pm$ 0.06 & - & 115.4 $\pm$  11.9 \\
\ N83 & 18.456 & -73.288 & 4.89 & 3.28 & 13.07 $\pm$ 0.38 & 0.26 & 0.12 $\pm$ 0.06 & - & 111.8 $\pm$  11.7 \\
\ N84C & 18.567 & -73.266 & 1.06 & 3.91 & 20.59 $\pm$ 0.43 & 0.59 & 0.18 $\pm$ 0.04 & - & 105.4 $\pm$  10.4 \\
\ N84AB & 18.681 & -73.322 & 4.89 & 3.19 & 11.26 $\pm$ 0.36 & 0.21 & 0.11 $\pm$ 0.07 & - & 117.7 $\pm$  12.8 \\
\ N85fg & 18.931 & -73.334 & 4.89 & 3.17 & 1.65 $\pm$ 0.30 & 0.25 & 0.72 $\pm$ 0.51 & - & 84 $\pm$  11.2 \\
  \hline
 \end{tabular}
  \vspace{1ex}
     
\end{table*}

\begin{table*}
  \renewcommand{\arraystretch}{1.35}
 \caption{ Measured fluxes $F_{\text{H}\alpha}$ and $F_{\text{H}\beta}$ of the MUSE SNR fields and their corresponding central wavelengths (CW). Also shown are dust surface density ($\Sigma_\text{d}$), visual extinction ($A_\text{V}$), and fraction of dust attenuating H$\alpha$ emission ($f_\text{d}$).}
 \begin{tabular}{llllllllll}
  \hline
   Field & CW$_{\rm H\alpha}$ &  CW$_{\rm H\beta}$  & $F_{\text{H}\alpha}$ & $F_{\text{H}\beta}$  &  $\Sigma_\text{d}$ &$A_\text{V}$ & $f_\text{d}$  \\ 
    &  ({\AA})&  ({\AA}) & ($10^{-12}$\,\text{erg}\,\text{s$^{-1}$}\,$\text{cm}^{-2})$ & ($10^{-12}$\,erg\,s$^{-1}$\,\text{cm}$^{-2}$) &  ($10^{-5}$\,$\text{g}$\,$\text{cm}^{-2}$) \\

  \hline
  
  \ LMC/N49 & 6567.48 & 4864.98 & 50.9 $\pm$ 2.5 & 11.9 $\pm$ 0.6 & 16.6 $\pm$ 1.2 & 0.9 $\pm$ 0.5 & 0.3 $\pm$ 0.1 \\
\ LMC/SNR 0540-69.3 & 6568.74 & 4866.24 & 4.4 $\pm$ 0.2 & 0.9 $\pm$ 0.1 & 16.8 $\pm$ 0.1 & 1.2 $\pm$ 0.5 & 0.3 $\pm$ 0.1 \\
\ SMC/1E0102–7219 & 6566.07 & 4863.57 & 3.3 $\pm$ 0.1 & 1.1 $\pm$ 0.1 & 6.0 $\pm$ 0.5 & 0.2 $\pm$ 0.1 & 0.2 $\pm$ 0.1 \\

  \hline
   \label{tab:muse_res}
 \end{tabular}
\end{table*}

\subsubsection{Extinction maps and de-reddening the H$\alpha$ emission}

Fig.~\ref{fig:extinction_map} (bottom row) shows the final maps of the effective extinction $\tau_\text{eff}$. In the LMC, we find on the average $\tau_\text{eff} = 0.08 \pm 0.01$ compared to $\tau_\text{eff} = 0.05$ $\pm$ 0.02 in the SMC. Maximum extinction occurs in the centre of 30~Dor with $\tau_\text{eff} = 0.75 \pm 0.11 $. We use the extinction map to de-redden the H$\alpha$ map and obtain the intrinsic H$\alpha$  intensity $I_{0}$ using $ I = \text{I}_\text{0}$ $e^{ -\tau_\text{eff} }$. 

Adopting the extinction map $\tau_\text{eff}$ to de-redden H$\alpha$ yields $\simeq$ 20 per cent obscuration for the LMC. This is higher than that in the SMC ($\simeq$ 4 per cent) and similar to the correction factor for the dust obscuration reported for M\,33 ($\simeq$ 13 per cent, \citealt{tab_2007}). Integration of the dust corrected H$\alpha$ map out to a radius of 5$\degr$ yields an intrinsic H$\alpha$ luminosity ${L}_{\text{H}\alpha}= (3.67 \pm 0.47) \times 10^{40}\,\text{erg}\,\text{s}^{-1}$ for the LMC. The SMC has a luminosity of ${L}_{\text{H}\alpha} = (6.87 \pm 1.06) \times 10^{39}\,\text{erg}\,\text{s}^{-1}$ out to a radius of 3.5$\degr$.

\section{Separation of Thermal and Non-thermal Radio Emission}  \label{sec:trt}

Separating the thermal and non-thermal radio emission is critical to study cosmic-ray energy loss mechanisms and the total magnetic fields. A pure non-thermal synchrotron spectral index map  can be obtained using the TRT method, in which the thermal (free-free) emission is obtained using the de-reddened H$\alpha$ emission as its template \citep{tab_2013}.

\subsection{From H$\alpha$ emission to radio free-free emission}

The H$\alpha$ intensity is related to the emission measure (EM) depending on the transparency of the ISM to Lyman continuum photons. In case of ( $\tau_{{Ly}\alpha}$ $\gg$ 1 ) condition \footnote{Optically thin condition is discussed in Section~\ref{ssec:recom}.} which is usually denoted as Case B recombination \citep{Osterbrock}, $I_{\text{H}\alpha}$ in erg\,cm$^{-2}$\,s$^{-1}$\,sr$^{-1}$ units is given by \citep{vallas}:  

\begin{equation}
\label{eqn:em_ha}
I_{\text{H}\alpha} = 9.41 \: \times \: 10^{-8} \: T_\text{e}^{-1.017} \: 10^{\small(\frac {-0.029} {T_\text{e}})} \: EM
\end{equation}

\noindent 
where $T_\text{e}$ is the electron temperature in units of 10$^{4}$\,K, and $EM$ in cm$^{-6}$\,pc. Equation~(\ref{eqn:em_ha}) is more than 1 per cent precise for the electron temperatures of 5,000 to 20,000\,K. \cite{oster} presents an expression of the free-free continuum optical thickness radiated from an ionized gas:

\begin{equation}
\tau_\text{c}= 0.08235 \: \times \: a \: T^{\small -1.35}_\text{e} \: \nu_\text{GHz} ^{-2.1} (1+0.08) \: EM
 \end{equation}

\noindent 
with $a \simeq 1$ and the continuum optical depth is corrected for singly ionized \ion{He} \text{atoms} by the factor (1+0.08). 

The brightness temperature $T_\text{b}$ of the radio continuum (free-free) emission in units of Kelvin is given by:

\begin{equation}
T_\text{b} = T_\text{e}(1 - e^{\tau_\text{c}})
 \end{equation}
 
\noindent 
The brightness temperature is converted to the radio flux density (Jy\,beam$^{-1}$) using a factor of 1.0, 78.2, and 919.3 for the LMC at 0.166\,GHz,  1.4\,GHz and 4.8\,GHz respectively. These factors are 1.2 and 88.4 for the SMC at 0.166\,GHz and 1.4\,GHz respectively. The electron temperature is the only free parameter in this conversion. Different methods of electron temperature determination \citep{dufour,dufour77,Vermeij,Peck97} indicate that 8000\,K$<{T}_\text{e}<$12000\,K in the MCs. However, we adopt $T_\text{e} = 10^{4}$\,K as this variation would not change the thermal emission by more than 20 per cent. 

\begin{figure*}
\centering

\includegraphics[width=0.8\columnwidth]{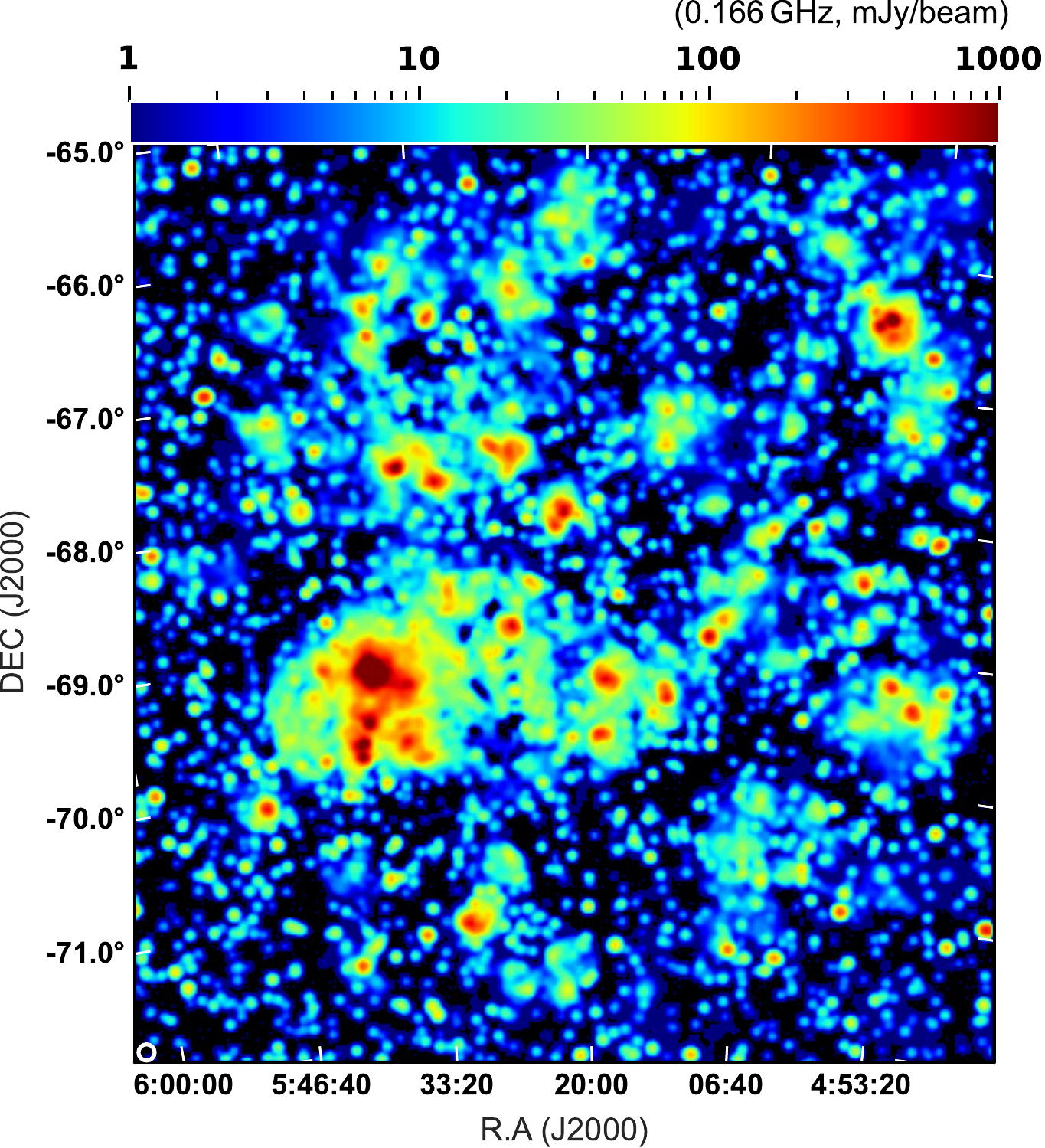}
\includegraphics[width=0.8\columnwidth]{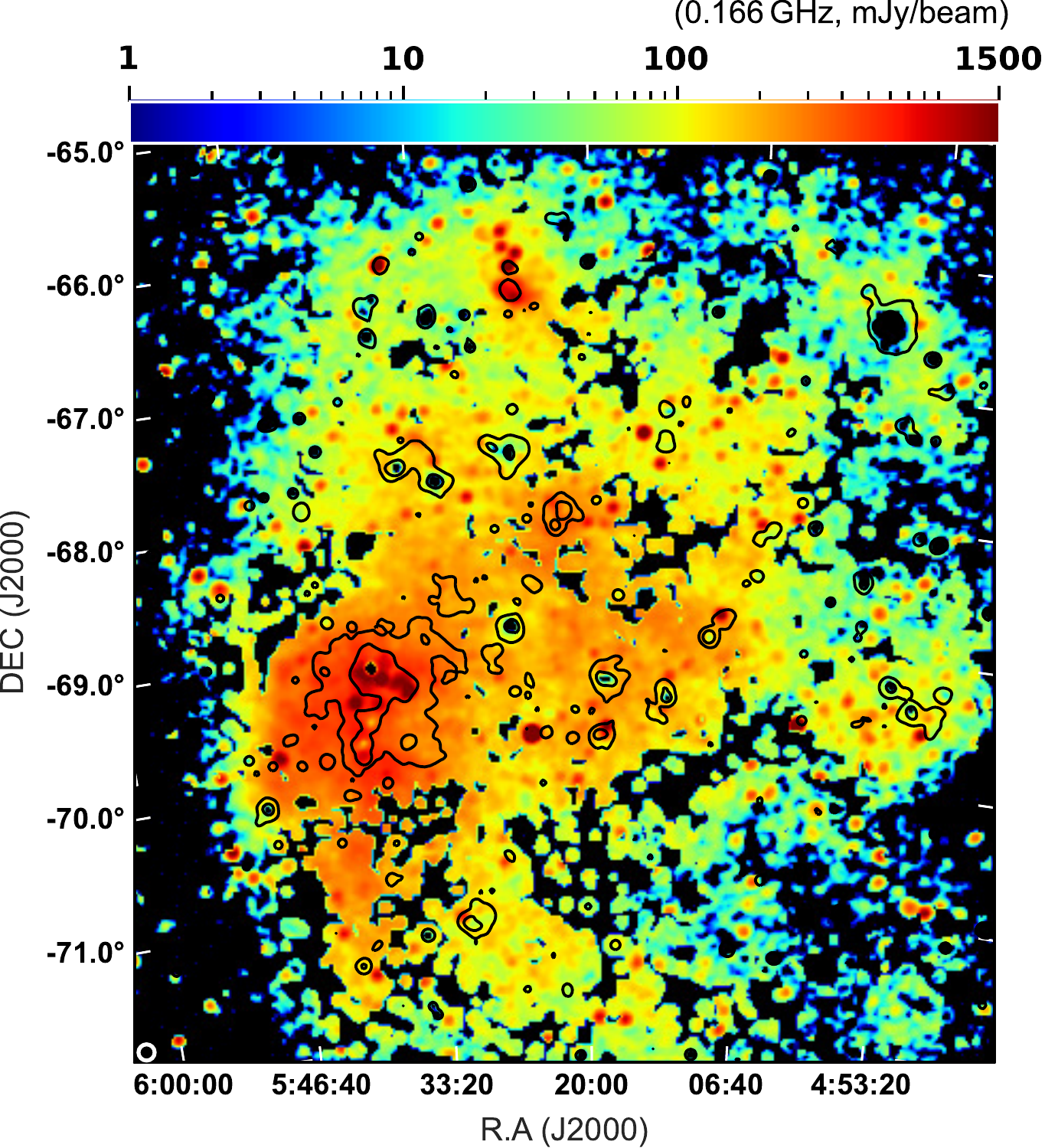}
\includegraphics[width=0.8\columnwidth]{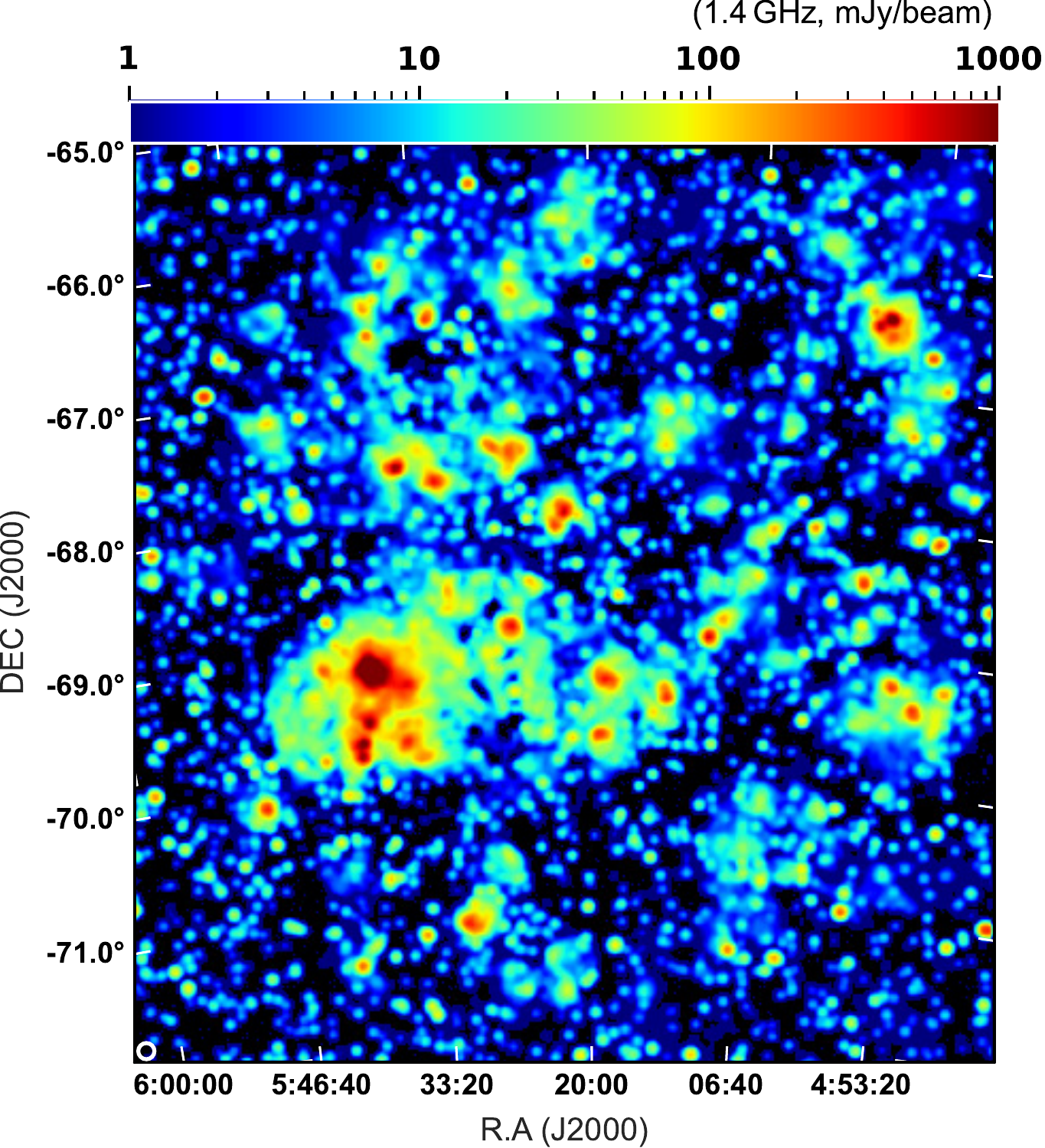}
\includegraphics[width=0.8\columnwidth]{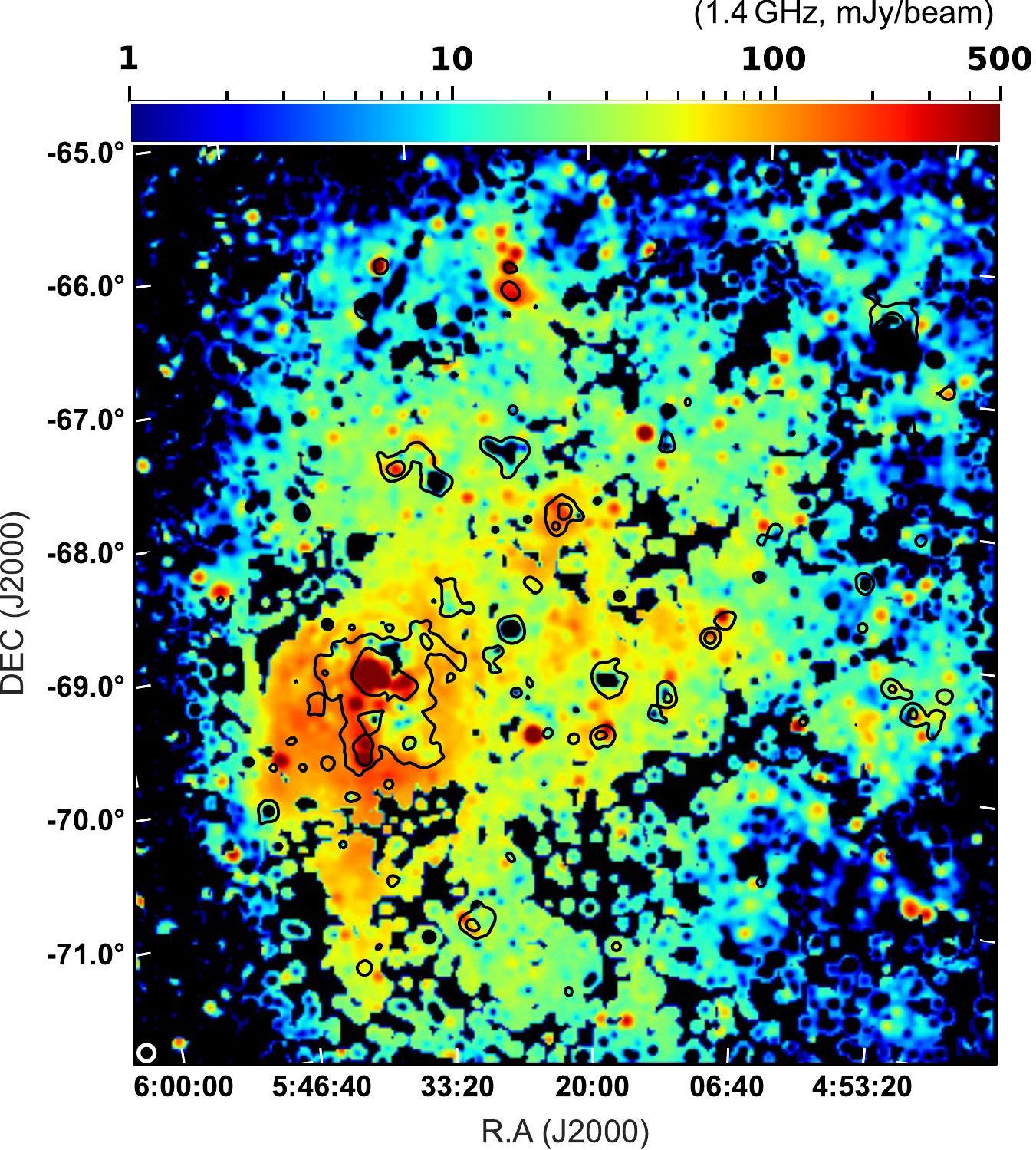}
\includegraphics[width=0.8\columnwidth]{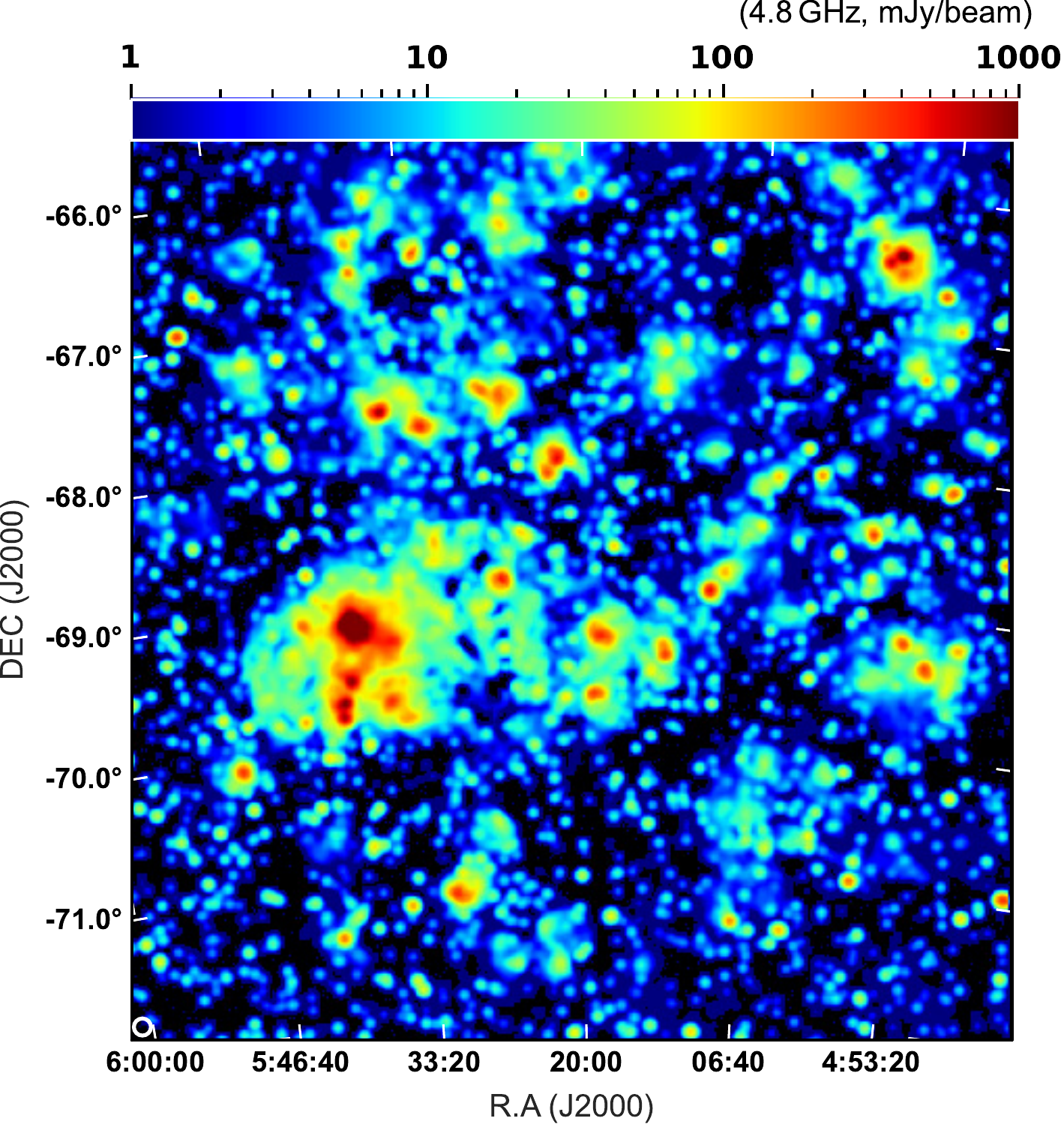}
\includegraphics[width=0.8\columnwidth]{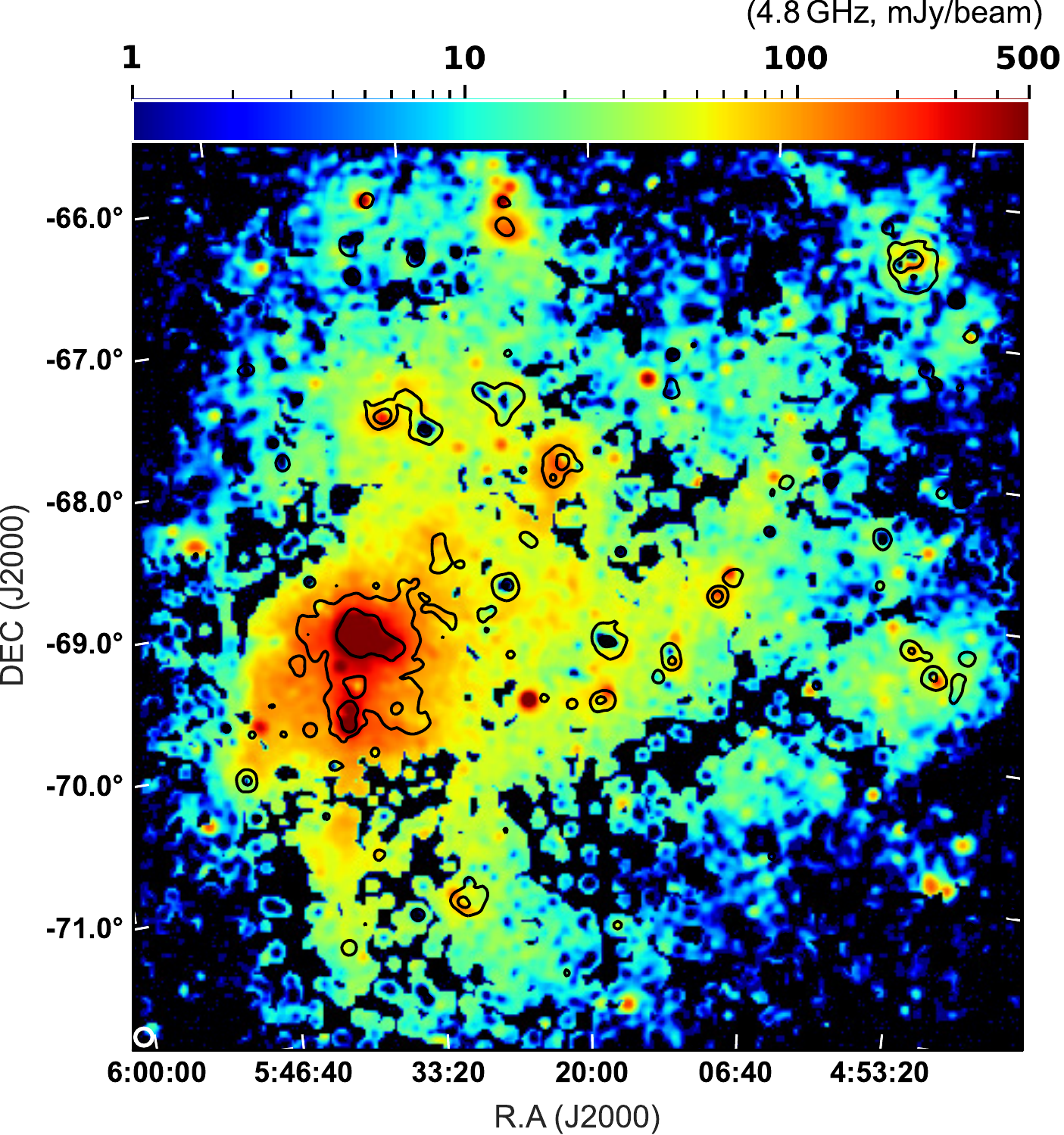}

 \caption{The LMC-- The thermal free-free (left) and the non-thermal synchrotron (right) emission maps at 0.166\,GHz (top), 1.4\,GHz (middle) and 4.8\,GHz (bottom). Colour bars show the intensities in units of mJy\,beam$^{-1}$ in log scale. The non-thermal maps are overlaid with contours of the thermal emission with levels of 40 and 100\,mJy\,beam$^{-1}$. The beam size of $221\arcsec$ is shown in the lower left corner of the maps. }
 
\label{fig:thermal_non_thermal_lmc}
\end{figure*}

\begin{figure*}
\centering

\includegraphics[width=0.8\columnwidth]{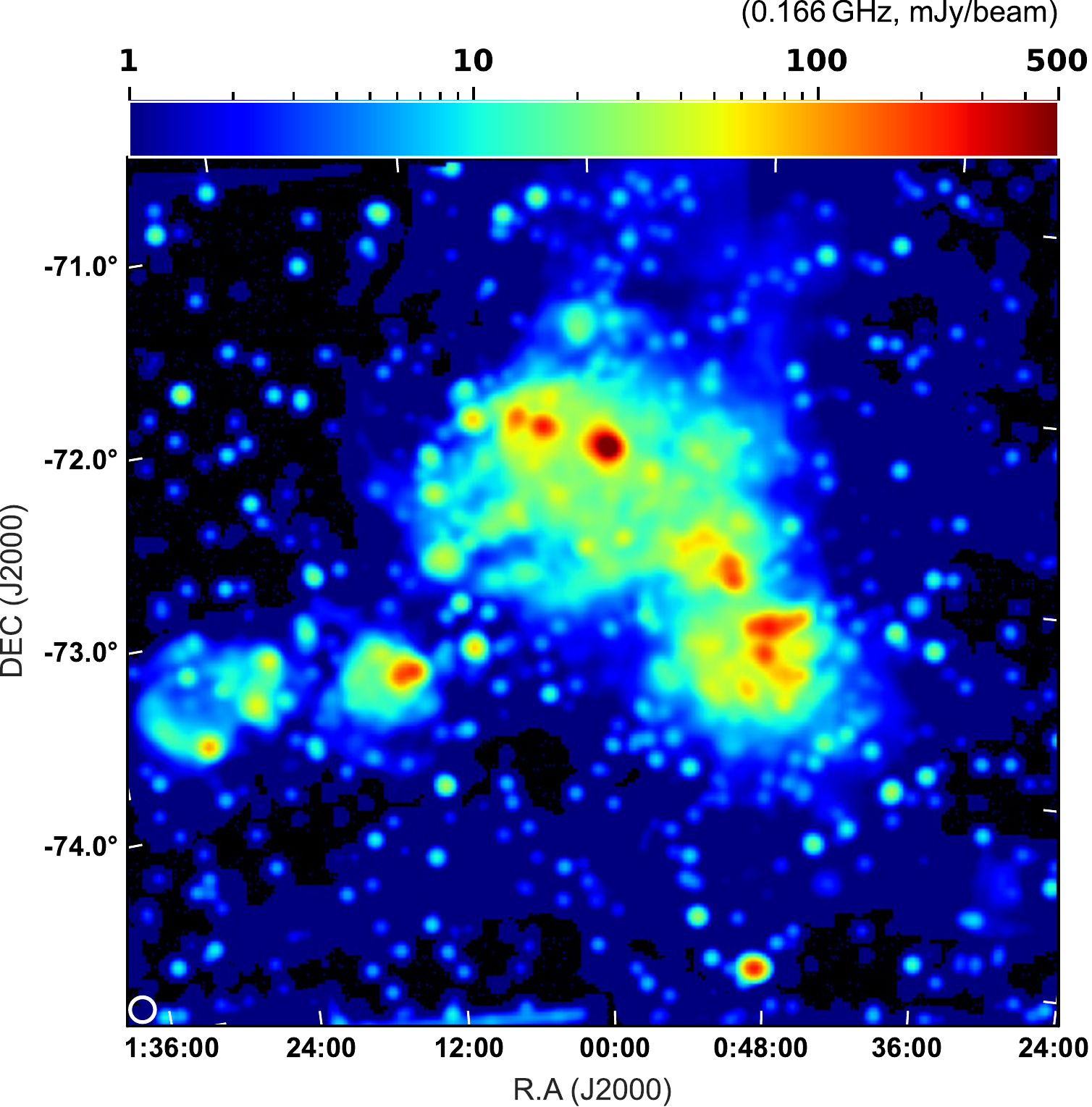}
\includegraphics[width=0.8\columnwidth]{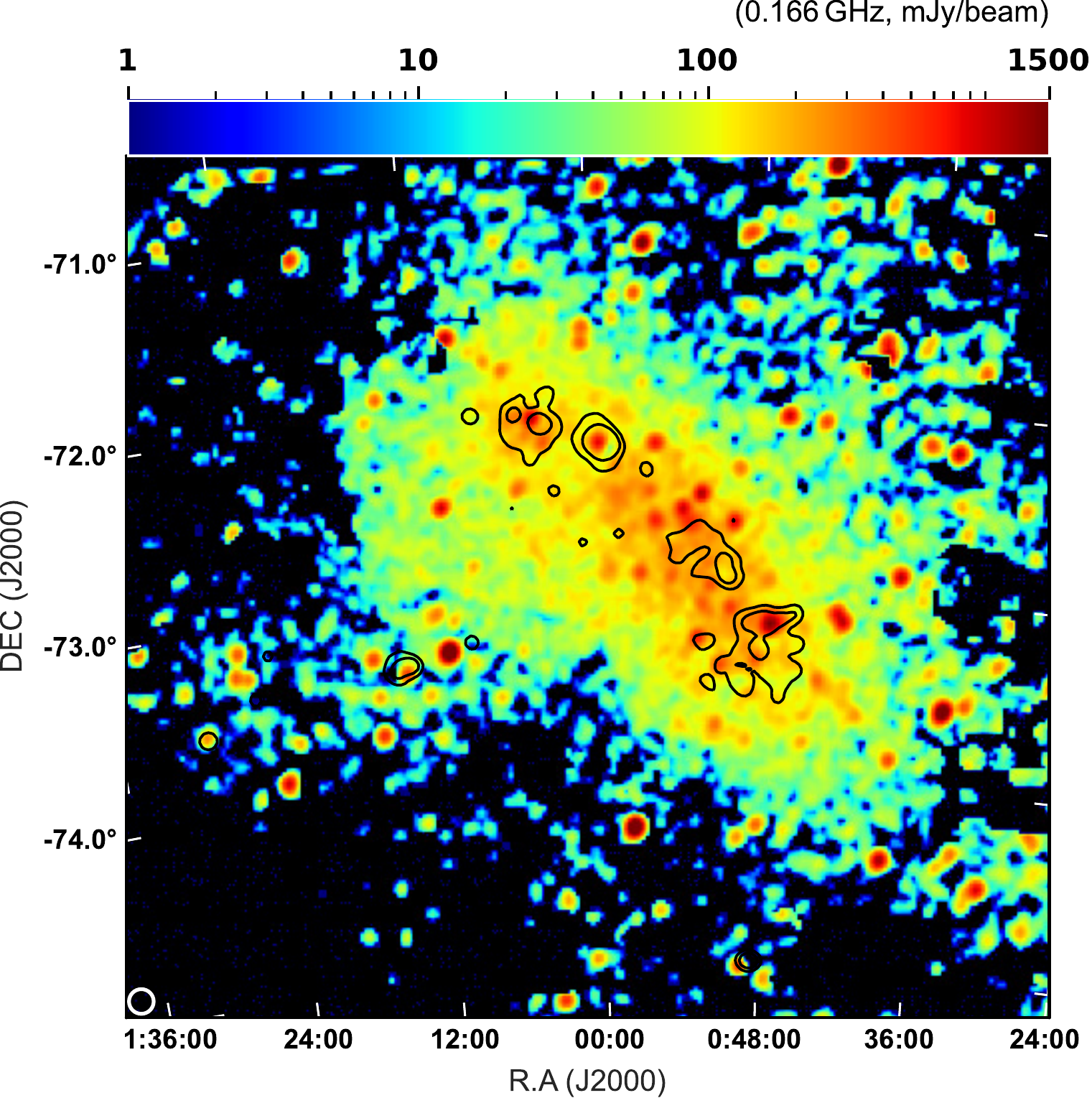}
\includegraphics[width=0.8\columnwidth]{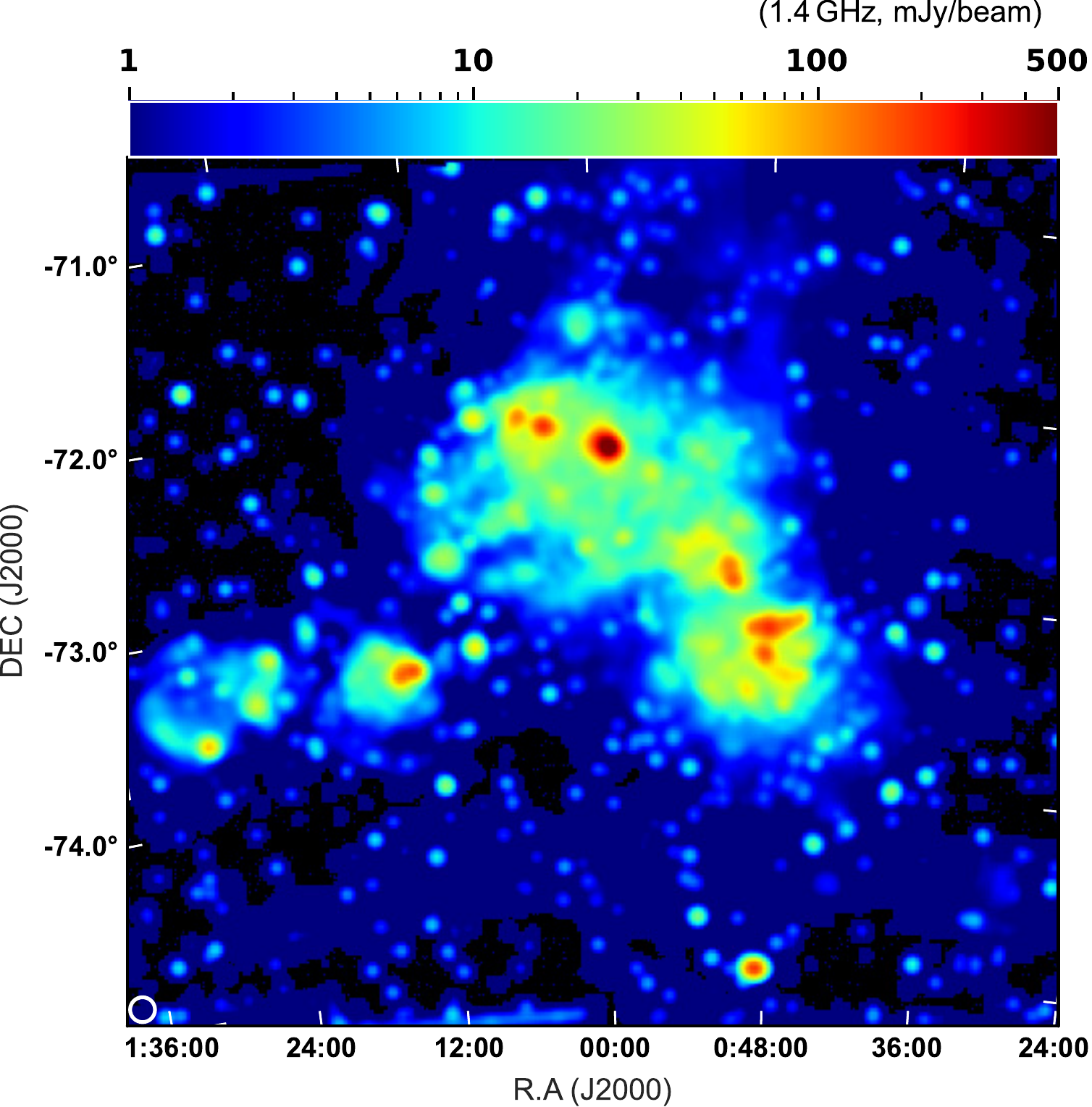}
\includegraphics[width=0.78\columnwidth]{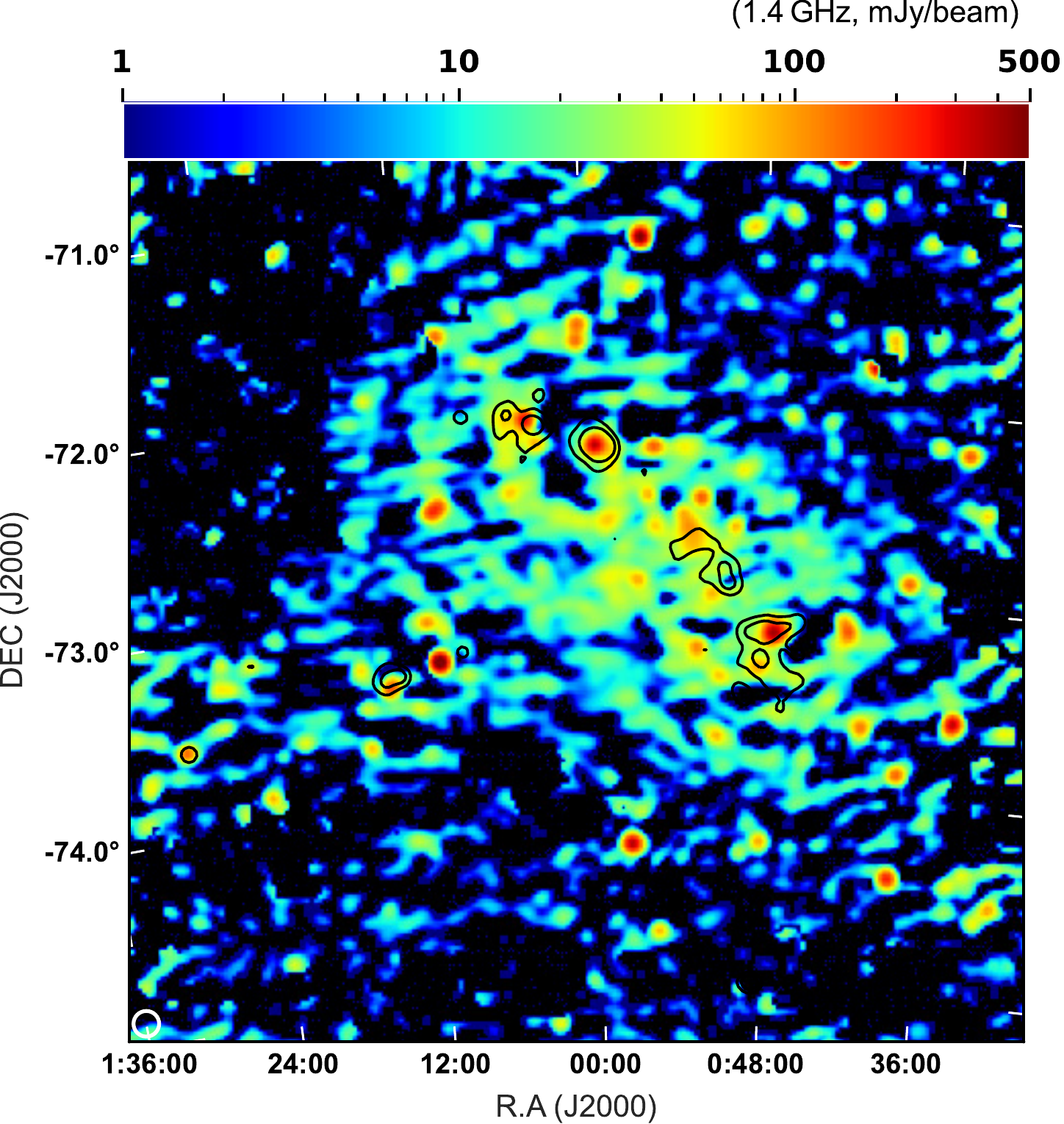}

\caption{The SMC-- The thermal free-free (left) and the non-thermal synchrotron (right) emission maps at 0.166\,GHz (top) and 1.4\,GHz (bottom). Colour bars show the intensities in units of mJy\,beam$^{-1}$ in log scale. The non-thermal maps are overlaid with contours of the thermal emission with levels of 40 and 100\,mJy\,beam$^{-1}$. The beam size of $235\arcsec$ is shown in the lower left corner of the maps. }

\label{fig:thermal_non_thermal_smc}
\end{figure*}

The free-free maps are presented in Figs.~\ref{fig:thermal_non_thermal_lmc},~\ref{fig:thermal_non_thermal_smc} (left panels) at frequencies of 0.166, 1.4, and 4.8\,GHz. The strongest thermal emission is visible in the MCs’ \ion{H}{ii} regions, in particular, the LMC \ion{H}{ii} region 30~Dor has a significant amount of free-free emission reaching 6\,Jy\,beam$^{-1}$ at 0.166\,GHz. The other LMC \ion{H}{ii} regions like N11 and N44 reach 0.5\,Jy\,beam$^{-1}$ as well as N66 in the SMC.

\subsection{Mapping synchrotron emission}  \label{ssec:nt_map}

The synchrotron emission maps are obtained at frequencies of 0.166, 1.4, and 4.8\,GHz by subtracting the thermal map from the observed RC map (Fig.~\ref{fig:thermal_non_thermal_lmc} and~\ref{fig:thermal_non_thermal_smc}). Strong synchrotron emission emerges from the brightest \ion{H}{ii} complex of the LMC, 30~Dor, possibly due to strong magnetic field and/or energetic CREs in this massive star-forming region. This source is the most prominent feature of the synchrotron emission from 0.166\,GHz to 4.8\,GHz. Extended diffuse synchrotron emission is also clearly visible around 30~Dor. At 1.4\,GHz, this region is affected by observational artefacts \citep{Huges2007} reducing the sensitivity to detect the diffuse emission at distances about 0.6\,kpc from the centre of 30~Dor in the north. In the SMC, an extended synchrotron emission is found along its bar. 
The synchrotron emission is also strong in other \ion{H}{ii} regions, which are classically considered as thermal sources such as \ion{H}{ii} N44 and N11 of the LMC, N66 and N19 in the SMC as indicated by contours of the thermal emission overlaid on the non-thermal maps in Figs.~\ref{fig:thermal_non_thermal_lmc},~\ref{fig:thermal_non_thermal_smc}. This non-thermal intensity is probably due to SNRs and strong non-thermal shocks in these young, massive star clusters and associations. Similar non-thermal emission was previously shown to exist in the \ion{H}{ii} regions of M33 \citep{tab_2007}. The SNRs as an intense source of energetic CREs show a strong magnetic field as previously indicated by \cite{klein_lmc}. The known SNRs \citep{bozzetto_2017,smc_snr} are well-matched with the brightest sources of our non-thermal maps.

\noindent 
\begin{figure*}
\centering
\includegraphics[width=0.8\columnwidth]{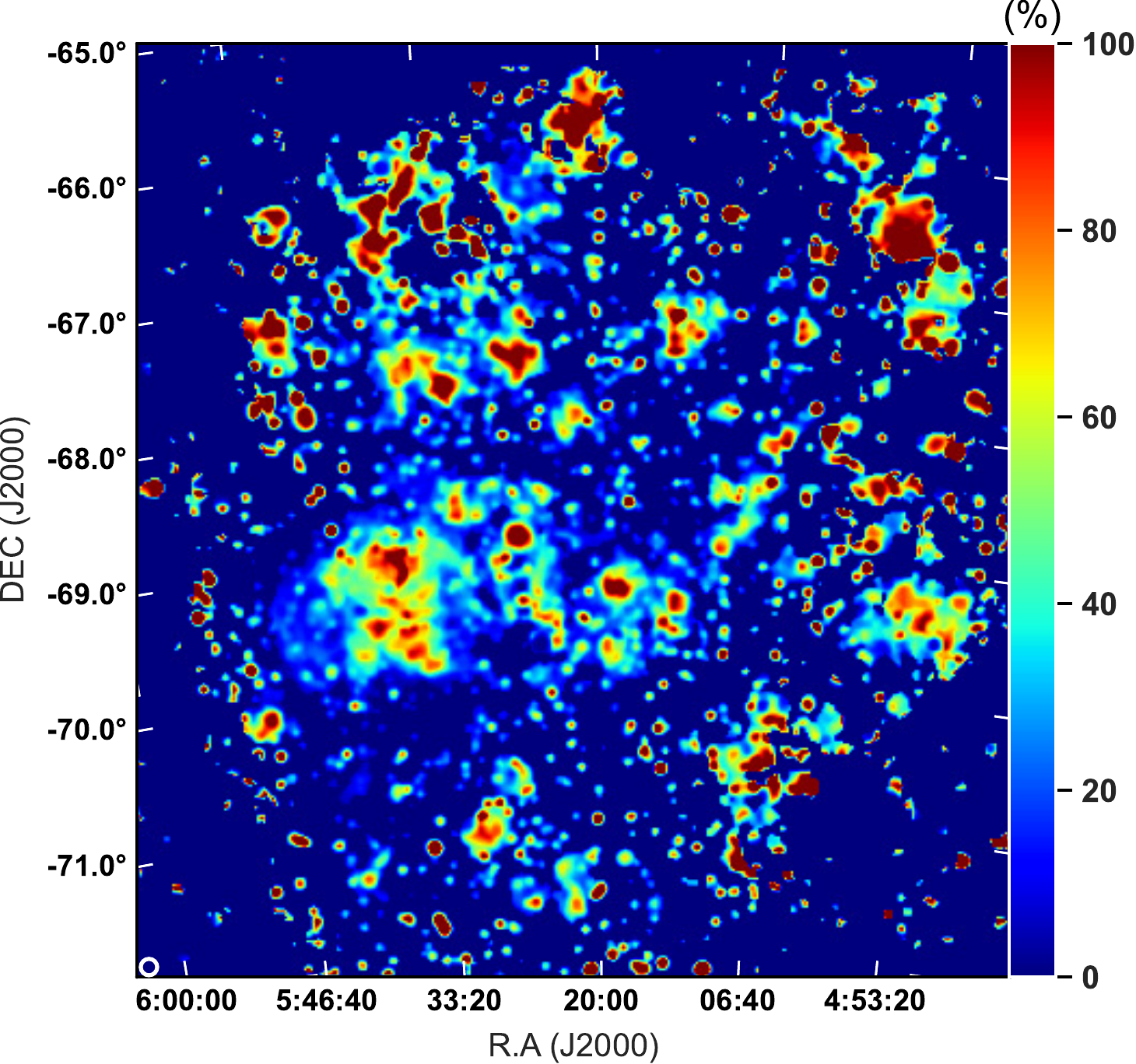}
\includegraphics[width=0.8\columnwidth]{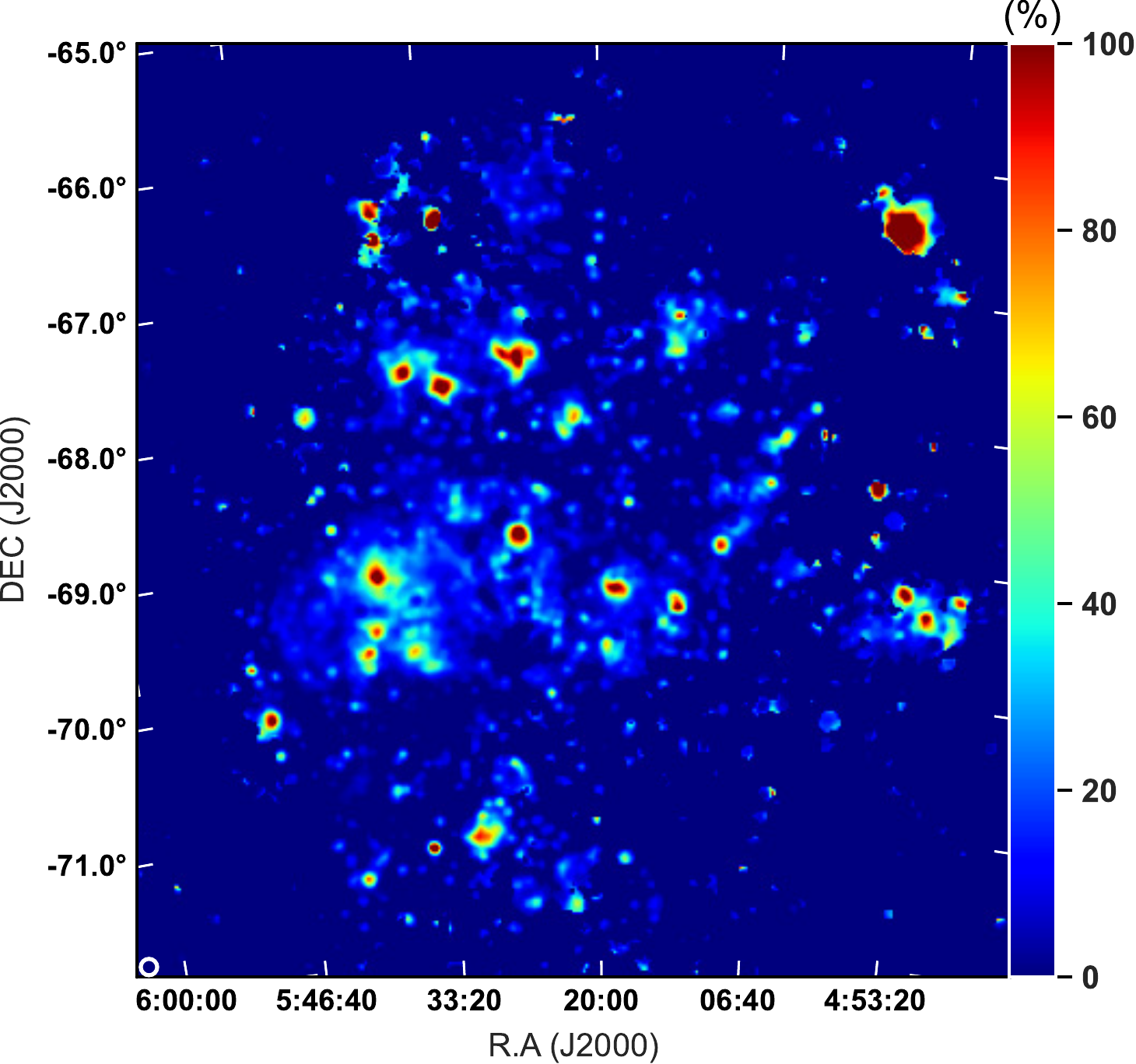}
\includegraphics[width=0.8\columnwidth]{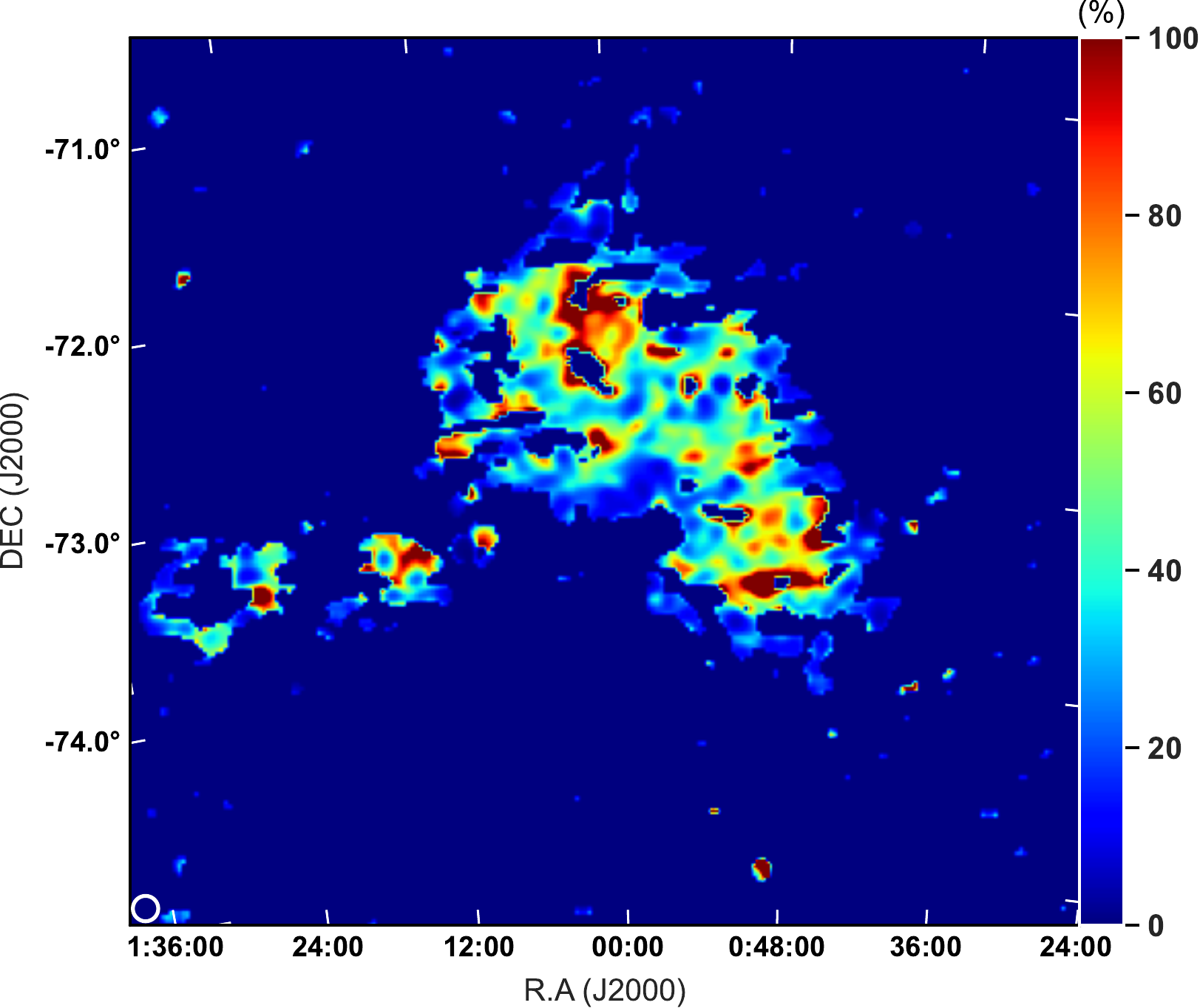}
\includegraphics[width=0.8\columnwidth]{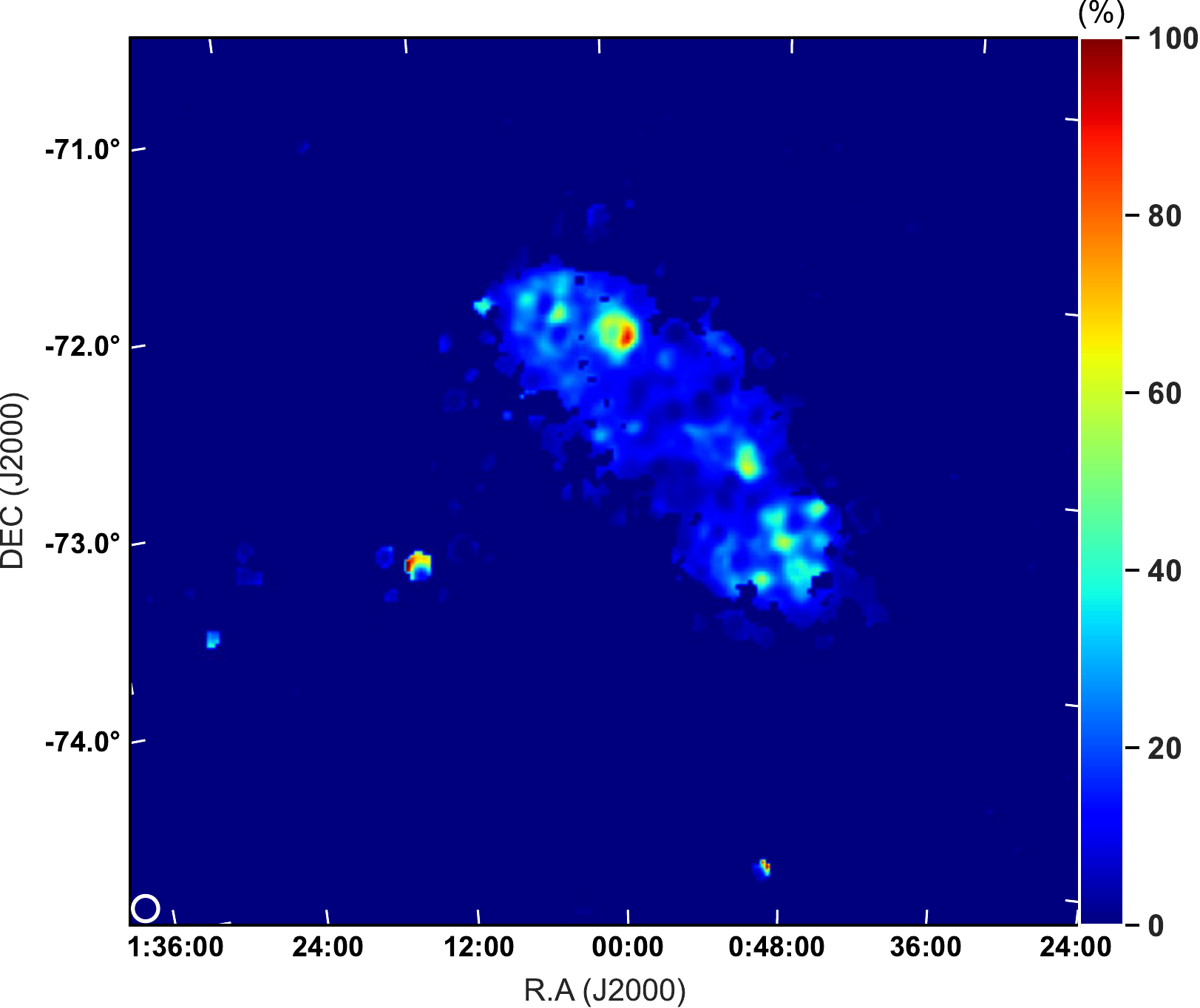}
 \caption{Maps of the thermal fraction for the LMC (top) and the SMC (bottom) at 0.166\,GHz (left) and 1.4\,GHz (right). Colour bars show the fractions in percentage. The beam sizes of 221\arcsec (for the LMC) and 235\arcsec (for the SMC)  are shown in the lower left corner of the maps. }
\label{fig:thermal_fraction_maps}
\end{figure*}

Table~\ref{tab:mcs_frac} shows the integrated flux densities of the synchrotron and observed RC emission. Integrations were performed from the LMC centre (05$^{\text{h}}$\,23$^{\text{m}}$\,34$^{\text{s}}$,\,-69$\degr$\,45$\myprime$22$\mydprime$)$_\text{J2000}$ out to the radius of $R= 5\degr$ and from the SMC centre (00$^{\text{h}}$\,52$^{\text{m}}$\,44$^{\text{s}}$,\,-72$\degr$\,49$\myprime$42$\mydprime$)$_\text{J2000}$ out to $R= 3\degr$. We note that negative map pixels are neglected in these calculations. Otherwise, these pixels reduce the integrated flux density of the observed radio continuum emission by at most 30-35 per cent at 0.166\,GHz in both the LMC and the SMC and at 1.4\,GHz in the SMC, resulting in an artificial increase in the thermal fraction by about 15 per cent (see Section~\ref{ssec:thermal_fraction_sec}). For comparison, the integrated flux densities are reported before and after subtracting background radio sources presented by \citet{lmc_source} and \citet{smc_bkg}. These sources account for 53 per cent, 14 per cent, and 4 per cent of the total RC flux at 0.166, 1.4, and 4.8\,GHz, respectively in the LMC, indicating that they are mostly steep radio sources ($\alpha < -1.1$). However, we note that as most of these sources are behind the face of the LMC, and hence mixed with the galaxies' extended emission, the resolution at which this subtraction is performed becomes important.

Table~\ref{tab:mcs_frac} shows that the source subtraction from the 1.4 and 4.8\,GHz maps results in lower galaxies' integrated fluxes (or higher fluxes of background sources at this frequency leading to their flatter spectra) at the GLEAM resolutions (221\arcsec and 235\arcsec) than at their own native resolutions ($\le 40\arcsec$ for the LMC and $98\arcsec$ for the SMC, see Table~\ref{tab:mcs_data}). This indicates that subtracting the sources from the GLEAM maps at 0.166\,GHz also removes a portion of the extended emission of the galaxies. Hence, studies of the spectral index at which the source subtracted maps/fluxes at different frequencies are used is actually sensitive to the resolution at which the subtraction is performed (see Section~\ref{sec:nt}).

\subsection{Thermal fraction} \label{ssec:thermal_fraction_sec}

The thermal faction $f_\text{th}$ maps are obtained by dividing the thermal emission by the observed RC at each frequency (Fig. \ref{fig:thermal_fraction_maps}). 
The prominent regions emitting strong free-free emission are \ion{H}{ii} regions with a median thermal fraction of $f^\text{0.166\,GHz}_\text{th}= (49 \pm 9)$ 
 for the LMC and $f^\text{0.166\,GHz}_\text{th}= (32 \pm 5)$ per cent for the SMC at 0.166\,GHz. The \ion{H}{ii} regions have a higher thermal fraction at 1.4\,GHz with a median $f^\text{1.4\,GHz}_\text{th}= (75 \pm 9)$ per cent for the LMC and $f^\text{1.4\,GHz}_\text{th}= (79 \pm 10)$ per cent for the SMC. Tables~\ref{tab:lmc_hII} and~\ref{tab:smc_hII} list $f^\text{0.166\,GHz}_\text{th}$ and $f^\text{1.4\,GHz}_\text{th}$ for the individual \ion{H}{ii} regions.

The global thermal fraction obtained above $5\sigma$ RMS level of the thermal emission is $f^\text{1.4\,GHz}_\text{th} = (30 \pm 4)$ per cent in the LMC and slightly higher $(35 \pm 7)$ per cent in the SMC at 1.4\,GHz. In both galaxies, the thermal fraction at 0.166\,GHz is less than 15 per cent. The lower thermal fraction at lower frequencies is expected due to a faster increase of the synchrotron emission than the thermal emission. 
The MCs’ global thermal fractions are higher compared with those of spiral galaxies such as M\,33, ($f_\text{th} \sim$ 18 per cent) and NGC\,6946 ($f_\text{th} \sim$ 7 per cent) at 1.4\,GHz \citep{trt2008,tab_2013}. However, they are lower than those estimated using the classical separation method, i.e., using a fixed spectral index, in the same galaxies \citep[e.g., $f^\text{1.4\,GHz}_\text{th}\simeq 55$ per cent in the LMC assuming $\alpha_{\text{n}}=-0.84$, ][]{klein_lmc}.
This assumption particularly leads to an excess thermal emission in \ion{H}{ii} regions. Previously, \cite{klein_lmc} indicated that 30~Dor complex is mostly thermal (or at least at 1.4\,GHz), however, we find strong non-thermal emission as well.  

\begin{figure}
\includegraphics[width=0.65\columnwidth,center]{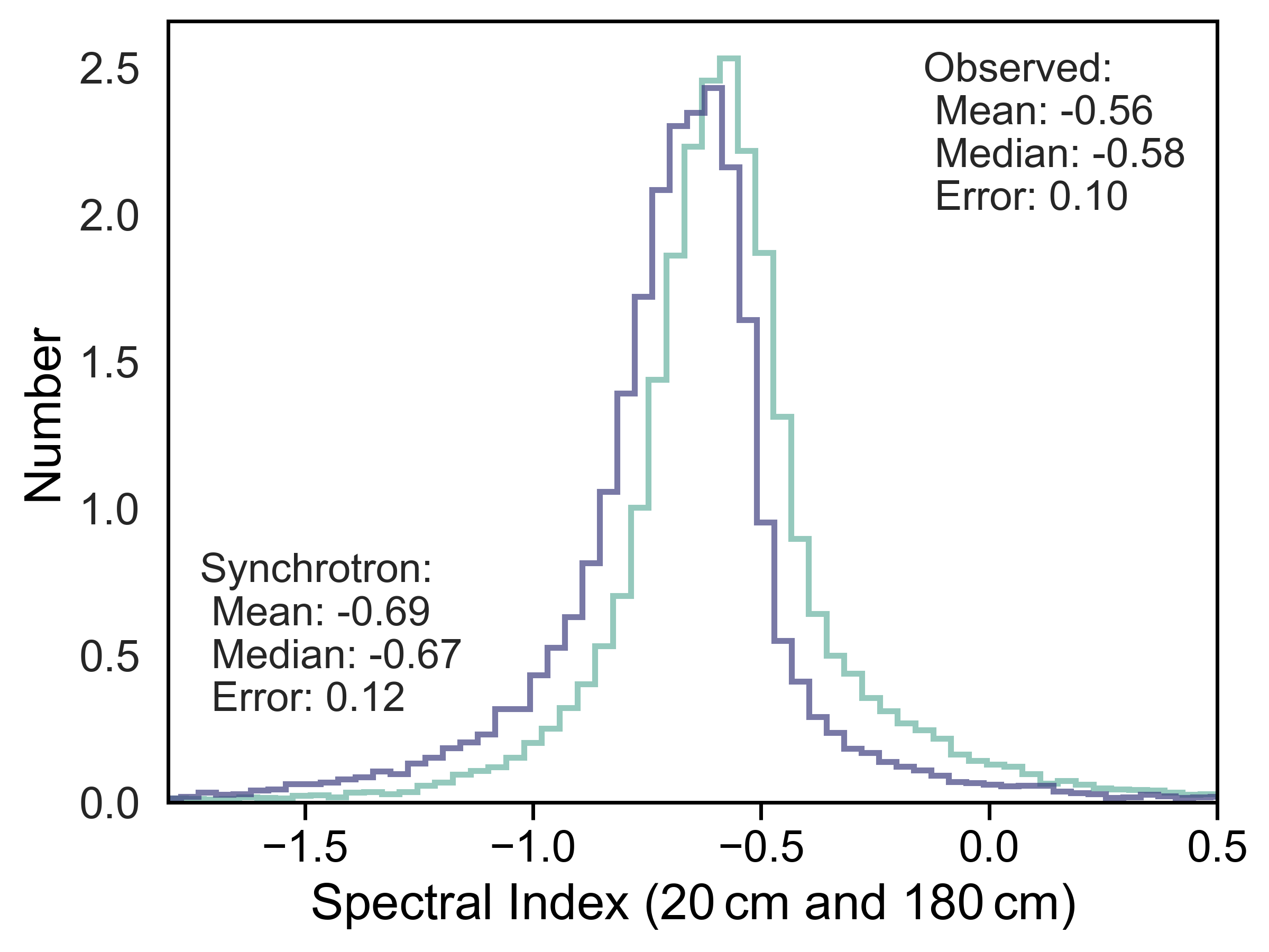}
\includegraphics[width=0.65\columnwidth,center]{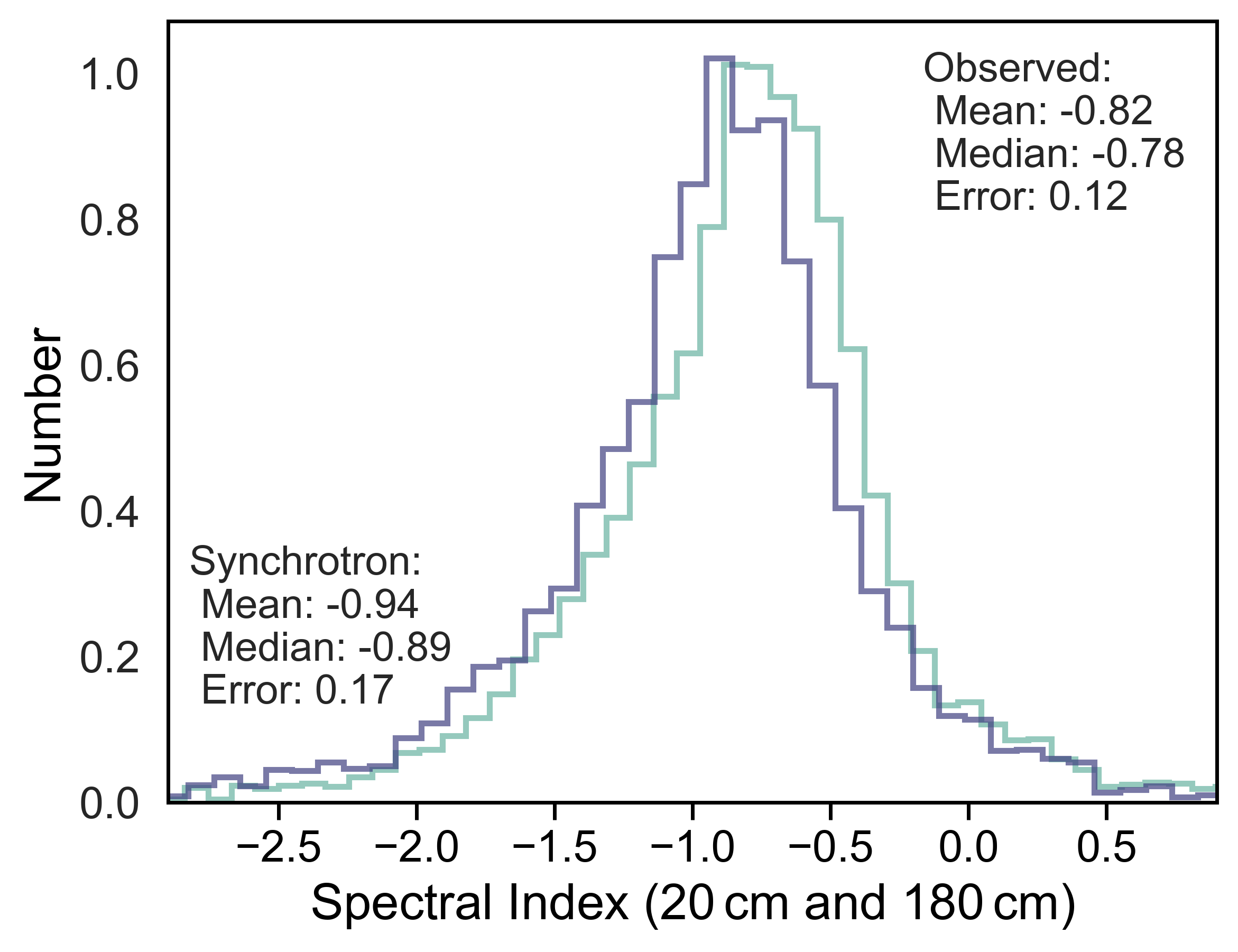}
 \caption{Histograms of the spectral index of the observed RC emission $\alpha$ (green) and its synchrotron component $\alpha_{\rm n}$ (purple) measured between 0.166\,GHz and 1.4\,GHz in the LMC (top) and the SMC (bottom). Vertical axes show the number of pixels in the spectral index maps corresponding to bins of equal width of $\alpha$ or $\alpha_{\rm n}$ values. The intensity maps were subtracted for background radio sources before deriving the $\alpha$ and $\alpha_{\rm n}$ maps. }
 \label{fig:histo_mcs}
\end{figure} 

\section{Non-thermal Spectral Index}   \label{sec:nt}

\noindent 
\begin{figure*}
\centering

\includegraphics[width=0.8\columnwidth]{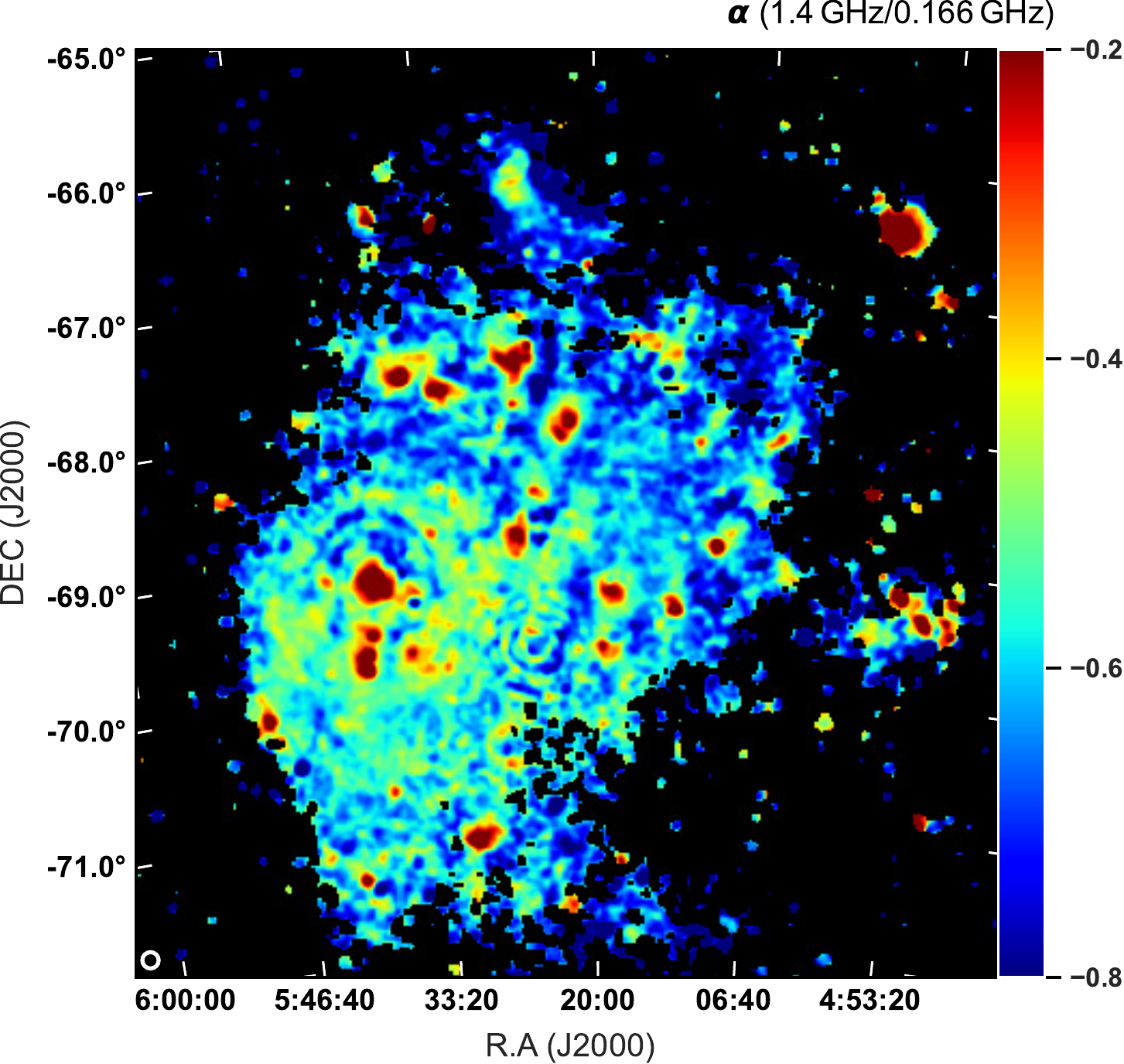}
\includegraphics[width=0.8\columnwidth]{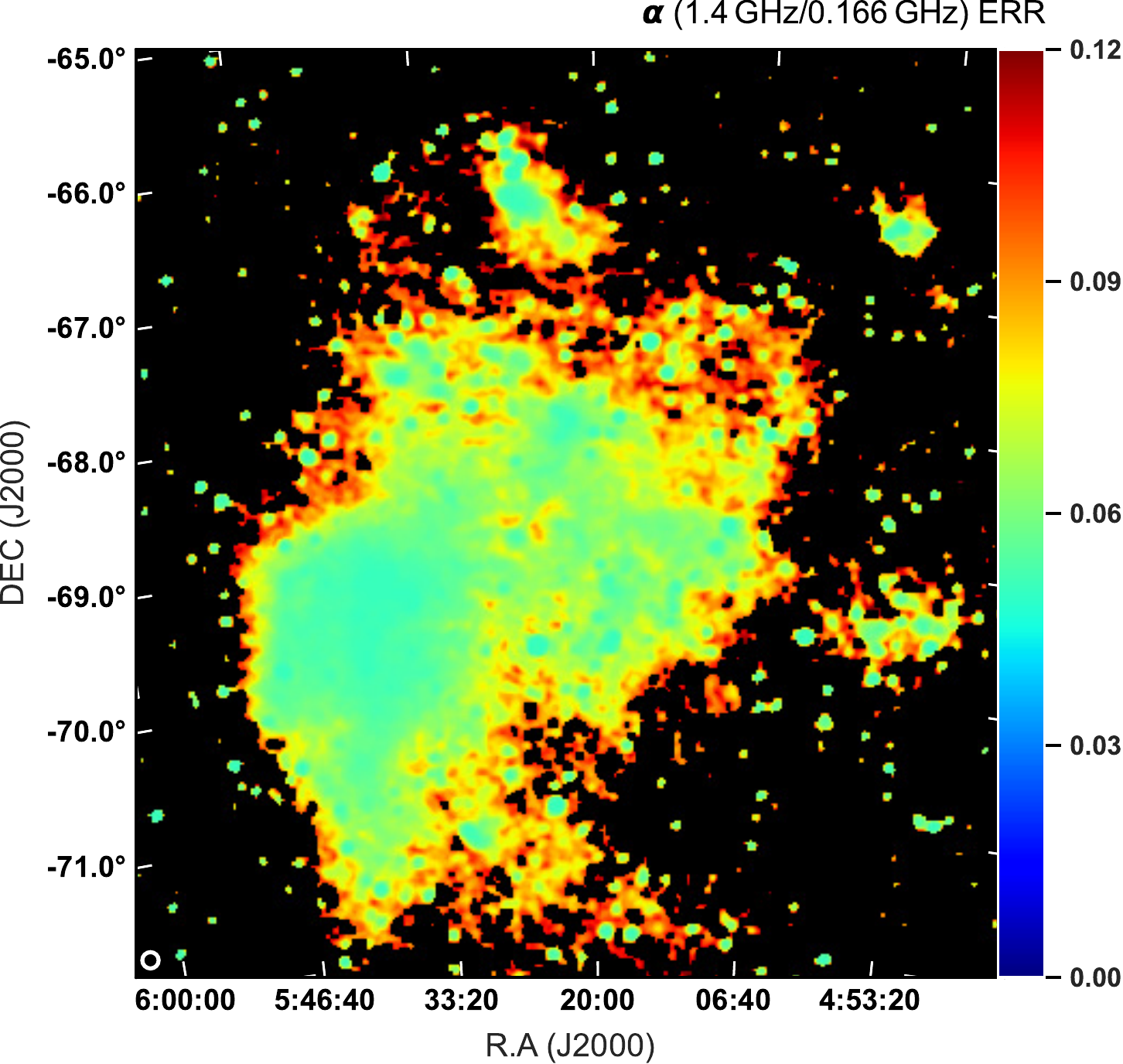}

\includegraphics[width=0.8\columnwidth]{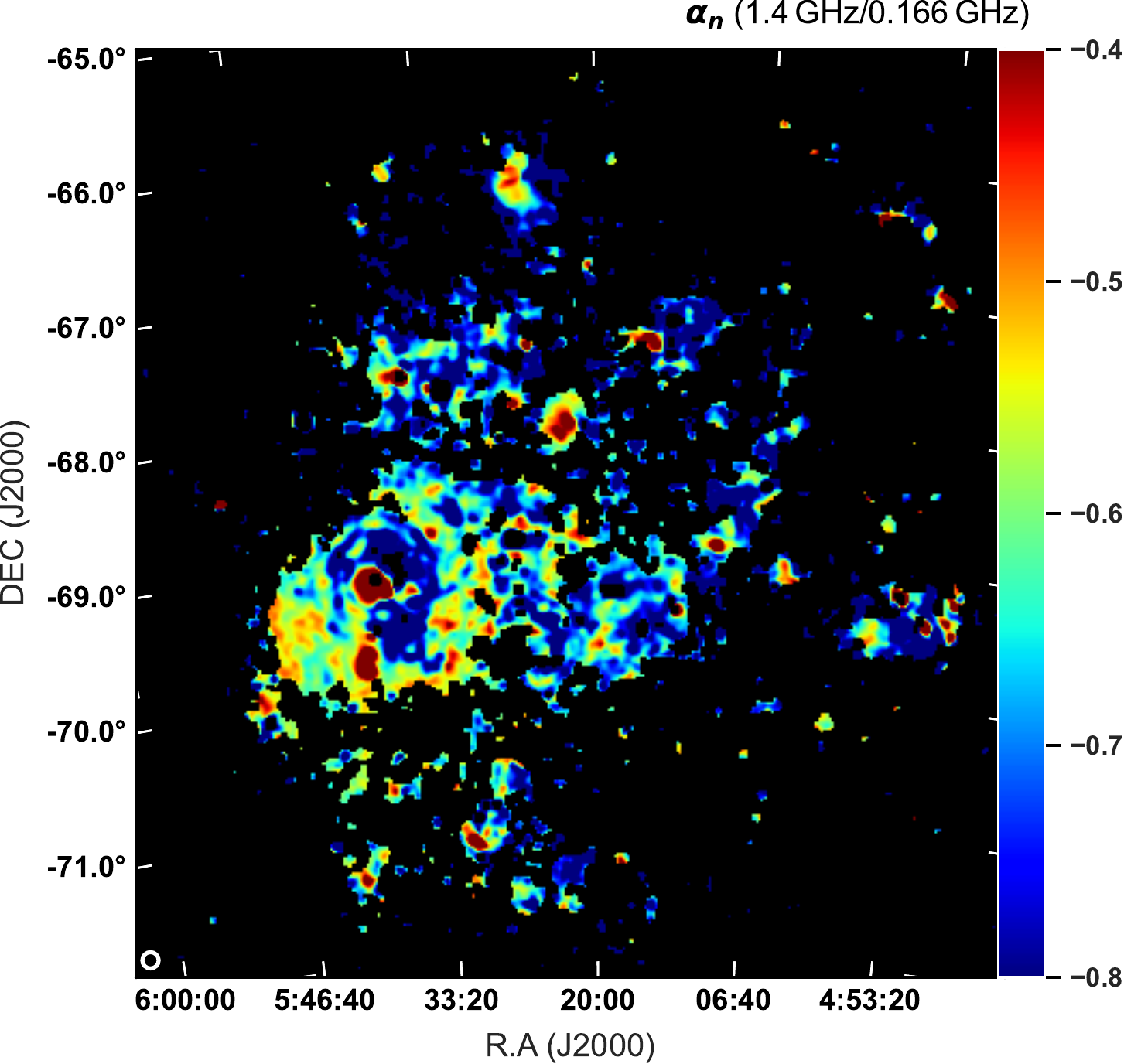}
\includegraphics[width=0.8\columnwidth]{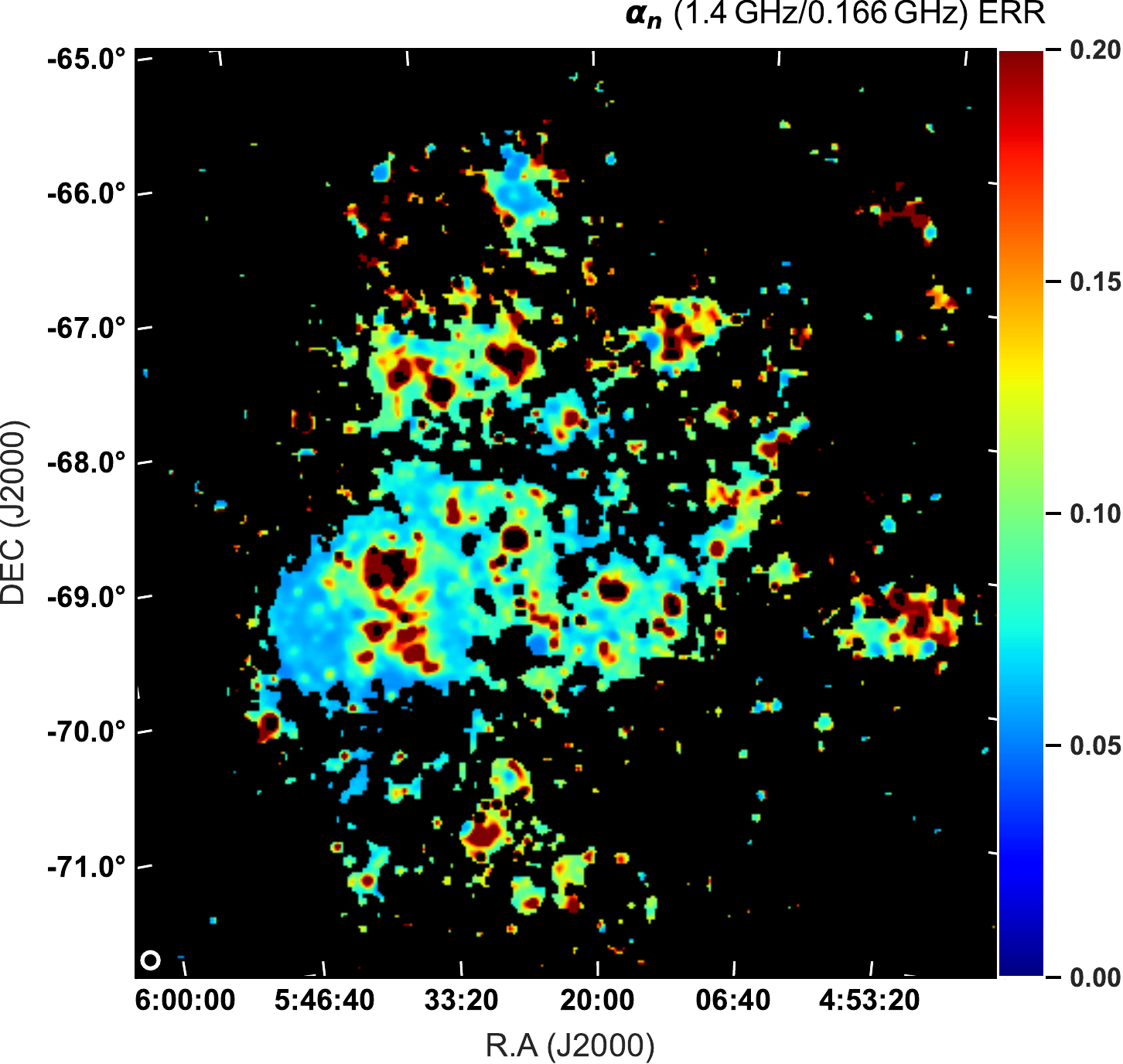}

 \caption{Maps of the spectral index (left) and its uncertainty (right) of the observed RC ($\alpha$, top) and synchrotron emission ($\alpha_{\rm n}$, bottom) measured between 0.166\,GHz and 1.4\,GHz for the LMC. The beam size of $221\arcsec$ is shown in the lower left corners.}
\label{fig:spectral_index_map_lmc}
\end{figure*}

\begin{figure*}
\centering

\includegraphics[width=0.8\columnwidth]{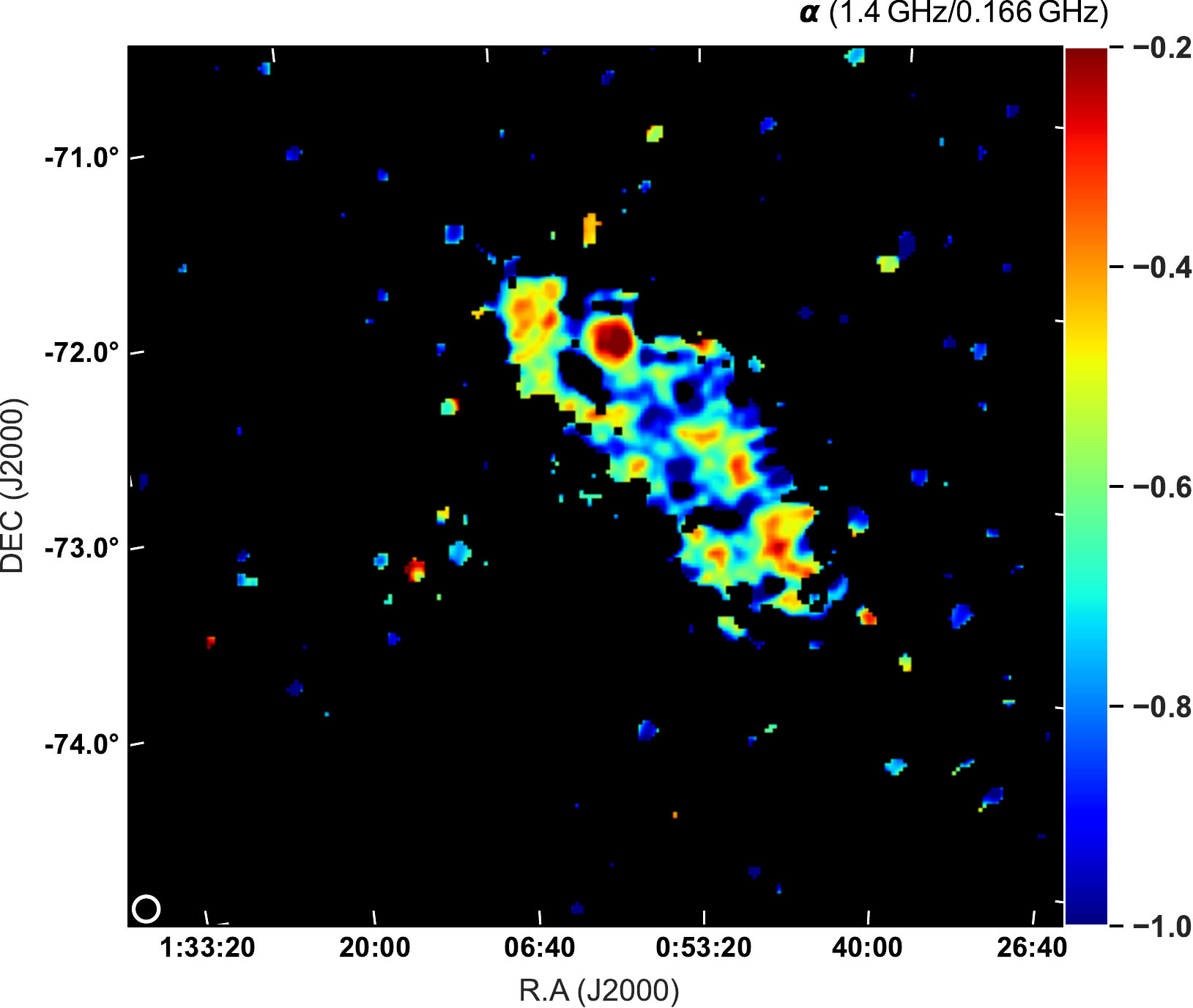}
\includegraphics[width=0.8\columnwidth]{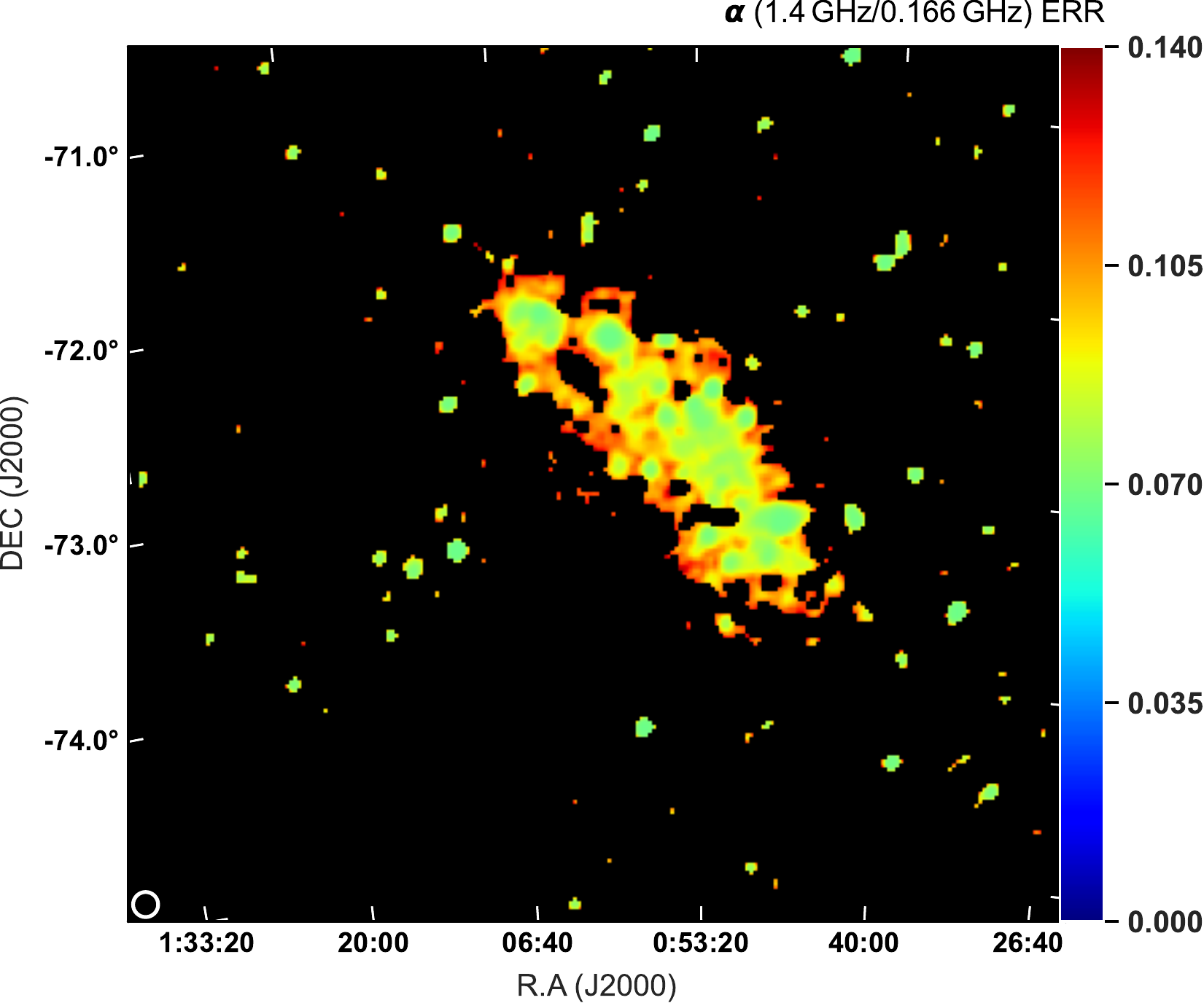}

\includegraphics[width=0.8\columnwidth]{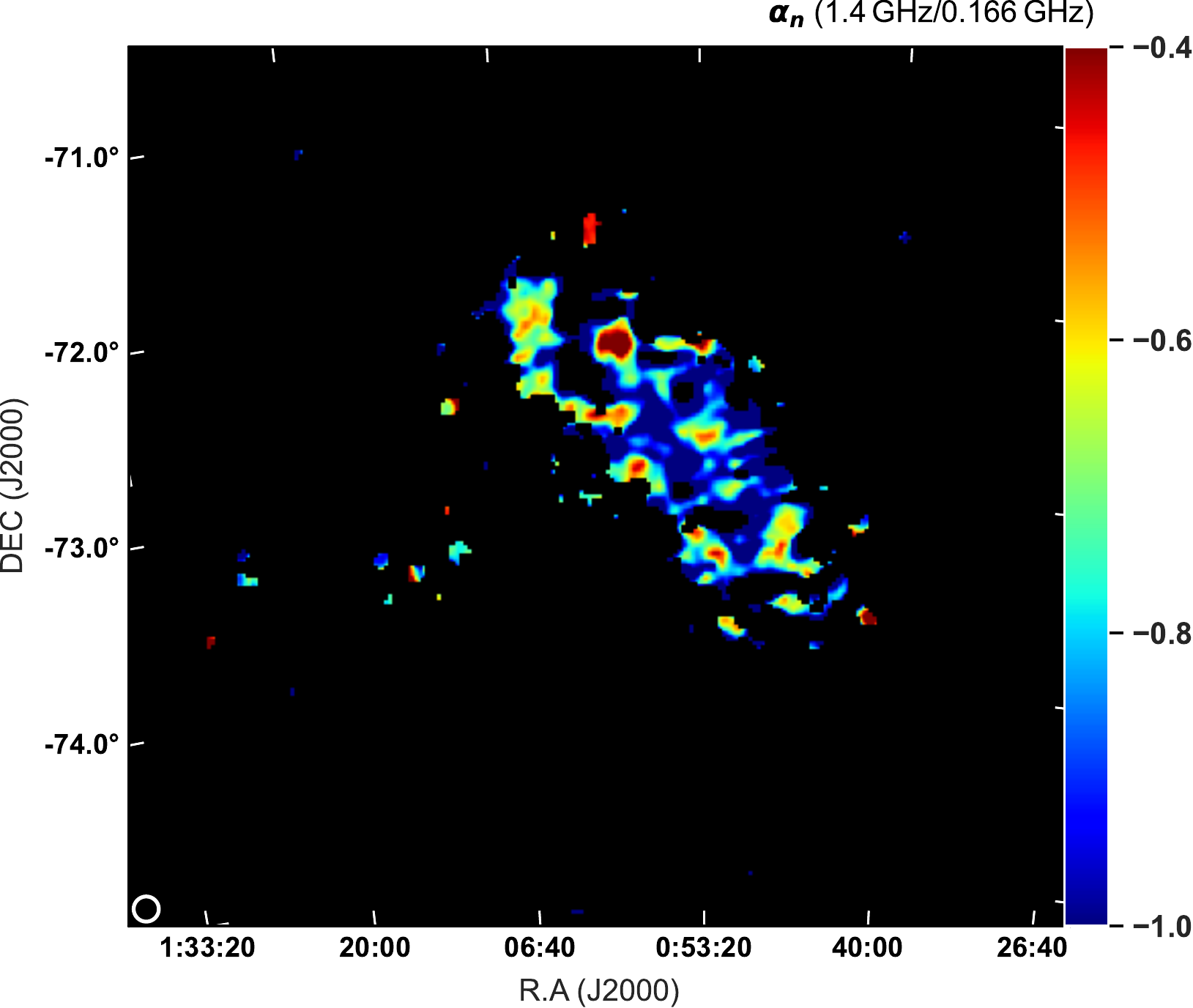}
\includegraphics[width=0.8\columnwidth]{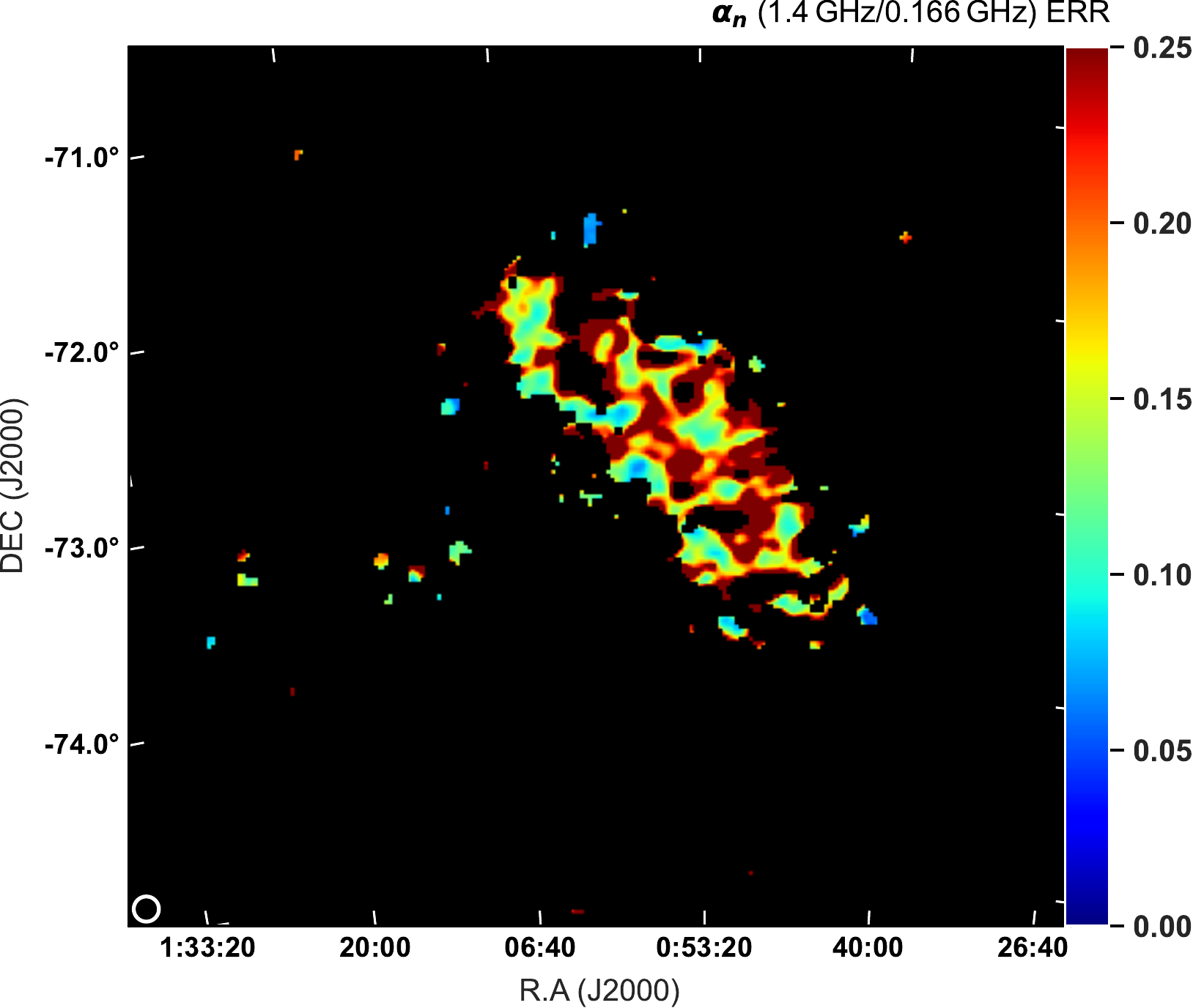}

 \caption{Maps of the spectral index (left) and its uncertainty (right) of the observed RC ($\alpha$, top) and synchrotron emission ($\alpha_{\rm n}$, bottom) measured between 0.166\,GHz and 1.4\,GHz for the SMC. The beam size of $235\arcsec$ is shown in the lower left corners.}
 
\label{fig:spectral_index_map_smc}
\end{figure*} 

The TRT separation technique used is independent of the synchrotron spectrum and free from any related assumption. Its products at several frequencies can be used to investigate variations in the non-thermal radio spectral index. The synchrotron emission is characterized by a power-law spectrum with an index $\alpha_\text{n}$, $S_\text{n}$ $\simeq \nu^{\alpha_\text{n}}$. We map $\alpha_\text{n}$ of the MCs between 0.166\,GHz and 1.4\,GHz using the non-thermal maps at the corresponding frequencies. This map as well as that of the observed spectral index $\alpha$, $S$ $\simeq \nu^{\alpha}$ are derived only for pixels with intensities higher than 5$\sigma$ RMS noise in the MCs. Both the LMC and the SMC show regions of steep spectrum ($\alpha<-0.8$, dark blue regions in Figs.~\ref{fig:spectral_index_map_lmc} and~\ref{fig:spectral_index_map_smc}), which mostly belong to diffuse parts of the ISM. In massive star-forming regions, the spectrum, even that of the pure synchrotron emission, is relatively flat ($ -0.7\lesssim \alpha_\text{n}\lesssim -0.4$), indicating the presence of high energy CREs in these regions. The flattest spectrum is found in the giant \ion{H}{ii} complex 30~Dor in the LMC.

 \begin{table}
    \begin{threeparttable}
 \caption{The effect of background source subtraction on the integrated flux densities of the observed RC ($S$) and the non-thermal ($S_\text{nt}$) emission for the LMC (${R}\leq 5\degr$) and the SMC (${R} \leq 3\degr$).    }
 \begin{tabularx}{\columnwidth}{lll}
  \hline
   \ $\nu$ & S & $S_\text{nt}$    \\
  (GHz) & (Jy) & (Jy) \\
 \hline
 LMC\\

 \ 0.166$^{a}$ & 1353 $\pm$ 138 & 1179 $\pm$ 130   \\ 
  \ 0.166$^{b}$ & 641$\pm$ 70& 467 $\pm$ 51   \\

 \ 1.4$^{a}$ & 449 $\pm$ 30 & 304 $\pm$ 22  \\
   \ 1.4$^{b}$ & 385 $\pm$ 26& 240 $\pm$ 18 \\
  \ 1.4$^{c}$ & 242 $\pm$ 16 & 97 $\pm$ 7 \\

 \ 4.8$^{a}$ &  361 $\pm$ 31 & 232 $\pm$ 21  \\
  \ 4.8$^{b}$ &  347 $\pm$ 29 & 218 $\pm$ 22    \\
 \ 4.8$^{c}$ &  231 $\pm$ 20 & 102 $\pm$ 9 \\

 \midrule
 
 SMC\\
 \ 0.166$^{a}$ & 260 $\pm$ 35 & 238 $\pm$ 33  \\
  \ 0.166$^{b}$ &  126 $\pm$ 17 & 104 $\pm$ 14  \\
 \ 1.4$^{a}$     &  49.7 $\pm$ 6.3 & 32 $\pm$ 4  \\
  \ 1.4$^{b}$    &  30.2 $\pm$ 3.9 & 12.5 $\pm$ 1.7  \\
  \ 1.4$^{c}$    &  18.9 $\pm$ 2.5 & 1.2 $\pm$ 0.2  \\

 \hline
  \label{tab:mcs_frac}
 \end{tabularx}
     \begin{tablenotes}
      \small
      \item {${a}-$ Background sources not subtracted. ${b}-$ Background sources subtracted at native resolutions (see Table~\ref{tab:mcs_data}). ${c}-$ Background sources subtracted after convolution to the GLEAM resolutions.}
      \end{tablenotes}
        \end{threeparttable}
\end{table}

 To obtain an average value of the spectral index, it is essential first to subtract the background radio sources at different frequencies. As discussed in Section~\ref{ssec:nt_map}, the final result can change depending on the resolution that subtraction performed. Hence, to keep the residual structures consistent at different frequencies, we have to subtract the sources at the same resolution as of the GLEAM data at 0.166\,GHz. In other words, at 1.4\,GHz, the maps are first convolved to the GLEAM resolution then subtracted for the sources. We note that this way, the spectral index measurement is accurate, but the ISM regions with a background source behind are excluded which is unavoidable due to the large beam of the GLEAM data.
Fig.~\ref{fig:histo_mcs} shows a histogram of $\alpha$ and $\alpha_\text{n}$ obtained using the background radio source subtracted intensity maps at 0.166\,GHz and 1.4\,GHz.
We found that both $\alpha$ and $\alpha_\text{n}$ are flatter in the LMC ($\alpha=-0.58\,\pm\,0.10$ and $\alpha_\text{n}= -0.67\,\pm\,0.12$) than in the SMC ($\alpha=-0.78\,\pm\,0.12$ and $\alpha_\text{n}= -0.89\,\pm\,0.17$) that is mainly due to the presence of 30~Dor and the fact that the LMC hosts more giant \ion{H}{ii} regions with flat spectrum than the SMC.

Using all multi-frequency RC maps and archival data available, we further obtain $\alpha_\text{n}$ globally over a wider range of frequencies by modeling the integrated RC spectrum from 0.02\,GHz to 8.5\,GHz in the LMC and 0.08\,GHz to 8.5\,GHz in the SMC. All maps are integrated out to the radius of 5$\degr$ for the LMC and 3$\degr$ for the SMC from their centres (see Table~\ref{tab:mc_details}). Tables~\ref{fig:lmc_fluxes} and~\ref{fig:smc_fluxes} summarize our integrated flux measurements \footnote{In Tables~\ref{fig:lmc_fluxes} and~\ref{fig:smc_fluxes}, the integrations refer to $>3\,\sigma$ level of the RC maps leading to slightly lower values than those listed in Table~\ref{tab:mcs_frac}.} as well as measurements from the literature. 

Assuming that the thermal emission is optically thin, the RC spectrum can be expressed as:

\begin{equation}
S_{\nu} = S^{\rm th}_{\nu} + S^{\rm nt}_{\nu}= A_1\,\nu^{-0.1} + A_2\,\nu^{\alpha_{\rm n}},   
\end{equation}
where $A_1$ and $A_2$ are constant scaling factors. To avoid dependencies on the units of the frequency space, this equation can be written as:
\begin{equation}
\label{eqn:mcmc}
S_{\nu} = A_1'\,(\frac{\nu}{\nu_0})^{-0.1} + A_2\, \nu_0^{\alpha_{\rm n}}\,(\frac{\nu}{\nu_0})^{\alpha_{\rm n}},   
\end{equation}

with $A_1'\,= \nu_0^{-0.1}\,A_1$ and $\nu_0$ a reference frequency. Following \cite{Tabatabaei_2017}, we fit this model to the observed data using the Markov chain Monte Carlo (MCMC) Bayesian method as it provides robust statistical constraints on the fit parameters $\alpha_\text{n}$, $A_1'$ and $A_2$. The priors and related model library are set by generating random combinations of the parameters sampled uniformly in wide ranges of parameter space \citep[$-1<A'_1, A_2<2000$ for the LMC and $-1<A'_1, A_2<400$ for the SMC\footnote{We note that negative values of flux density are not physically motivated but are included to assess the robustness of the final results.} and $0<\alpha_\text{n}<2.2$][]{Tabatabaei_2017}. The parameters $A_1'$ and $A_2$ are set to fit the integrated flux densities in unites of Jy.  

\begin{figure*}
\centering
\includegraphics[width=0.93\columnwidth]{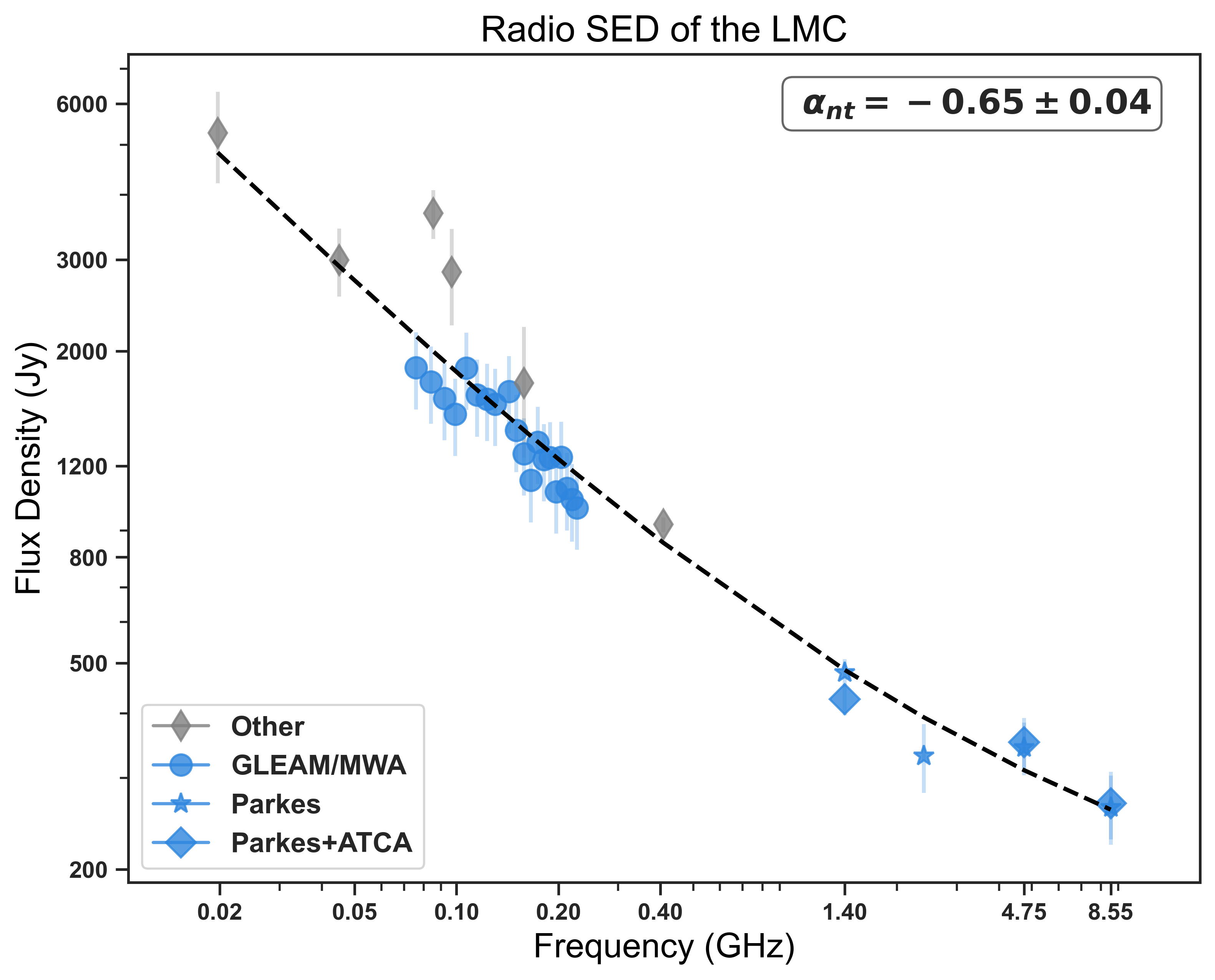}
\includegraphics[width=0.93\columnwidth]{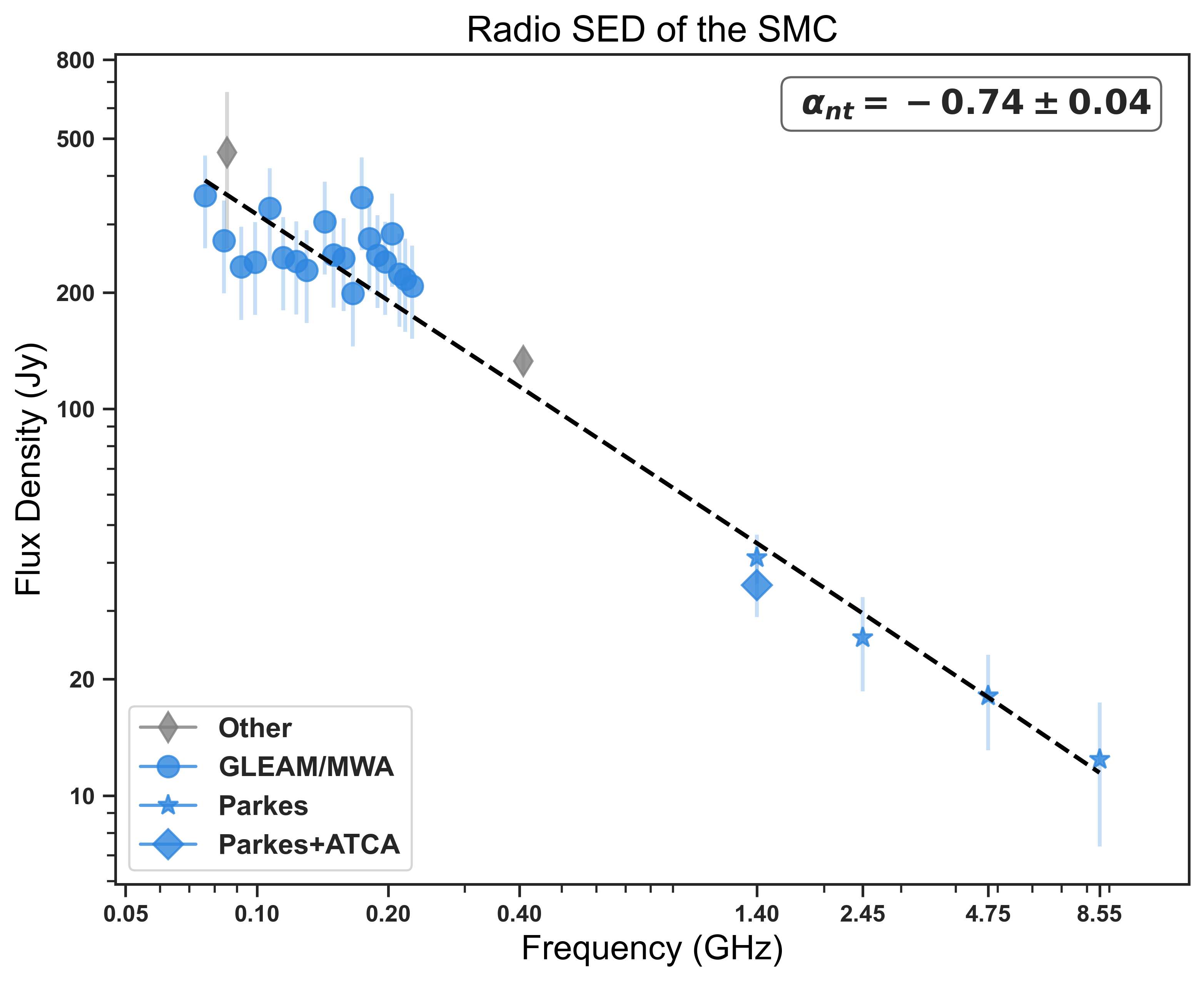}
\caption{The radio continuum spectrum from about 0.02\,GHz to 8.5\,GHz for the LMC (left) and the SMC (right). }
\label{fig:seds_plot}
\end{figure*}

\begin{figure*}
\centering
\includegraphics[width=0.6\textwidth]{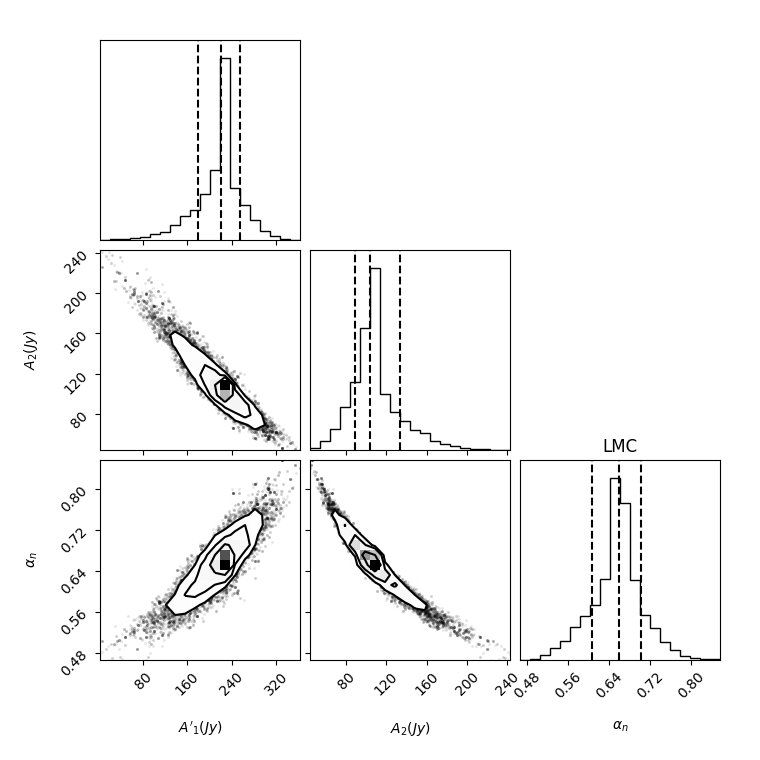}
\caption{ Bayesian corner plots for the parameters A$^{\prime}_{1}$ and A$_{2}$ in equation~(\ref{eqn:mcmc}) showing the posterior PDF and their 0.16, 0.5, and 0.86 percentiles for the LMC. The uncertainty contours show that the posteriors have the highest probability to occur within the confidence intervals indicated. }
\label{fig:bays}
\end{figure*}

Figs.~\ref{fig:seds_plot} and~\ref{fig:bays} show the resulting fit to the integrated flux densities and the posterior probability distribution function (PDF) of the parameters, respectively. 

  \begin{table}
 \caption{Mean physical properties of the magneto-ionic plasma in the MCs.  }
  \begin{tabular}{llll}
  \hline
   \ Object & \,\,\,$B_{\rm tot}$ & \,\,\,\,\,\,$\langle n_{\rm e}\rangle$ & \,\,\,\,\,\,$\beta$   \\
  &($\mu$G) & (${\rm cm^{-3}}$) &  \\
 \hline
 \ LMC & $10.1\,\pm\,1.7$ &  $0.06\,\pm\,0.01$ &  $0.06\,\pm\,0.01$ \\
  
 \ SMC &$5.5\,\pm\,1.3$ &  $0.020\,\pm\,0.019$ &  $0.06\,\pm\,0.06$ \\
  
 \hline
  \label{tab:phys}
 \end{tabular}
\end{table}

We obtain a global non-thermal spectral index $\alpha_\text{n} = -0.65 \pm 0.04$ in the LMC and $-0.74 \pm 0.04$ in the SMC using a reference frequency of $\nu_0=0.166$\,GHz. These results agree with the median of the $\alpha_\text{n}$ distribution in the maps presented above. Moreover, we find a good match with the results of \cite{2018for} studying the integrated RC SEDs. Our fitting process suggests a curve line for the LMC, which agrees with the best-fit model suggested by \cite{2018for}. The curvature shows that both the thermal and non-thermal components are indeed necessary to explain the RC spectrum in the LMC. The non-thermal emission is dominated at lower frequencies and the thermal emission at higher frequencies. In the SMC, however, the non-thermal emission dominates the RC spectrum at the selected frequency range that also agrees with \cite{2018for}.

\begin{table*}
    \begin{threeparttable}
\parbox{.45\linewidth}{
\caption{Integrated flux density ($S_{\nu}$) and uncertainty in flux density ($\sigma$) of the LMC.}
        \label{fig:lmc_fluxes}
\begin{tabular}{llll}
  \hline
  $\nu$ & $S_{\nu}$  &  $\sigma$ & Reference   \\
    (MHz) & (Jy)  &  (Jy)    \\

    \hline
20 & 5270 & 1054 & \cite{lowfreq}$^1$ \\
45 & 2997 & 450  &  \cite{alvarezs} \\
76 & 1857 & 315.7 & \cite{2018for}$^1$  \\
84 & 1743.3 & 296.4  & \cite{2018for}$^1$\\
85 & 3689 & 400 & \cite{millss}$^{2}$  \\
92 & 1619.7 & 275.3 &  \cite{2018for}$^1$\ \\
97 & 2839 & 600 & \cite{millss}$^{2}$ \\
99 & 1511.3 & 256.9 &  \cite{2018for}$^1$\\
107 & 1853.7 & 315.1 &   \cite{2018for}$^1$\\
115 & 1646.4 & 279.9 &  \cite{2018for}$^1$\\
123 & 1614.1 & 274.4 &  \cite{2018for}$^1$\\
130 & 1580.3 & 268.6 &  \cite{2018for}$^1$ \\
143 & 1670.3 & 284 &  \cite{2018for}$^1$\\
150 & 1406 & 239 &   \cite{2018for}$^1$\\
158 & 1267 & 215.4 &  \cite{2018for}$^1$\\
158 & 1736 & 490  & \cite{millss}$^{2}$ \\
166 & 1125.7 & 191.4 &   \cite{2018for}$^1$\\
174 & 1334.2 & 226.8 &  \cite{2018for}$^1$\\\
181 & 1235.4 & 210 &  \cite{2018for}$^1$ \\
189 & 1245.9 & 211.8 &   \cite{2018for}$^1$\\
197 & 1070.2 & 181.9 &  \cite{2018for}$^1$\\
204 & 1247.8 & 212.1 &  \cite{2018for}$^1$\\
212 & 1086 & 184.6 & \cite{2018for}$^1$\\
219 & 1033.6 & 175.7 &   \cite{2018for}$^1$\\
227 & 997 & 169.5 &   \cite{2018for}$^1$\\
408 & 925 & 30  &   \cite{klein_lmc}\\
1400 & 478.9 & 30 &  \cite{klein_lmc}$^1$\\
1400 & 426.1 & 30 &  \cite{Huges2007}$^1$\\
2400 & 331 & 50 &   \cite{lmc_24}$^1$\\
4750 & 351.5 & 40 &  \cite{Haynes91}$^1$\\
4750 & 343.5 & 40 &   \cite{Dickel2005}$^1$\\
8550 & 268.3 & 40 &   \cite{Haynes91}$^1$\\
8550 & 263 & 40 &   \cite{Dickel2005}$^1$\\
  \hline

\end{tabular}

    \begin{tablenotes}
      \small
      \item Notes.\\
      $^1$Re-integrated for ${R}\leq 5\degr$.\\ 
      $^2$Revised by \cite{klein_lmc}.
      \end{tablenotes}

}
\hfill
\parbox{.45\linewidth}{
\caption{Integrated flux density ($S_{\nu}$) and uncertainty in flux density ($\sigma$) of the SMC.}
   \label{fig:smc_fluxes}
\begin{tabular}{llll}
  \hline
  $\nu$ & $S_{\nu}$  &  $\sigma$ & Reference   \\
    (MHz) & (Jy)  &  (Jy)    \\
    \hline
76 & 356.2 & 96.2 & \cite{2018for}$^{1}$ \\
84 & 272.5 & 73.6 &  \cite{2018for}$^{1}$ \\
85.5 & 460 & 200 & \cite{millss}$^{2}$ \\
92 & 233 & 62.9 & \cite{2018for}$^{1}$ \\
99 & 239.6 & 64.7 & \cite{2018for}$^{1}$ \\
107 & 330.2 & 89.1 &  \cite{2018for}$^{1}$ \\
115 & 246.5 & 66.5 & \cite{2018for}$^{1}$ \\
123 & 240.4 & 64.9 & \cite{2018for}$^{1}$ \\
130 & 228.4 & 61.7  & \cite{2018for}$^{1}$ \\
143 & 304.7 & 82.3 &  \cite{2018for}$^{1}$ \\
150 & 250.3 & 67.6 &  \cite{2018for}$^{1}$ \\
158 & 245 & 66.2 &  \cite{2018for}$^{1}$ \\
166 & 198.8 & 53.7 &  \cite{2018for}$^{1}$ \\
174 & 352.1 & 95.1 & \cite{2018for}$^{1}$ \\
181 & 275.8 & 74.5 &  \cite{2018for}$^{1}$ \\
189 & 249.6 & 67.4 & \cite{2018for}$^{1}$ \\
197 & 240  & 64.8 & \cite{2018for}$^{1}$ \\
204 & 283.6 & 76.6 &  \cite{2018for}$^{1}$ \\
212 & 223.4 & 60.3 &  \cite{2018for}$^{1}$ \\
219 & 216.7 & 58.5 &  \cite{2018for}$^{1}$ \\
227 & 208.1 & 56.2 & \cite{2018for}$^{1}$ \\
408 & 133 & 10  & \cite{smc_l} \\
1400 & 35 & 6 &  \cite{wong2011}$^{1}$ \\
1400 & 41.2 & 6 &  \cite{Haynes91}$^{1}$ \\
2450 & 25.6 & 7  & \cite{Haynes91}$^{1}$ \\
4750 & 18.1 & 5  &  \cite{Haynes91}$^{1}$ \\
8550 & 12.4 & 5 &  \cite{Haynes91}$^{1}$ \\
\hline

\end{tabular}

    \begin{tablenotes}
      \small
      \item Notes.\\
      $^1$Re-integrated for ${R}\leq 3\degr$.\\ 
        $^{2}$Revised by \cite{smc_l}.  
      \end{tablenotes}           
}
    \end{threeparttable}

\end{table*}

\section{Discussion} \label{sec:dis}
 After mapping the extinction and de-reddening the H$\alpha$ emission, we presented the distribution of the thermal and non-thermal components of the radio continuum emission across the MCs. In this section, we first derive total magnetic field strength, compare the thermal and magnetic energy densities, and investigate a possible correlation between the magnetic field and massive star formation. Moreover, we compare different cooling mechanisms of the CREs in the MCs.
  
\subsection{Magnetic field strength} \label{ssec:b_s}

Assuming equipartition between the energy densities of the magnetic field and cosmic-rays (CRs) and using synchrotron intensity (${I}_\text{n}$), the total magnetic field strength can be derived \citep[e.g.,][]{trt2008,beck2005}:

\begin{equation}
B_\text{tot} = {C}(\alpha_\text{n}, {K}, {L}) \: {I}_\text{n}^{\frac{1}{\alpha_\text{n}+3}}
\end{equation}

\begin{figure*}
\centering
\includegraphics[width=0.80\columnwidth]{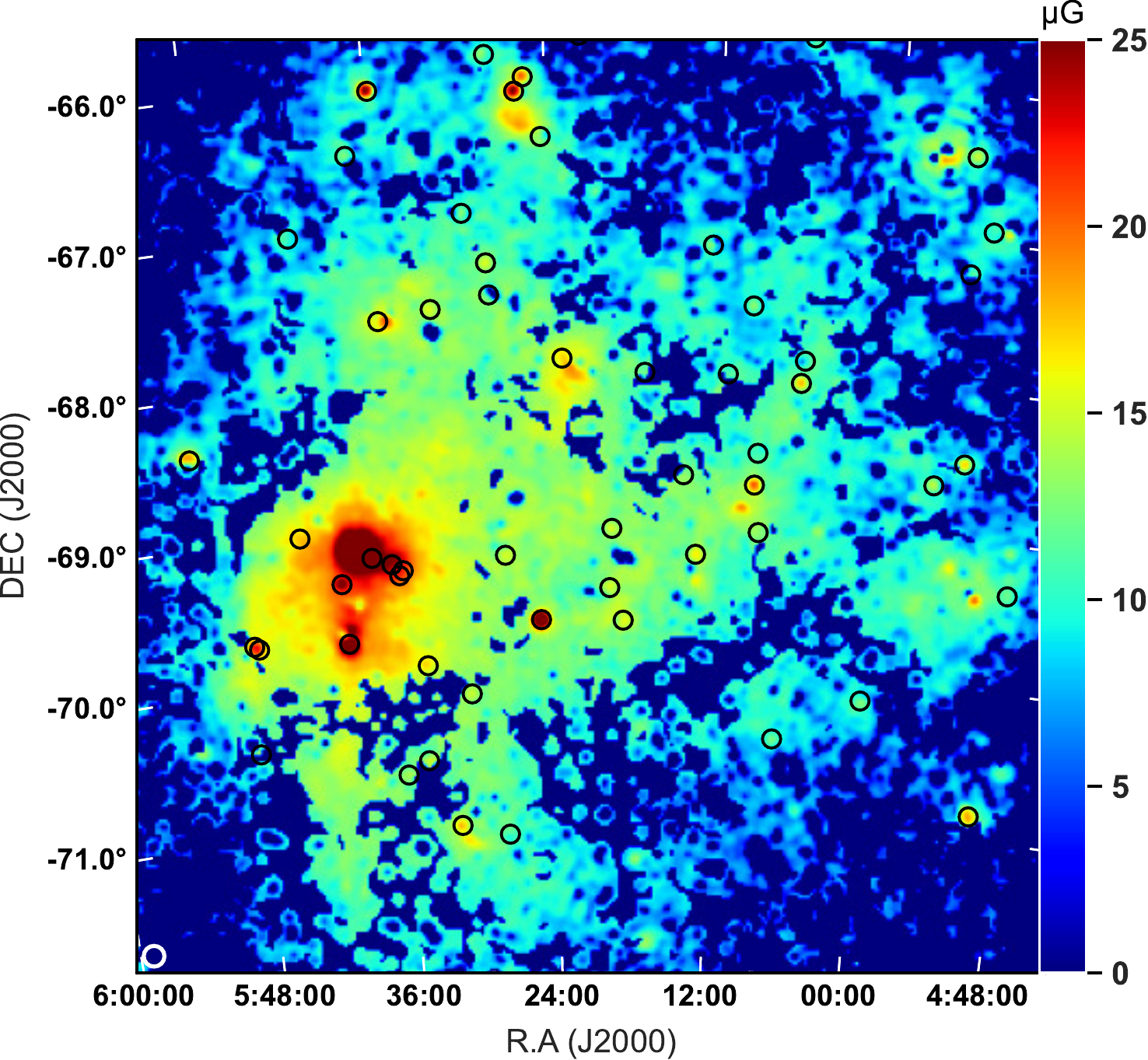}
\includegraphics[width=0.83\columnwidth]{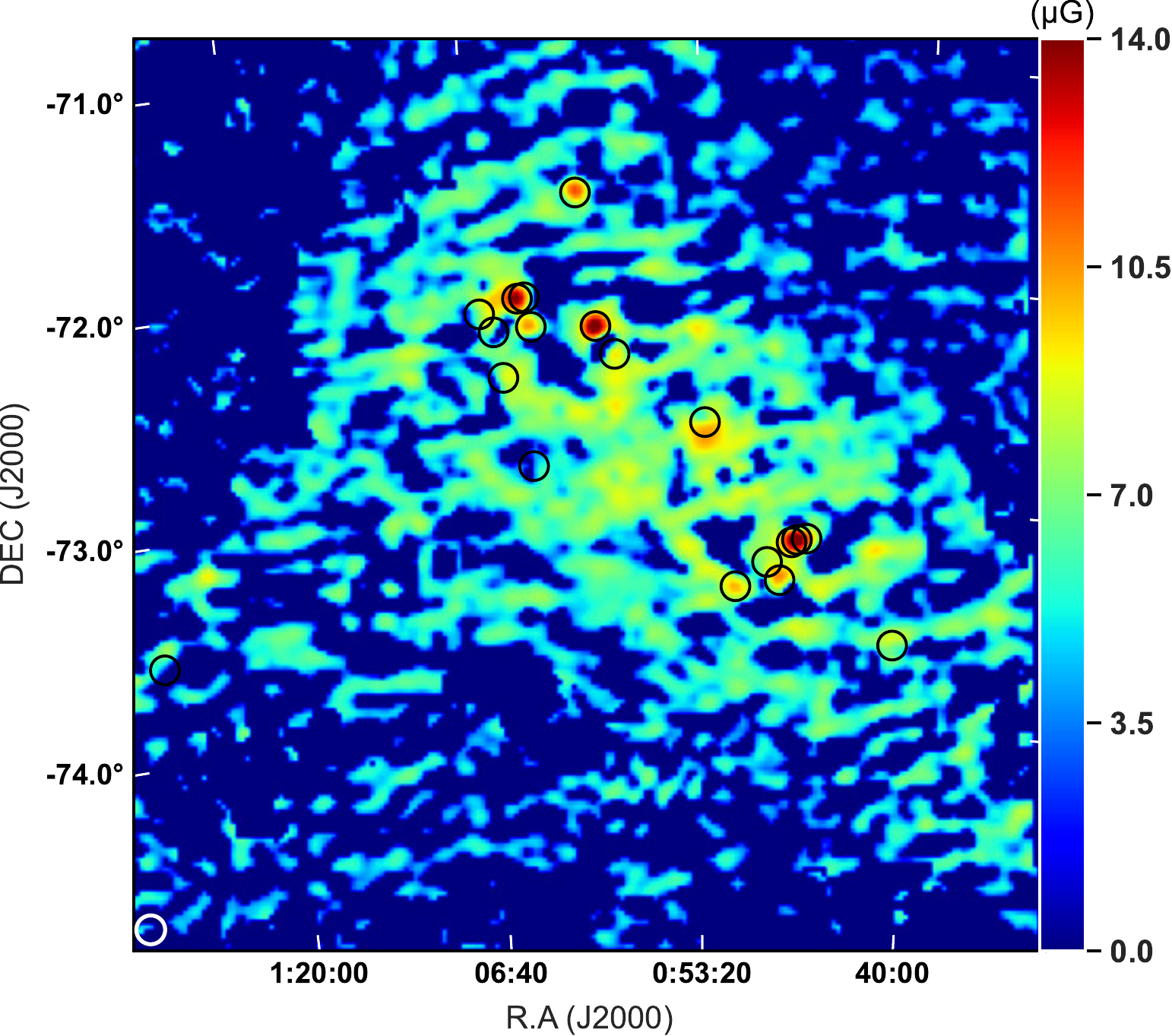}
 \caption{ Total magnetic field strength of the LMC (left) and the SMC (right) derived using the background source subtracted non-thermal intensity maps at 4.8\,GHz in the LMC and at 1.4\,GHz in the SMC. Colour bars show the field strength in units of $\mu$G. The beam sizes of 221\arcsec (for the LMC) and 235\arcsec (for the SMC) are indicated in the lower left corners. Black circles show the position of confirmed SNRs in the LMC and the SMC. \citep{bozzetto_2017,smc_snr}.} 
\label{fig:magnetic_field}
\end{figure*}

 \noindent 
Where $C$ is the function of the non-thermal spectral index, $K$ the ratio between the number densities of cosmic-ray protons and electrons, and $L$ is the synchrotron emitting medium's pathlength. Assuming a fixed ${K}=100$ \citep{beck2005}, yields maps of the total magnetic field strength (Fig.~\ref{fig:magnetic_field}).

Using our non-thermal maps, the mean non-thermal spectral index $\alpha_\text{n}=-0.67$,  and  a synchrotron pathlength of ${L} = 530$\,pc \citep{Gaensler2005} in the LMC, we derive a mean total magnetic field $B_\text{tot}=10.1$ $\pm$ 1.7\,$\mu$G  with errors the uncertainties in the non-thermal intensity and $\alpha_\text{n}$ propagated. This is higher than the previous equipartition estimate given by \cite{klein_lmc}. The magnetic field is maximum in 30~Dor, $B_\text{tot} =  45 \pm 7$\,$\mu$G, but few other \ion{H}{ii} regions such as N48B and N63A also show the presence of strong magnetic fields. We note that the synchrotron intensity measures the total magnetic field perpendicular to the line of sight. The magnetic field along the line of sight is about 4.3\,$\mu$G as estimated by \cite{Gaensler2005} using the Faraday rotation measures. Our equipartition assumption agrees with \cite{Mao2012}  who showed that this condition holds in the LMC using energy density of cosmic-ray with an ordinary cosmic-ray proton to electron ratio $K=100$ and an upper limit of $B_\text{tot}=7$\,$\mu$G.

For the SMC, large uncertainties are found with measurements of $L$ in literature. Assuming a 1\,kpc thick-disk for this galaxy, ${L}=1\,\text{kpc} \times \text{cos(i)}^{-1}=2\,\text{kpc}$, we obtain total magnetic field strength $B_\text{tot}=5.5$ $\pm$ 1.3\,$\mu$G using $\alpha$$_\text{n}=-0.89$ that agrees with \cite{smc_l}. However, adoption of a lower inclination $\text{i} = 40\degr$ from \cite{Stanimirovi2004}, decreases the magnetic field strength by 10 per cent in the SMC. The total magnetic field strength is higher than that \cite{Mao_2008} presented. It also appears that the magnetic field of the SMC is much more dominated by diffuse synchrotron emission more than the LMC. Previous studies also show (total) magnetic field strength in dwarf galaxies is about three times weaker than in spiral galaxies \citep{dwarf_mag}.

We note that assuming revised minimum energy formula following \cite{beck2005}, does not change total magnetic field strength significantly (less than 5 per cent). The total magnetic field strength obtained from TRT’s non-thermal emission is higher than using “classical" non-thermal maps. Using a fixed radio spectral index for separation in all regions (including \ion{H}{ii} regions) causes lower non-thermal emission and hence underestimates magnetic field strength.

\subsection{Thermal vs magnetic energy density}
The corrected H$\alpha$ emission or the thermal free-free emission is an ideal tracer of the density of thermal electrons $n_\text{e}$ \citep{Condon}. The emission measure $EM$ of the thermal emission is related to $n_\text{e}$ as $EM=\int{ n_{e}^2.\,dl}=\langle n_\text{e}^2\rangle\,L$, with $L$ the line-of-sight pathlength of the ionized medium\footnote{It is assumed to be the same as of the magneto-ionic medium (Section~\ref{ssec:b_s}).}. The volume-averaged electron density along the line of sight is given by $\langle n_\text{e} \rangle=\sqrt{f\langle n_\text{e}^2\rangle}$, with $f$ the volume filling factor describing the fluctuations in $n_\text{e}$ \citep[$f\simeq$5 per cent following][]{Gaensler_8, Ehle}.

A large variation in $\langle n_\text{e} \rangle$ is obtained in the LMC ranging from 0.01\,$\text{cm}^{-3}$ in weak diffuse regions to higher than 1\,$\text{cm}^{-3}$ in dense areas of 30~Dor with a mean of 0.045\,$\text{cm}^{-3}$ (median 0.033\,$\text{cm}^{-3}$). Considering only 30~Dor, we obtain $\langle n_e \rangle=0.30 \pm 0.01$ on average. Assuming $L=530$\,pc, the mean volume-averaged electron density density increases to 0.062\,$\text{cm}^{-3}$ that agrees with the distribution modeling of \cite{yao}.

For the SMC, we obtain $\langle n_\text{e} \rangle=0.020 \pm 0.019\,\text{cm}^{-3}$ with error the standard deviation. This agrees with \cite{Mao_2008} deriving a mean electron density of $0.039\,\text{cm}^{-3}$ using the pulsar dispersion measure technique.

To address the energy balance in the magneto-ionic ISM of the MCs, the thermal energy density ($E_{th}=\frac{3}{2} \langle n_{e} \rangle kT_e$) is compared with the magnetic energy density ($E_{\textrm B}=B^2/8\pi$) for the warm ionized gas $T_{e} \simeq 10^4$\,K. The energy density of the hot ionized gas with $T_{e} \simeq 10^6$\,K and an electron density of $\simeq 0.01 \langle n_{e} \rangle$ is about the same order of magnitude as the warm ionized gas energy density assuming the pressure equilibrium between the warm and hot ionized gas \citep[e.g.,][]{Ferrire}. On average, the resulting magnetic energy density is larger than the thermal energy density by about one order of magnitude in both the LMC and the SMC. This means that the ionized ISM is a low-beta plasma ($\beta\equiv\,E_{th}/E_{B}<1$) in these galaxies. In other words, the ionized ISM is magnetically confined to electron densities of $\langle n_{e} \rangle = 0.06 \pm 0.01$ in the LMC and $\langle n_{e} \rangle = 0.020 \pm 0.019$ in the SMC. Moreover, the total non-thermal pressure inserted from both CREs and magnetic fields (it is twice the magnetic pressure in case of equipartition) dominates the thermal pressure in the ionized phase of the ISM. We summarize magneto-ionic plasma physical properties of the MCs in Table~\ref{tab:phys}.

 \subsection{Magnetic field--star formation rate correlation}
 Studies show that the magnetic field scales with recent SFR globally in galaxies following a power-law index of about $\gamma = 0.3$ \citep[e.g.,][]{Heesen2014,dwarf_mag,Tabatabaei_2017}. That is linked to the amplification of the magnetic field in star-forming regions \citep{CR_driven,small_scale}. However, deviations are found on local scales \citep[e.g.,][]{mag_in_4254,tab_2013}. It is hence important to dissect the role of the diffuse ISM and spatial resolution in these studies. 
Thanks to their proximity, the MCs are ideal for testing this correlation at different resolutions down to $\sim$\,50\,pc. It is shown that the 70\,$\mu$m infrared emission can well trace the recent star formation (over the last 100\,Myr) in the MCs via the following relation \cite{70sfr} that is calibrated for the \ion{H}{ii} regions in the MC:
  
 \begin{equation}
 \left( \frac {{\text{SFR}}_{70\,\mu m}} {\text{M}_{\sun} \: \text{y}r^{-1} } \right) = 9.7 \: \times 10^{-44} \: \left( \frac{{L}_{70\,\mu m}}{\text{erg}\, \text{s}^{-1}} \right) 
 \end{equation}

\begin{figure*}
\centering
\includegraphics[width=0.4\textwidth]{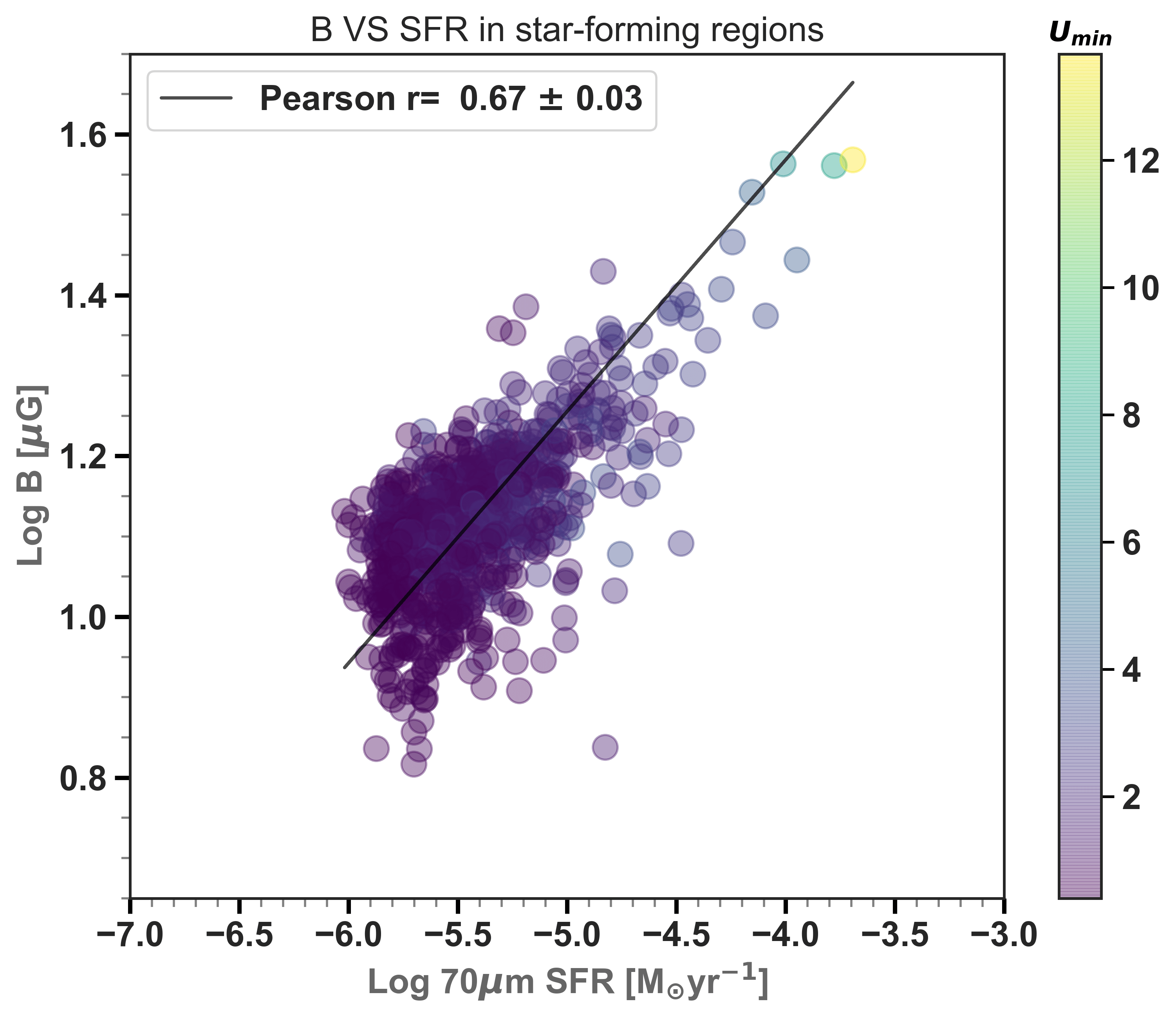}
\includegraphics[width=0.4\textwidth]{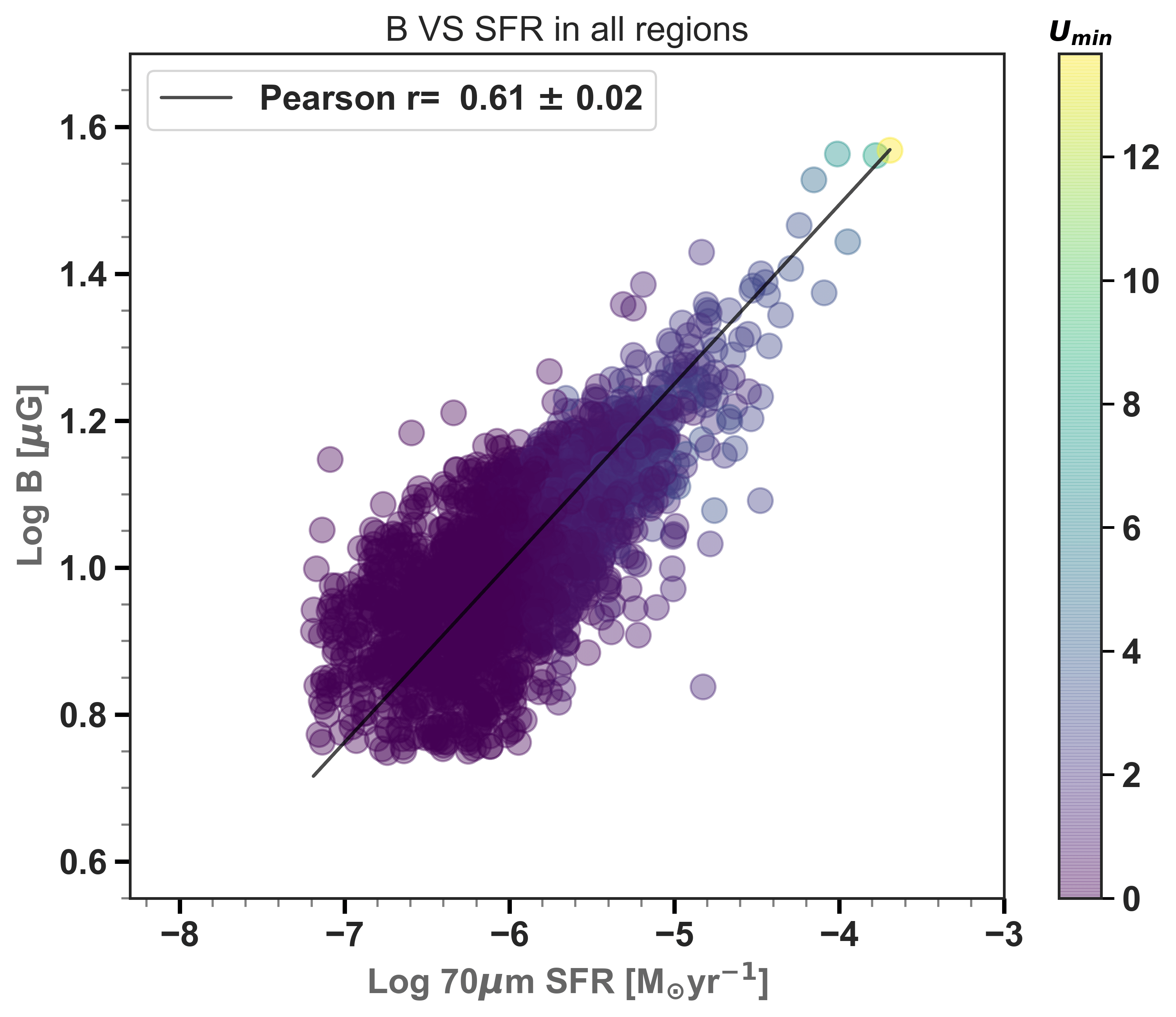}
\includegraphics[width=0.4\textwidth]{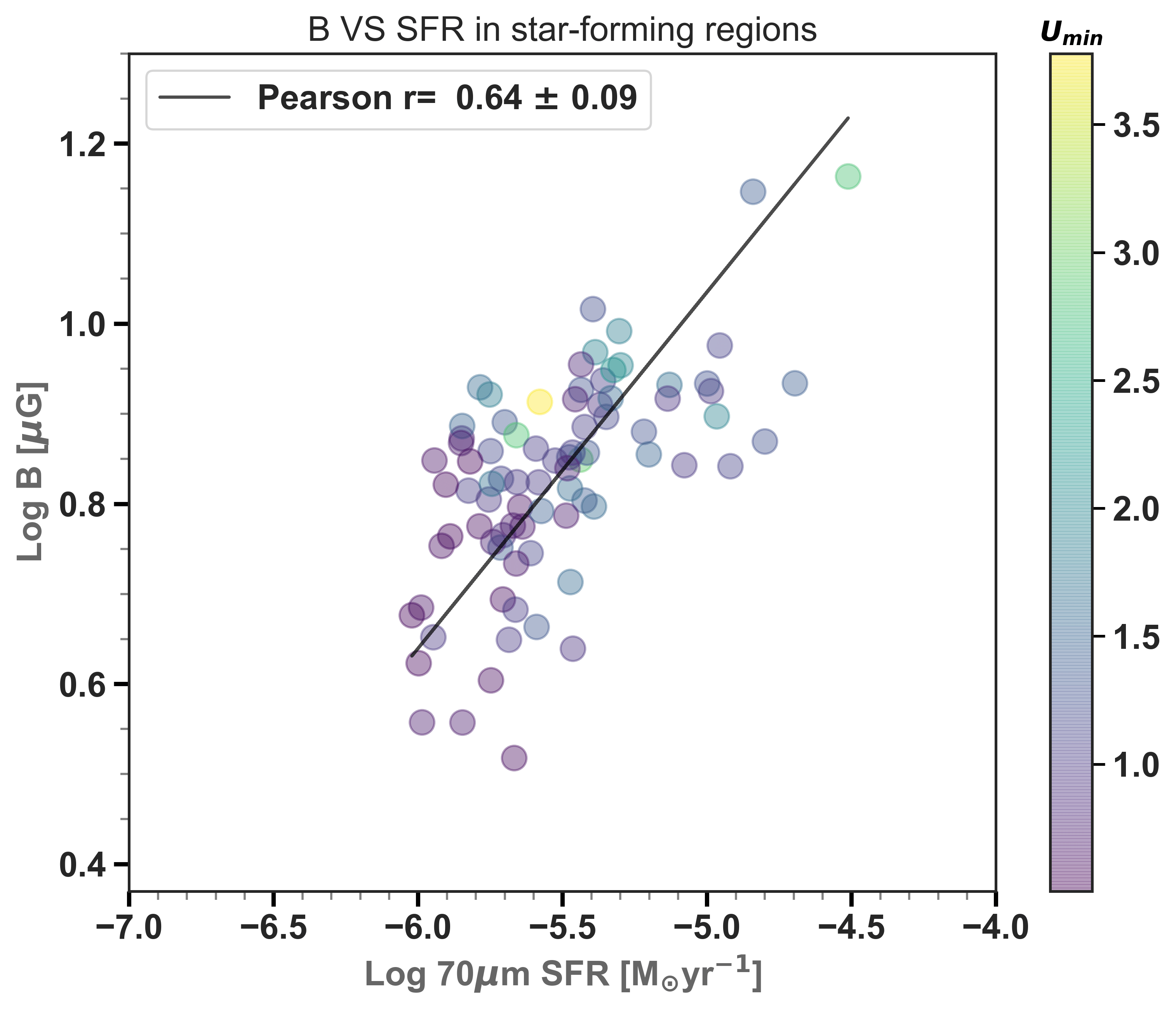}
\includegraphics[width=0.4\textwidth]{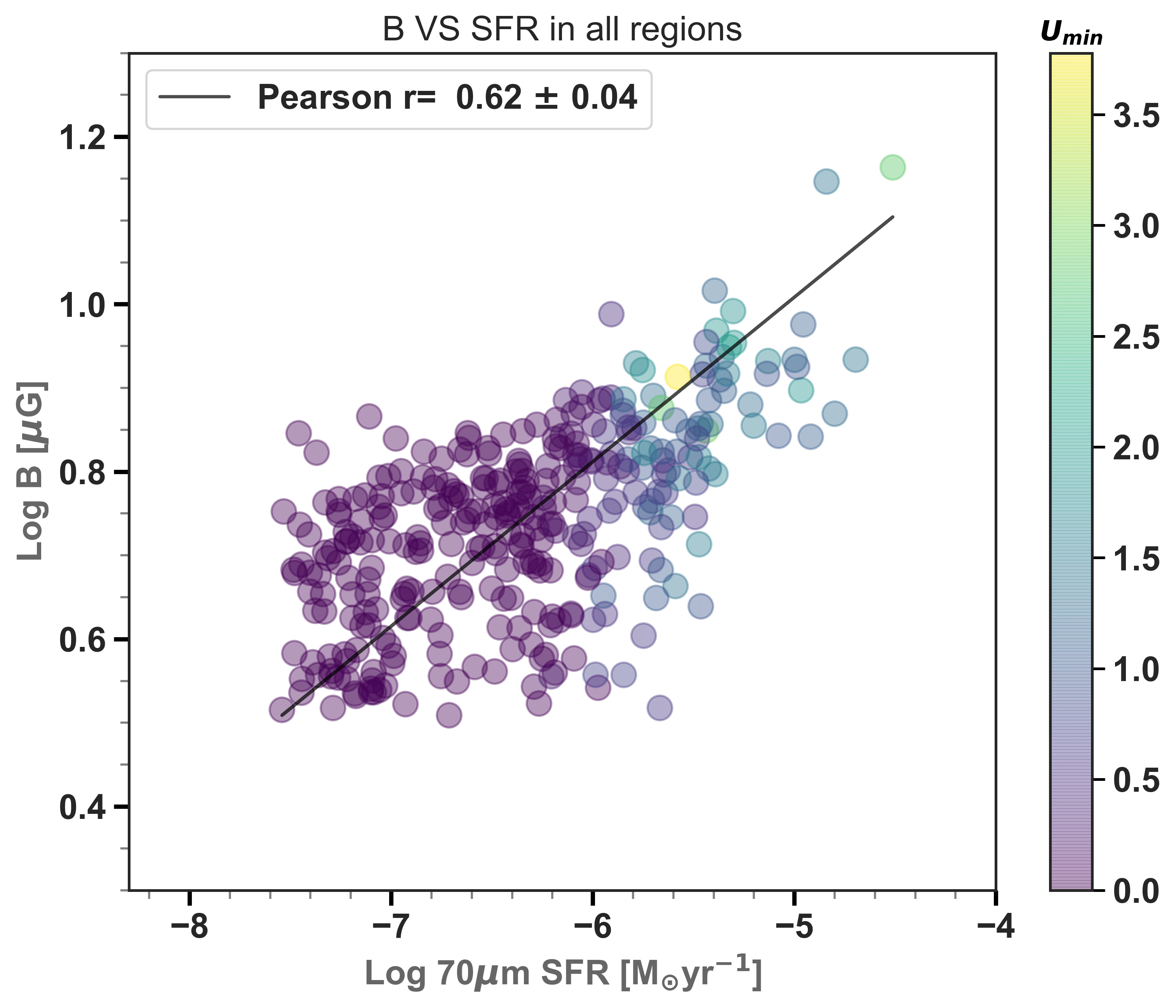}

\caption{ Correlation between the total magnetic field strength B and the 70$\mu$m emission as a SFR tracer for the LMC (top) and the SMC (bottom) in star-forming regions (left) as well as all regions above the $3\sigma$ rms level (right). Solid line shows the bisector fit and the bars indicate $U_\text{min}$ in units of $U_{\sun}$.}
 
\label{fig:b_sfr_plt}
\end{figure*}

We investigate the B--SFR correlation by separating the diffuse ISM from star-forming regions at different resolutions. Using $U_\text{min}$ maps from \cite{Chastenet_2019}, we consider star-forming regions in the LMC ($U_{\text{min}} >0.4$) and the SMC ($U_{\text{min}} >0.5$). The correlation between magnetic field strength and 70\,$\mu$m SFR is obtained through a beam independent spacing, yielding Pearson correlation coefficient r $\sim$ 0.6 in the LMC and the SMC. The B vs SFR can best be explained by a power-law relation with an index $\gamma$ fitted using the bisector OLS \cite{Isobe} regression (Table~\ref{tab:b_sfr_table}). Fig.~\ref{fig:b_sfr_plt} present the correlation of B with SFR in spatial resolution of 53\,pc for the LMC and 71\,pc for the SMC.

We find that B and SFR traced by the 70\,$\mu$m are correlated in the SF regions at all selected resolutions. In the LMC, the power-law exponent agrees with the theoretical turbulent amplification value \citep{bsfr}. However, in the SMC, it is slightly lower at the original resolution. As expected, no correlation holds in the diffuse ISM at the native resolution. However, it is interesting to note that this correlation is increased at lower resolutions due to mixing with the SF regions.
The exponent of $\gamma_\text{LMC} = 0.24 \pm 0.01$ and $\gamma_\text{SMC} = 0.20 \pm 0.01$ obtained is in excellent agreement with global studies of Magellanic-type galaxies with slope $\gamma = 0.25 $ $\pm$ 0.02 given by \citep{mag_in_mcs}.
This experiment shows that, in local studies, the power-law exponent can become flatter than the theoretical value because of contamination by the diffuse ISM.  
  
\begin{table*}
  \begin{threeparttable}
  
  \renewcommand{\arraystretch}{1.35}
 \caption{ Correlation between the magnetic field strength and SFR in different regimes of radiation field (SF, ISRF) at different linear resolutions.}
 \label{tab:b_sfr_table}
 \begin{tabular}{llllllllll}
  \hline
      Y  &  $\gamma^\text{total}$  &  $r_\text{p}^\text{total}$ &  $t(n)$$^\text{total}$ &  $\gamma^\text{SF}$  & $r_\text{p}^\text{SF}$ & $t(n)$$^\text{SF}$  & $\gamma^\text{ISRF}$   & $r_\text{p}^\text{ISRF}$ & $t(n)$$^\text{ISRF}$   \\
  \hline
  LMC\\
   B (53\,pc) &  0.24 $\pm$  0.01  & 0.61  $\pm$ 0.02 & 159.70(2166) &  0.31  $\pm$ 0.01 &   0.67  $\pm$ 0.03 & 116.83(843) &  -  & 0.39 $\pm$ 0.03 & 177.11(1281)   \\
   
    B (150\,pc) &  0.28 $\pm$  0.01  & 0.57  $\pm$ 0.04 & 51.95(355) &  0.29  $\pm$ 0.03 &   0.56  $\pm$ 0.08 & 42.07(120) &  0.30  $\pm$ 0.02  & 0.49 $\pm$ 0.06 & 60.16(235)   \\

     B (300\,pc) &  0.31 $\pm$  0.02  & 0.78  $\pm$ 0.08 & 22.59(63) &  0.31  $\pm$ 0.04 &   0.70  $\pm$ 0.15 & 23.28(26) &  0.30  $\pm$ 0.05 & 0.50 $\pm$ 0.15 & 25.36(37)   \\

\hline

  SMC\\
   B (71\,pc) &   0.20 $\pm$ 0.01 & 0.62 $\pm$ 0.04 & 65.18(328) & 0.40 $\pm$ 0.04 &  0.64 $\pm$ 0.09 & 32.96(82) &  -  &  0.38 $\pm$ 0.06 &  70.49(246) \\
   
   B (150\,pc) &  0.24 $\pm$ 0.01 &  0.59  $\pm$ 0.07  & 38.53(156) &  0.43 $\pm$ 0.11 & 0.53 $\pm$ 0.21 & 20.07(19) & 0.28 $\pm$ 0.03 & 0.51 $\pm$ 0.08   & 41.46(132) \\   
  \hline
 \end{tabular}
         \begin{tablenotes}
      \small
      \item { The linear fit obtained in logarithmic scale (Log Y = $\gamma$ Log X ) using the bisector least square fit.} \end{tablenotes}
      
  \end{threeparttable}
 
\end{table*}

 \subsection{Cooling of cosmic-ray electrons}

   \begin{figure*}
\centering
\includegraphics[width=0.8\columnwidth]{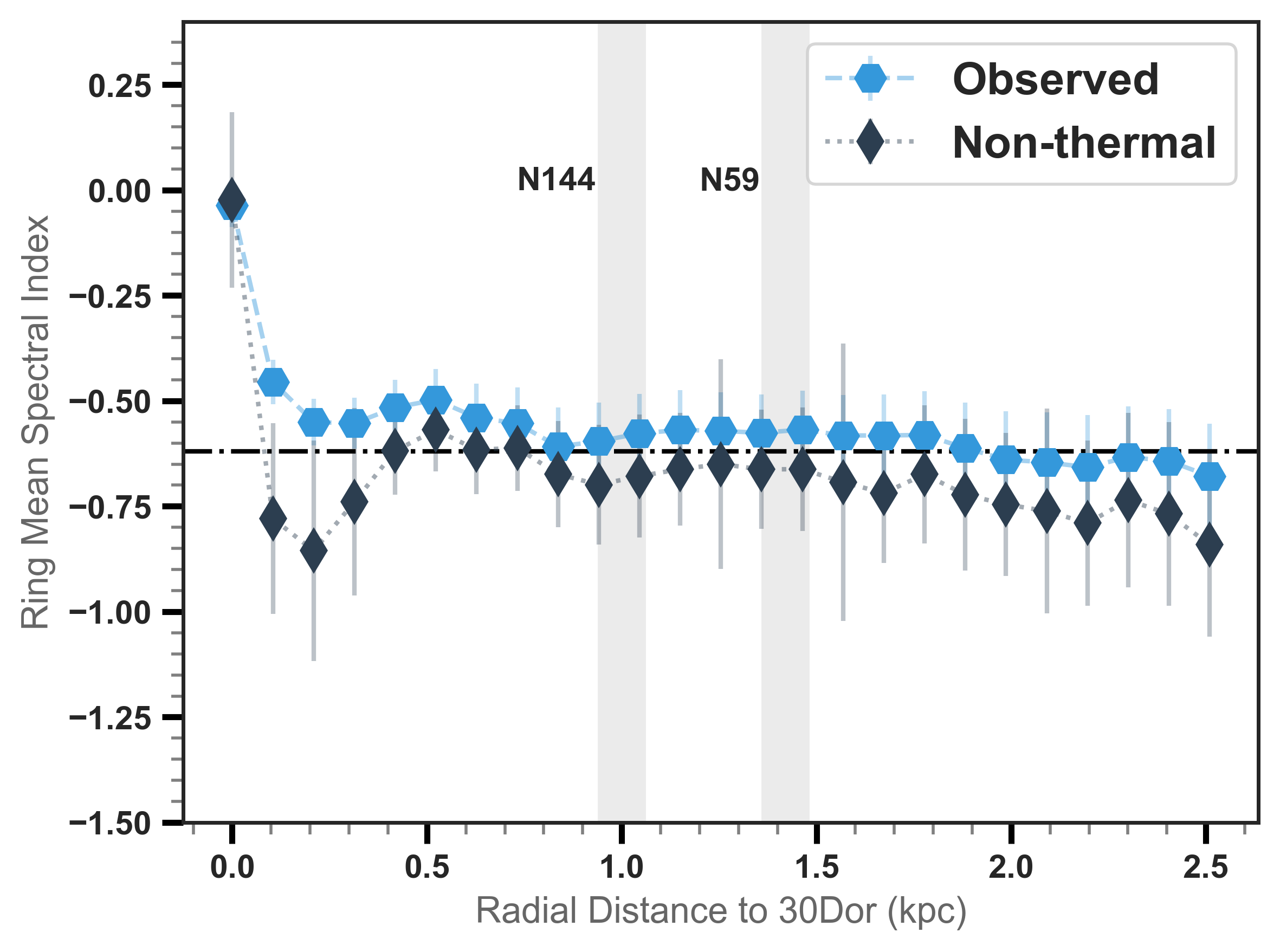}
\includegraphics[width=0.8\columnwidth]{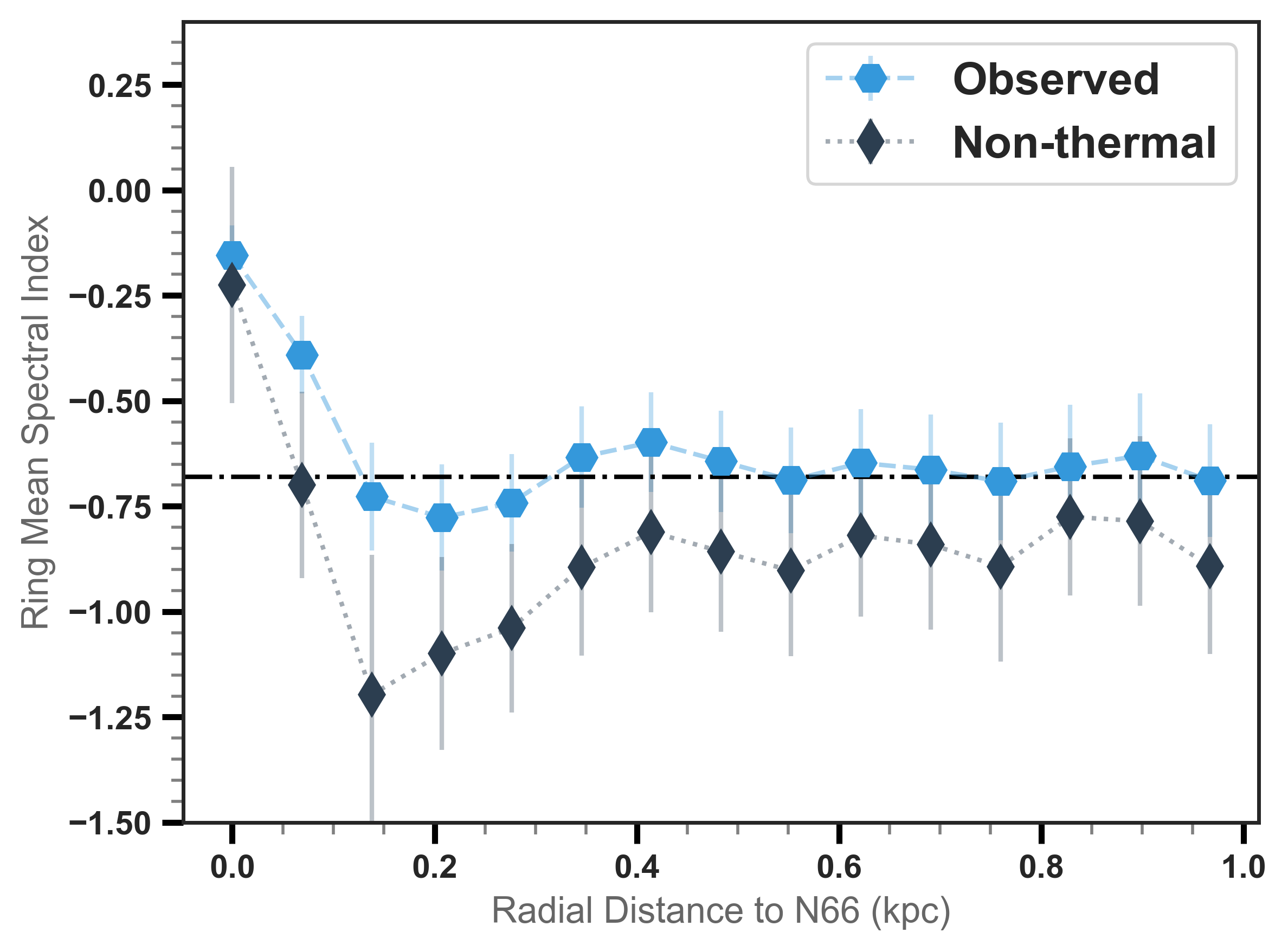}

\includegraphics[width=0.8\columnwidth]{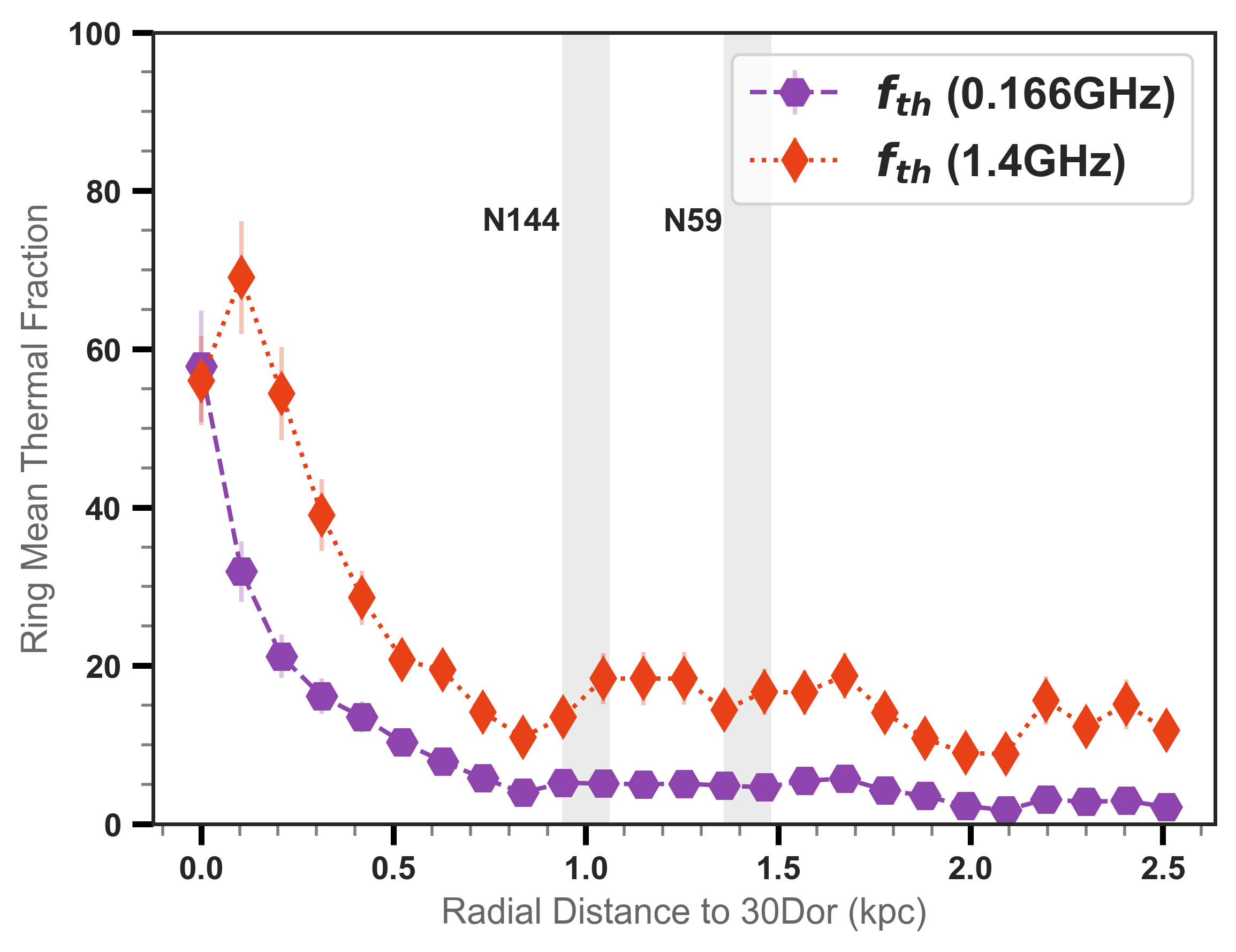}
\includegraphics[width=0.8\columnwidth]{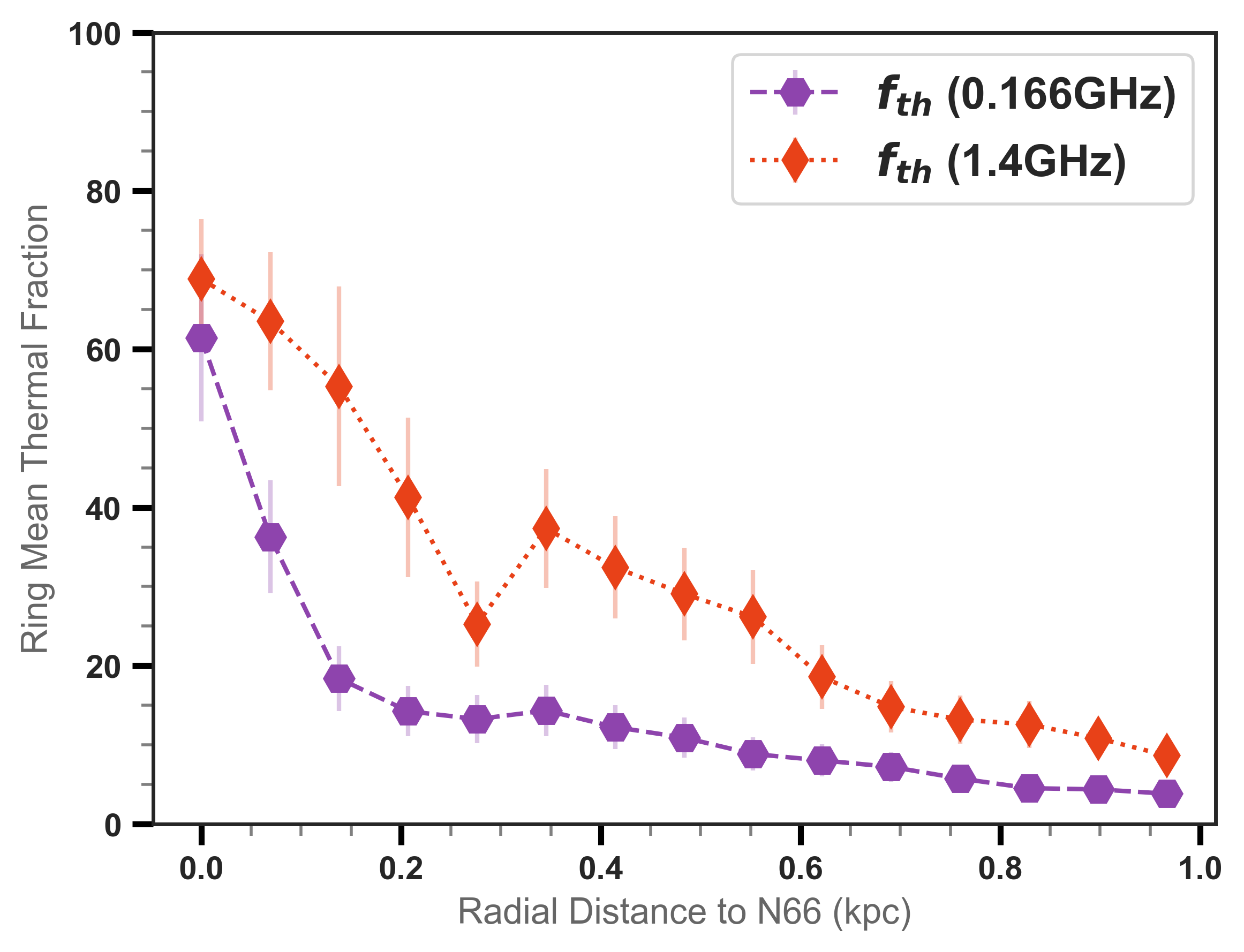}

\caption{ Radial profiles of the spectral index (top) and thermal fraction (bottom) in 30~Dor (left) and N66 (right). The dashed horizontal lines in the spectral index profiles indicate the average galaxy values. The vertical grey bands show the position of the 30~Dor's neighbouring \ion{H}{ii} regions N144 and N59. }
 
\label{fig:ring_CR}
\end{figure*}

The power law energy distribution of CREs $dN/dE$ $\propto$ $E^{-\beta}$ where $\beta$ is related to the observed spectral index with $\alpha = -(2\beta-1)$ could change due to various energy loss mechanisms. The non-thermal spectral index becomes steeper due to different energy losses of the CREs’ diffusion from primary sources (e.g., \ion{H}{ii} regions). In order to clarify the most effective energy loss mechanism of CREs in \ion{H}{ii} regions, we study two well-known \ion{H}{ii} regions in the MCs. We measure the radial profile of the average observed/non-thermal spectral index and thermal fraction centred on (05$^{\text{h}}$\,38$^{\text{m}}$\,42$^{\text{s}}$,\,-69$\degr$\,06$\myprime$03$\mydprime$)$_\text{J2000}$ and (0$^{\text{h}}$\,59$^{\text{m}}$\,05$^{\text{s}}$,\,-72$\degr$\,10$\myprime$\,41$\mydprime$)$_\text{J2000}$ in 30~Dor and N66 \ion{H}{ii} complex respectively using beam independent rings of 7.2$\arcmin$ radius in 30~Dor and 3.91$\arcmin$ in N66. Fig.~\ref{fig:ring_CR} shows the radial profile of the spectral index and thermal fraction from the centre of 30~Dor and N66. The spectral index profile of 30~Dor starts with very flat spectrum $\alpha$ $\sim$\,0 in the centre but it drops rapidly moving away from the centre. At a distance of ${R}=$0.3\,kpc,  $\alpha_{\text{n}}$ reaches to a steep value ($\sim -0.85$) and again flattens to about the average galaxy value. The flattening of the synchrotron spectrum around ${R}=0.5$\,kpc is most likely due to the ring-like artefact in the observed RC map at 1.4\,GHz \citep{hughes_2011} which also leads to an artificially higher thermal fraction.

The  spectral index of 30~Dor starts to fall off after ${R}>0.5$\,kpc due to the diffusion of CREs, but it has again two slight flattening and higher thermal fractions around $R\sim$1\,kpc and $ R \sim$ 1.4\,kpc. This can be interpreted as re-acceleration of CREs due to feedback and shock waves from the two nearby \ion{H}{ii} regions, N144 and N59. Overall, 30~Dor radio spectral index would decrease slightly to the galaxy average value $\alpha = -0.64$ at the scale of $ R \sim$ 2\,kpc. Beyond ${R}>2$\,kpc, the non-thermal spectral index $\alpha_{\text{n}} \sim -0.85$, never attaining to characteristic values of CREs synchrotron losses \citep[i.e., $\alpha_\text{n} <-1$, ][]{CRE}. Similar to 30~Dor, N66 \ion{H}{ii} complex of the SMC also shows flat values of $\alpha=-0.1$ in the core (${R}<0.1$\,kpc). The non-thermal spectral index $\alpha_\text{n}$ decreases to stepper values of $\alpha_\text{n}<-1$ for ${R}<0.3$\,kpc of N66. Such a steep $\alpha_\text{n}$ could be interpreted as CREs energy loss due to synchrotron loss mechanism. However, the measurement suffers from large uncertainties due to the low sensitivity of the radio continuum data at 1.4\,GHz. After ${R}>0.3$\,kpc, the observed spectral index increases to the mean value of the galaxy, and $\alpha_\text{n}$ does not reach to steeper values than $<-1$, indicating that the synchrotron energy loss mechanism is not an efficient cooling mechanism in N66 as well. In order to assess different CREs cooling processes in the core of these \ion{H}{ii} regions (${R}<0.2$\,kpc), we estimate their cooling timescale using the equations presented in \cite{FIR_cooling}:

 \begin{equation}
 \begin{split}
 \left( \frac {\tau_\text{syn}}{\text{yr}} \right) \approx 4.5 \times 10^{7} \: \left( \frac{B}{10 \mu \text{G}} \right)^{-3/2} \: \nu^{-1/2}_{\text{GHz}} \\
 \left( \frac {\tau_\text{ion}}{\text{yr}} \right) \approx 2.1 \times 10^{8} \: \left( \frac{B}{10\mu \text{G}} \right)^{-1/2} \: \nu^{1/2}_{\text{GHz}} \: \left( \frac {n_\text{eff}}{\text{cm}^{3}} \right)^{-1}  \\ 
 \left( \frac {\tau_\text{IC}}{\text{yr}} \right) \approx 1.8 \times 10^{8} \: \left( \frac{B}{10 \mu \text{G}} \right)^{1/2} \: \nu^{-1/2}_{\text{GHz}} \: \text{U}^{-1}_{\text{rad},-12}  \\
 \left( \frac {\tau_\text{brem}}{\text{yr}} \right) \approx 3.7 \times 10^{7} \: \left( \frac {n_\text{eff}}{\text{cm}^{3}} \right)^{-1} \\
  \left( \frac {\tau_\text{diff}}{\text{yr}} \right) \approx 2.6 \times 10^{7} \: \left( \frac {\text{E}}{3\text{GeV}} \right)^{-1/2} \\
 \end{split}
 \end{equation}
  
\noindent 
Where the effective ISM number density experienced by CRs $n_\text{eff} = f_\text{ISM} <n>$ and mean ISM number density within the CREs confinement volume $<n> = \Sigma_\text{gas}/2h$. We note that CRs traverse in high-density clumpy ISM which is confined by clumps magnetic field experiences a higher $f_\text{ISM}$>1 in comparison with a low-density diffuse ISM. We assume $f_\text{ISM}=1$ for the MC.

However, taking higher $f_\text{ISM}>1$ for the compact \ion{H}{ii} regions of the MCs, decreases the cooling timescale of bremsstrahlung and ionization losses, making them more effective cooling mechanisms for CR electrons and positrons by propagating away from \ion{H}{ii} regions. We assume the scale height of CR $ {h} = 1$\,kpc. The radiation energy density $U_\text{rad}$ is obtained from \cite{draine2007} and corrected for contributions of the cosmic microwave background radiation energy density $U_\text{CMB}= 4.17$ $\times$ 10$^{-13}$\,erg\,cm$^{-3}$ and dust emission energy density $U_\text{FIR} = 5$ $\times$ 10$^{-13}$\,erg\,cm$^{-3}$ \citep{tab_2013,draine2011}. In order to determine the diffusion timescale, the most energetic CR electrons emit at a critical frequency $\nu_{c}$ obtained from \cite{CREs_30dor}:

 \begin{equation}
 \left( \frac {E} {\text{GeV}} \right) = 8.8 \: \left( \frac{\nu_{c}} {\text{GHz}} \right) ^{1/2} \: \left( \frac{B} {\mu G} \right) ^{-1/2}
 \end{equation}
  
 \noindent 
Out to the radius of 200\,pc from 30~Dor central point, total magnetic field strength map yields an average $B_\text{tot} = 26.43$\,$\mu$G and corresponding CR electron energy $\sim 3.8$\,GeV at 4.8\,GHz. We found the average number density of the ISM experienced by the CRs $n_\text{eff} \sim$ 0.94\,cm$^{-3}$ and average total radiation energy density $U_\text{rad} = 5.06$\,$U_{\sun}$ in 30~Dor. These estimation suggests that synchrotron escape is the dominant cosmic-ray loss mechanism in 30~Dor core $\tau_\text{syn}$ $\sim 4.7$\,Myr and there is a strong competition between Inverse-Compton cooling timescale $\tau_\text{IC}$ $\sim 26.3$\,Myr and diffuse cooling $\tau_\text{diff}$ $\sim 23.2$\,Myr for CR, while the ionization cooling has a less cooling effect on CR with a characteristic time scale $\tau_\text{ion}$ $\sim 300.8$\,Myr.

Integrating out to the radius of 100\,pc from the SMC N66 complex, we found a mean $B_\text{tot} = 8.50$\,$\mu$G, CR electron energy $\sim 3.5$ GeV at 1.4\,GHz, and $n_\text{eff} \sim$ 0.73\,cm$^{-3}$. The diffuse cooling is the most effective cooling mechanisms with $\tau_\text{diff}$ $\sim 23.8$\,Myr for CR in this region. Furthermore, it shows a strong competition between synchrotron escape, $\tau_\text{syn}$ $\sim 48.5$ Myr and Inverse-Compton cooling, $\tau_\text{diff}$ $\sim 45.7$\,Myr. Although Ionization loss is the weakest cooling mechanism in the N66, however, taking the contribution of $H_{2}$ in the total gas surface density will decrease its cooling time scale.
  
\subsection{Recombination scenario}  \label{ssec:recom}
Determining extinction and free-free emission depends on the optical depth of Lyman continuum photons and whether the optically thick condition known as Case B recombination or the optically thin scenario known as case A recombination is assumed. Although the theoretical Balmer decrement ratio $j_{\text{H}\alpha}$/$j_{\text{H}\beta}=2.86$ is almost the same for both scenarios under the assumption of $T_{\text{e}}=10,000$\,K \citep{Osterbrock}, but variation of EM is significant depending on the electron density as well as the optical depth of Lyman continuum photons \citep{vallas}. \cite{Pellegrini_2012} estimated that the escape fraction of Lyman continuum photons is at least $f_{\text{esc}}=0.42$ and $f_{\text{esc}}=0.40$ in the \ion{H}{ii} regions of the LMC and SMC, respectively. Similar results were obtained by \cite{Kennicutt95}, i.e., $f_{\text{esc}}=0.51$,  based on diffuse to total H$\alpha$ emission flux for the LMC \citep{Oey1997}. Adopting the case A recombination increases the thermal emission flux by about 49-51 per cent for both the LMC and the SMC at 0.166\,GHz and 1.4\,GHz. Consequently, the global magnetic field strength decreases by about 16 per cent.

\section{Conclusion} \label{sec:con}
 Using multi-frequency observations in the radio, FIR, and optical domains, we presented the most precise picture of the distribution of the thermal and non-thermal RC emission in the MCs ever with no use of any assumption about the synchrotron spectral index or extinction. The highly resolved and sensitive Spitzer and Herschel observations have allowed mapping dust mass in galaxies in general. However, using these maps to determine extinction needs a priori assumption about the distribution of ionized gas and dust along the line of sight. Combining the dust mass maps with the data of the Balmer decrements, we introduced a new method to determine the true fraction of dust that actually attenuates H$\alpha$ emission along the line of sight, $f_\text{d}$. This fraction is determined for the first time in the MCs in both \ion{H}{ii} regions and more diffuse regions in the ISM using the MUSE IFU observations. We also presented a calibration relation to map $f_\text{d}$ based on its anti-correlation with the neutral gas surface density. Using the de-reddened H$\alpha$ emission, we derived the free-free emission at 0.166, 1.4, and 4.8\,GHz and obtained maps of the non-thermal emission and the pure synchrotron spectral index in the MCs. The latter two maps allowed us to map the strength of the total magnetic field at a spatial resolution of $\sim$ 3.7$\arcmin$ in the LMC and $\sim$ 3.9$\arcmin$ in the SMC. Our results are summarized as follows.

(i) The dust optical depth $\tau_\text{dust}$ is optically thin ($\tau_\text{dust}<1$) to the H$\alpha$ photons in the MCs’s diffuse ISM. Whereas total dust mass surface density indicates $\tau_\text{dust}$ ranges between $2-5$ in \ion{H}{ii} regions of the MCs. Only a fraction of this dust mass plays a role in the attenuation of optical waves.

(ii) $f_\text{d}$ is lower in dense star-forming regions ($ \sim 0.1$) than in the more diffuse ISM ($> 0.2$).
Based on the anti-correlation between $f_\text{d}$ and total gas surface density, we find that the mean $f_\text{d}$ over the entire LMC is about 0.3 that agrees with the Milky Way value.  

(iii) The non-thermal fraction is significant at 0.166\,GHz with an average value of 80-90 per cent in the diffuse ISM of these galaxies and in massive star-forming regions such as N157. Supernova remnants contribute the most to our 0.166\,GHz non-thermal emission map with an average value of non-thermal fraction $>85$ per cent. Furthermore, thermal fraction at 1.4\,GHz is estimated to be $f_\text{th}=30$ per cent in the LMC and 35 per cent in the SMC. 

(iv)  The synchrotron spectrum steepens from massive star-forming regions $\alpha_\text{n} >-0.4$ to the diffuse ISM $\alpha_\text{n} <-0.7$. This indicates energy loss and cooling of CREs as they propagate away from their birthplaces in SF regions. After injection, CREs can also experience re-acceleration in complexes of SF regions flattening their spectrum as observed in 30~Dor. Comparing different mechanisms, we find that the synchrotron energy loss dominates cooling of CREs in 30~Dor core ($\text{R}<0.2$\,kpc). Using non-thermal maps between 0.166\,GHz and 1.4\,GHz, we found the median synchrotron spectral index $\alpha_\text{n} = -0.67 \pm 0.12$ and $\alpha_\text{n} = -0.89 \pm 0.17$ in the LMC and the SMC respectively. Using this synchrotron spectral index and equipartition assumption, we found an average total magnetic field strength ${B}_\text{tot}\simeq10.1 $ $\mu$G in the LMC and 5.5 $\mu$G in the SMC.

(v) This study shows that the ISM is dominated by a low-beta plasma given that the thermal energy density is smaller than the non-thermal energy density (E$_{\text{B}}$+E$_{\text{CR}}$) by more than one order of magnitude.

 It is worth mentioning that the present generation of the interferometric radio surveys of the MCs are limited to either low sensitivity caused by instrumental artefacts (particularly at 4.8\,GHz and 1.4\,GHz) or low resolution (particularly at 0.2\,GHz). The limitations in sensitivity can prevent studying the physics of the low-surface brightness ISM which constitutes a major part of low-mass and irregular galaxies such as the MCs. Designing the next generation of the RC surveys, it is important to attain a much larger dynamic range. Technically, limitations in dynamic range can be caused by instrumental artefacts or incomplete calibration of instrumental response \citep{Braun}. As discussed in Section~\ref{ssec:nt_map} and Section~\ref{sec:nt}, the resolution at which background radio sources are subtracted can affect the integrated flux density measurements as well as the spectral index analysis in the MCs. Hence, higher resolution observations are required (particularly at low-frequencies) to disentangle the ISM from the external sources more accurately. The upcoming surveys with the SKA-low can ideally overcome this issue in the MCs. Therefore, following these observational limitations, the main contribution of this paper is in setting a methodology for the unbiased separation of thermal vs non-thermal emission for the upcoming surveys of the MCs and nearby galaxies with the SKA and its pathfinders such as ASKAP and MeerKAT.

\section*{Data availability}
The data underlying this article will be shared on reasonable requests.

\section*{Acknowledgements}
The authors thank Miroslav D. Filipovic for providing the catalog of the background radio sources in the LMC and also his helpful comments improving the paper. HH thanks the School of Astronomy at the Institute for Research in Fundamental Sciences for its financial support. 




\bibliographystyle{mnras}
\bibliography{hassani_mcs} 

\begin{thebibliography}{}
\makeatletter
\relax
\def\mn@urlcharsother{\let\do\@makeother \do\$\do\&\do\#\do\^\do\_\do\%\do\~}
\def\mn@doi{\begingroup\mn@urlcharsother \@ifnextchar [ {\mn@doi@}
  {\mn@doi@[]}}
\def\mn@doi@[#1]#2{\def\@tempa{#1}\ifx\@tempa\@empty \href
  {http://dx.doi.org/#2} {doi:#2}\else \href {http://dx.doi.org/#2} {#1}\fi
  \endgroup}
\def\mn@eprint#1#2{\mn@eprint@#1:#2::\@nil}
\def\mn@eprint@arXiv#1{\href {http://arxiv.org/abs/#1} {{\tt arXiv:#1}}}
\def\mn@eprint@dblp#1{\href {http://dblp.uni-trier.de/rec/bibtex/#1.xml}
  {dblp:#1}}
\def\mn@eprint@#1:#2:#3:#4\@nil{\def\@tempa {#1}\def\@tempb {#2}\def\@tempc
  {#3}\ifx \@tempc \@empty \let \@tempc \@tempb \let \@tempb \@tempa \fi \ifx
  \@tempb \@empty \def\@tempb {arXiv}\fi \@ifundefined
  {mn@eprint@\@tempb}{\@tempb:\@tempc}{\expandafter \expandafter \csname
  mn@eprint@\@tempb\endcsname \expandafter{\@tempc}}}

\bibitem[\protect\citeauthoryear{{Alvarez}, {Aparici}  \& {May}}{{Alvarez}
  et~al.}{1987}]{alvarezs}
{Alvarez} H.,  {Aparici} J.,   {May} J.,  1987, \aap, \href
  {https://ui.adsabs.harvard.edu/abs/1987A&A...176...25A} {176, 25}

\bibitem[\protect\citeauthoryear{{Bacon} et~al.,}{{Bacon}
  et~al.}{2010}]{muse_2010}
{Bacon} R.,  et~al., 2010, {The MUSE second-generation VLT instrument}.
p. 773508, \mn@doi{10.1117/12.856027}

\bibitem[\protect\citeauthoryear{{Beck} \& {Krause}}{{Beck} \&
  {Krause}}{2005}]{beck2005}
{Beck} R.,  {Krause} M.,  2005, \mn@doi [Astronomische Nachrichten]
  {10.1002/asna.200510366}, \href
  {https://ui.adsabs.harvard.edu/abs/2005AN....326..414B} {326, 414}

\bibitem[\protect\citeauthoryear{{Biermann} \& {Strom}}{{Biermann} \&
  {Strom}}{1993}]{CRE}
{Biermann} P.~L.,  {Strom} R.~G.,  1993, \aap, \href
  {https://ui.adsabs.harvard.edu/abs/1993A&A...275..659B} {275, 659}

\bibitem[\protect\citeauthoryear{Bowman et~al.,}{Bowman et~al.}{2013}]{mwa_1}
Bowman J.~D.,  et~al., 2013, \mn@doi [Publications of the Astronomical Society
  of Australia] {10.1017/pas.2013.009}, 30, e031

\bibitem[\protect\citeauthoryear{Bozzetto et~al.,}{Bozzetto
  et~al.}{2017}]{bozzetto_2017}
Bozzetto L.~M.,  et~al., 2017, \mn@doi [The Astrophysical Journal Supplement
  Series] {10.3847/1538-4365/aa653c}, 230, 2

\bibitem[\protect\citeauthoryear{{Braun}}{{Braun}}{2013}]{Braun}
{Braun} R.,  2013, \mn@doi [\aap] {10.1051/0004-6361/201220257}, \href
  {https://ui.adsabs.harvard.edu/abs/2013A&A...551A..91B} {551, A91}

\bibitem[\protect\citeauthoryear{{Brocklehurst}}{{Brocklehurst}}{1971}]{Brocklehurst}
{Brocklehurst} M.,  1971, \mn@doi [\mnras] {10.1093/mnras/153.4.471}, \href
  {https://ui.adsabs.harvard.edu/abs/1971MNRAS.153..471B} {153, 471}

\bibitem[\protect\citeauthoryear{{Calzetti}, {Armus}, {Bohlin}, {Kinney},
  {Koornneef}  \& {Storchi-Bergmann}}{{Calzetti} et~al.}{2000}]{Calzetti}
{Calzetti} D.,  {Armus} L.,  {Bohlin} R.~C.,  {Kinney} A.~L.,  {Koornneef} J.,
   {Storchi-Bergmann} T.,  2000, \mn@doi [\apj] {10.1086/308692}, \href
  {https://ui.adsabs.harvard.edu/abs/2000ApJ...533..682C} {533, 682}

\bibitem[\protect\citeauthoryear{{Caplan} \& {Deharveng}}{{Caplan} \&
  {Deharveng}}{1985}]{Caplan85}
{Caplan} J.,  {Deharveng} L.,  1985, \aaps, \href
  {https://ui.adsabs.harvard.edu/abs/1985A&AS...62...63C} {62, 63}

\bibitem[\protect\citeauthoryear{{Caplan}, {Ye}, {Deharveng}, {Turtle}  \&
  {Kennicutt}}{{Caplan} et~al.}{1996}]{Caplan96}
{Caplan} J.,  {Ye} T.,  {Deharveng} L.,  {Turtle} A.~J.,   {Kennicutt} R.~C.,
  1996, \aap, \href {https://ui.adsabs.harvard.edu/abs/1996A&A...307..403C}
  {307, 403}

\bibitem[\protect\citeauthoryear{{Cardelli}, {Clayton}  \& {Mathis}}{{Cardelli}
  et~al.}{1989}]{Cardelli}
{Cardelli} J.~A.,  {Clayton} G.~C.,   {Mathis} J.~S.,  1989, \mn@doi [\apj]
  {10.1086/167900}, \href
  {https://ui.adsabs.harvard.edu/abs/1989ApJ...345..245C} {345, 245}

\bibitem[\protect\citeauthoryear{Chastenet et~al.,}{Chastenet
  et~al.}{2019}]{Chastenet_2019}
Chastenet J.,  et~al., 2019, \mn@doi [The Astrophysical Journal]
  {10.3847/1538-4357/ab16cf}, 876, 62

\bibitem[\protect\citeauthoryear{{Chy{\.z}y}}{{Chy{\.z}y}}{2008}]{mag_in_4254}
{Chy{\.z}y} K.~T.,  2008, \mn@doi [\aap] {10.1051/0004-6361:20078688}, \href
  {https://ui.adsabs.harvard.edu/abs/2008A&A...482..755C} {482, 755}

\bibitem[\protect\citeauthoryear{{Chy{\.z}y}, {We{\.z}gowiec}, {Beck}  \&
  {Bomans}}{{Chy{\.z}y} et~al.}{2011}]{dwarf_mag}
{Chy{\.z}y} K.~T.,  {We{\.z}gowiec} M.,  {Beck} R.,   {Bomans} D.~J.,  2011,
  \mn@doi [\aap] {10.1051/0004-6361/201015393}, \href
  {https://ui.adsabs.harvard.edu/abs/2011A&A...529A..94C} {529, A94}

\bibitem[\protect\citeauthoryear{{Condon}}{{Condon}}{1992}]{Condon}
{Condon} J.~J.,  1992, \mn@doi [\araa] {10.1146/annurev.aa.30.090192.003043},
  \href {https://ui.adsabs.harvard.edu/abs/1992ARA&A..30..575C} {30, 575}

\bibitem[\protect\citeauthoryear{{Crawford}, {Filipovic}, {de Horta}, {Wong},
  {Tothill}, {Draskovic}, {Collier}  \& {Galvin}}{{Crawford}
  et~al.}{2011}]{Crawford2011}
{Crawford} E.~J.,  {Filipovic} M.~D.,  {de Horta} A.~Y.,  {Wong} G.~F.,
  {Tothill} N.~F.~H.,  {Draskovic} D.,  {Collier} J.~D.,   {Galvin} T.~J.,
  2011, \mn@doi [Serbian Astronomical Journal] {10.2298/SAJ1183095C}, \href
  {https://ui.adsabs.harvard.edu/abs/2011SerAJ.183...95C} {183, 95}

\bibitem[\protect\citeauthoryear{{Dickel}, {McIntyre}, {Gruendl}  \&
  {Milne}}{{Dickel} et~al.}{2005}]{Dickel2005}
{Dickel} J.~R.,  {McIntyre} V.~J.,  {Gruendl} R.~A.,   {Milne} D.~K.,  2005,
  \mn@doi [\aj] {10.1086/426916}, \href
  {https://ui.adsabs.harvard.edu/abs/2005AJ....129..790D} {129, 790}

\bibitem[\protect\citeauthoryear{Dickel, Gruendl, McIntyre  \& Amy}{Dickel
  et~al.}{2010}]{dickel_2010}
Dickel J.~R.,  Gruendl R.~A.,  McIntyre V.~J.,   Amy S.~W.,  2010, \mn@doi [The
  Astronomical Journal] {10.1088/0004-6256/140/5/1511}, 140, 1511

\bibitem[\protect\citeauthoryear{Dickinson, Davies  \& Davis}{Dickinson
  et~al.}{2003}]{di2013}
Dickinson C.,  Davies R.~D.,   Davis R.~J.,  2003, \mn@doi [Monthly Notices of
  the Royal Astronomical Society] {10.1046/j.1365-8711.2003.06439.x}, 341, 369

\bibitem[\protect\citeauthoryear{{Draine}}{{Draine}}{2011}]{draine2011}
{Draine} B.~T.,  2011, {Physics of the Interstellar and Intergalactic Medium}

\bibitem[\protect\citeauthoryear{{Draine} \& {Li}}{{Draine} \&
  {Li}}{2007}]{draine2007}
{Draine} B.~T.,  {Li} A.,  2007, \mn@doi [\apj] {10.1086/511055}, \href
  {https://ui.adsabs.harvard.edu/abs/2007ApJ...657..810D} {657, 810}

\bibitem[\protect\citeauthoryear{{Dufour}}{{Dufour}}{1975}]{dufour}
{Dufour} R.~J.,  1975, \mn@doi [\apj] {10.1086/153330}, \href
  {https://ui.adsabs.harvard.edu/abs/1975ApJ...195..315D} {195, 315}

\bibitem[\protect\citeauthoryear{{Dufour} \& {Harlow}}{{Dufour} \&
  {Harlow}}{1977}]{dufour77}
{Dufour} R.~J.,  {Harlow} W.~V.,  1977, \mn@doi [\apj] {10.1086/155513}, \href
  {https://ui.adsabs.harvard.edu/abs/1977ApJ...216..706D} {216, 706}

\bibitem[\protect\citeauthoryear{{Ehle} \& {Beck}}{{Ehle} \&
  {Beck}}{1993}]{Ehle}
{Ehle} M.,  {Beck} R.,  1993, \aap, \href
  {https://ui.adsabs.harvard.edu/abs/1993A&A...273...45E} {273, 45}

\bibitem[\protect\citeauthoryear{Ferrari}{Ferrari}{1998}]{Ferrire}
Ferrari A.,  1998, \mn@doi [Annual Review of Astronomy and Astrophysics]
  {10.1146/annurev.astro.36.1.539}, 36, 539

\bibitem[\protect\citeauthoryear{{Filipovic}, {Haynes}, {White}, {Jones},
  {Klein}  \& {Wielebinski}}{{Filipovic} et~al.}{1995}]{filp}
{Filipovic} M.~D.,  {Haynes} R.~F.,  {White} G.~L.,  {Jones} P.~A.,  {Klein}
  U.,   {Wielebinski} R.,  1995, \aaps, \href
  {https://ui.adsabs.harvard.edu/abs/1995A&AS..111..311F} {111, 311}

\bibitem[\protect\citeauthoryear{{Filipovic}, {White}, {Haynes}, {Jones},
  {Meinert}, {Wielebinski}  \& {Klein}}{{Filipovic}
  et~al.}{1996a}]{Filipovic96}
{Filipovic} M.~D.,  {White} G.~L.,  {Haynes} R.~F.,  {Jones} P.~A.,  {Meinert}
  D.,  {Wielebinski} R.,   {Klein} U.,  1996a, \aaps, \href
  {https://ui.adsabs.harvard.edu/abs/1996A&AS..120...77F} {120, 77}

\bibitem[\protect\citeauthoryear{{Filipovic}, {White}, {Haynes}, {Jones},
  {Meinert}, {Wielebinski}  \& {Klein}}{{Filipovic} et~al.}{1996b}]{lmc_24}
{Filipovic} M.~D.,  {White} G.~L.,  {Haynes} R.~F.,  {Jones} P.~A.,  {Meinert}
  D.,  {Wielebinski} R.,   {Klein} U.,  1996b, \aaps, \href
  {https://ui.adsabs.harvard.edu/abs/1996A&AS..120...77F} {120, 77}

\bibitem[\protect\citeauthoryear{{Filipovic}, {Jones}, {White}, {Haynes},
  {Klein}  \& {Wielebinski}}{{Filipovic} et~al.}{1997}]{Filipovic97}
{Filipovic} M.~D.,  {Jones} P.~A.,  {White} G.~L.,  {Haynes} R.~F.,  {Klein}
  U.,   {Wielebinski} R.,  1997, \mn@doi [\aaps] {10.1051/aas:1997317}, \href
  {https://ui.adsabs.harvard.edu/abs/1997A&AS..121..321F} {121, 321}

\bibitem[\protect\citeauthoryear{{Filipovic}, {Haynes}, {White}  \&
  {Jones}}{{Filipovic} et~al.}{1998}]{dis_mcs}
{Filipovic} M.~D.,  {Haynes} R.~F.,  {White} G.~L.,   {Jones} P.~A.,  1998,
  \mn@doi [\aaps] {10.1051/aas:1998417}, \href
  {https://ui.adsabs.harvard.edu/abs/1998A&AS..130..421F} {130, 421}

\bibitem[\protect\citeauthoryear{{Filipovi{\'c}} et~al.,}{{Filipovi{\'c}}
  et~al.}{2021}]{lmc_source}
{Filipovi{\'c}} M.~D.,  et~al., 2021, arXiv e-prints, \href
  {https://ui.adsabs.harvard.edu/abs/2021arXiv210710967F} {p. arXiv:2107.10967}

\bibitem[\protect\citeauthoryear{{For} et~al.,}{{For} et~al.}{2018}]{2018for}
{For} B.~Q.,  et~al., 2018, \mn@doi [\mnras] {10.1093/mnras/sty1960}, \href
  {https://ui.adsabs.harvard.edu/abs/2018MNRAS.480.2743F} {480, 2743}

\bibitem[\protect\citeauthoryear{Gaensler, Haverkorn, Staveley-Smith, Dickey,
  McClure-Griffiths, Dickel  \& Wolleben}{Gaensler et~al.}{2005}]{Gaensler2005}
Gaensler B.~M.,  Haverkorn M.,  Staveley-Smith L.,  Dickey J.~M.,
  McClure-Griffiths N.~M.,  Dickel J.~R.,   Wolleben M.,  2005, Science, 307
  5715, 1610

\bibitem[\protect\citeauthoryear{{Gaensler}, {Madsen}, {Chatterjee}  \&
  {Mao}}{{Gaensler} et~al.}{2008}]{Gaensler_8}
{Gaensler} B.~M.,  {Madsen} G.~J.,  {Chatterjee} S.,   {Mao} S.~A.,  2008,
  \mn@doi [\pasa] {10.1071/AS08004}, \href
  {https://ui.adsabs.harvard.edu/abs/2008PASA...25..184G} {25, 184}

\bibitem[\protect\citeauthoryear{Gordon, Clayton, Misselt, Landolt  \&
  Wolff}{Gordon et~al.}{2003}]{Gordon_2003}
Gordon K.~D.,  Clayton G.~C.,  Misselt K.~A.,  Landolt A.~U.,   Wolff M.~J.,
  2003, \mn@doi [The Astrophysical Journal] {10.1086/376774}, 594, 279

\bibitem[\protect\citeauthoryear{{Gordon} et~al.,}{{Gordon}
  et~al.}{2011}]{sage_smc}
{Gordon} K.~D.,  et~al., 2011, \mn@doi [\aj] {10.1088/0004-6256/142/4/102},
  \href {https://ui.adsabs.harvard.edu/abs/2011AJ....142..102G} {142, 102}

\bibitem[\protect\citeauthoryear{{Graczyk} et~al.,}{{Graczyk}
  et~al.}{2020}]{smc_distance_new}
{Graczyk} D.,  et~al., 2020, arXiv e-prints, \href
  {https://ui.adsabs.harvard.edu/abs/2020arXiv201008754G} {p. arXiv:2010.08754}

\bibitem[\protect\citeauthoryear{{Gressel}, {Elstner}, {Ziegler}  \&
  {R{\"u}diger}}{{Gressel} et~al.}{2008}]{small_scale}
{Gressel} O.,  {Elstner} D.,  {Ziegler} U.,   {R{\"u}diger} G.,  2008, \mn@doi
  [\aap] {10.1051/0004-6361:200810195}, \href
  {https://ui.adsabs.harvard.edu/abs/2008A&A...486L..35G} {486, L35}

\bibitem[\protect\citeauthoryear{{Haynes}, {Klein}, {Wielebinski}  \&
  {Murray}}{{Haynes} et~al.}{1986}]{mc_parks}
{Haynes} R.~F.,  {Klein} U.,  {Wielebinski} R.,   {Murray} J.~D.,  1986, \aap,
  \href {https://ui.adsabs.harvard.edu/abs/1986A&A...159...22H} {159, 22}

\bibitem[\protect\citeauthoryear{{Haynes} et~al.,}{{Haynes}
  et~al.}{1991}]{Haynes91}
{Haynes} R.~F.,  et~al., 1991, \aap, \href
  {https://ui.adsabs.harvard.edu/abs/1991A&A...252..475H} {252, 475}

\bibitem[\protect\citeauthoryear{{Heesen}, {Brinks}, {Leroy}, {Heald}, {Braun},
  {Bigiel}  \& {Beck}}{{Heesen} et~al.}{2014}]{Heesen2014}
{Heesen} V.,  {Brinks} E.,  {Leroy} A.~K.,  {Heald} G.,  {Braun} R.,  {Bigiel}
  F.,   {Beck} R.,  2014, \mn@doi [\aj] {10.1088/0004-6256/147/5/103}, \href
  {https://ui.adsabs.harvard.edu/abs/2014AJ....147..103H} {147, 103}

\bibitem[\protect\citeauthoryear{{Helou}, {Soifer}  \&
  {Rowan-Robinson}}{{Helou} et~al.}{1985}]{helou}
{Helou} G.,  {Soifer} B.~T.,   {Rowan-Robinson} M.,  1985, \mn@doi [\apjl]
  {10.1086/184556}, \href
  {https://ui.adsabs.harvard.edu/abs/1985ApJ...298L...7H} {298, L7}

\bibitem[\protect\citeauthoryear{Hughes}{Hughes}{2011}]{hughes_2011}
Hughes A.,  2011, PhD thesis, Swinburne University of Technology

\bibitem[\protect\citeauthoryear{Hughes, Wong, Ekers, Staveley-Smith,
  Filipovic, Maddison, Fukui  \& Mizuno}{Hughes et~al.}{2006}]{hughes2006}
Hughes A.,  Wong T.,  Ekers R.,  Staveley-Smith L.,  Filipovic M.,  Maddison
  S.,  Fukui Y.,   Mizuno N.,  2006, \mn@doi [Monthly Notices of the Royal
  Astronomical Society] {10.1111/j.1365-2966.2006.10483.x}, 370, 363

\bibitem[\protect\citeauthoryear{Hughes, Staveley-Smith, Kim, Wolleben  \&
  Filipović}{Hughes et~al.}{2007}]{Huges2007}
Hughes A.,  Staveley-Smith L.,  Kim S.,  Wolleben M.,   Filipović M.,  2007,
  \mn@doi [Monthly Notices of the Royal Astronomical Society]
  {10.1111/j.1365-2966.2007.12466.x}, 382, 543

\bibitem[\protect\citeauthoryear{Hurley-Walker et~al.,}{Hurley-Walker
  et~al.}{2016}]{Hurley-Walker2016}
Hurley-Walker N.,  et~al., 2016, \mn@doi [Monthly Notices of the Royal
  Astronomical Society] {10.1093/mnras/stw2337}, 464, 1146

\bibitem[\protect\citeauthoryear{{Isobe}, {Feigelson}, {Akritas}  \&
  {Babu}}{{Isobe} et~al.}{1990}]{Isobe}
{Isobe} T.,  {Feigelson} E.~D.,  {Akritas} M.~G.,   {Babu} G.~J.,  1990,
  \mn@doi [\apj] {10.1086/169390}, \href
  {https://ui.adsabs.harvard.edu/abs/1990ApJ...364..104I} {364, 104}

\bibitem[\protect\citeauthoryear{{Jameson} et~al.,}{{Jameson}
  et~al.}{2016}]{all_gas_mcs}
{Jameson} K.~E.,  et~al., 2016, \mn@doi [\apj] {10.3847/0004-637X/825/1/12},
  \href {https://ui.adsabs.harvard.edu/abs/2016ApJ...825...12J} {825, 12}

\bibitem[\protect\citeauthoryear{{Joseph} et~al.,}{{Joseph}
  et~al.}{2019}]{askap_smc}
{Joseph} T.~D.,  et~al., 2019, \mn@doi [\mnras] {10.1093/mnras/stz2650}, \href
  {https://ui.adsabs.harvard.edu/abs/2019MNRAS.490.1202J} {490, 1202}

\bibitem[\protect\citeauthoryear{{Jurusik}, {Drzazga}, {Jableka}, {Chy{\.z}y},
  {Beck}, {Klein}  \& {We{\.z}gowiec}}{{Jurusik} et~al.}{2014}]{mag_in_mcs}
{Jurusik} W.,  {Drzazga} R.~T.,  {Jableka} M.,  {Chy{\.z}y} K.~T.,  {Beck} R.,
  {Klein} U.,   {We{\.z}gowiec} M.,  2014, \mn@doi [\aap]
  {10.1051/0004-6361/201323060}, \href
  {https://ui.adsabs.harvard.edu/abs/2014A&A...567A.134J} {567, A134}

\bibitem[\protect\citeauthoryear{{Kennicutt}, {Bresolin}, {Bomans}, {Bothun}
  \& {Thompson}}{{Kennicutt} et~al.}{1995}]{Kennicutt95}
{Kennicutt} Robert~C. J.,  {Bresolin} F.,  {Bomans} D.~J.,  {Bothun} G.~D.,
  {Thompson} I.~B.,  1995, \mn@doi [\aj] {10.1086/117304}, \href
  {https://ui.adsabs.harvard.edu/abs/1995AJ....109..594K} {109, 594}

\bibitem[\protect\citeauthoryear{Kim, Staveley-Smith, Dopita, Freeman, Sault,
  Kesteven  \& McConnell}{Kim et~al.}{1998}]{atca}
Kim S.,  Staveley-Smith L.,  Dopita M.~A.,  Freeman K.~C.,  Sault R.~J.,
  Kesteven M.~J.,   McConnell D.,  1998, \mn@doi [The Astrophysical Journal]
  {10.1086/306030}, 503, 674

\bibitem[\protect\citeauthoryear{{Kim}, {Staveley-Smith}, {Dopita}, {Sault},
  {Freeman}, {Lee}  \& {Chu}}{{Kim} et~al.}{2003}]{lmc_HI_merged}
{Kim} S.,  {Staveley-Smith} L.,  {Dopita} M.~A.,  {Sault} R.~J.,  {Freeman}
  K.~C.,  {Lee} Y.,   {Chu} Y.-H.,  2003, \mn@doi [\apjs] {10.1086/376980},
  \href {https://ui.adsabs.harvard.edu/abs/2003ApJS..148..473K} {148, 473}

\bibitem[\protect\citeauthoryear{{Klein}, {Wielebinski}, {Haynes}  \&
  {Malin}}{{Klein} et~al.}{1989}]{klein_lmc}
{Klein} U.,  {Wielebinski} R.,  {Haynes} R.~F.,   {Malin} D.~F.,  1989, \aap,
  \href {https://ui.adsabs.harvard.edu/abs/1989A&A...211..280K} {211, 280}

\bibitem[\protect\citeauthoryear{{Klein}, {Haynes}, {Wielebinski}  \&
  {Meinert}}{{Klein} et~al.}{1993}]{mcs_bb}
{Klein} U.,  {Haynes} R.~F.,  {Wielebinski} R.,   {Meinert} D.,  1993, \aap,
  \href {https://ui.adsabs.harvard.edu/abs/1993A&A...271..402K} {271, 402}

\bibitem[\protect\citeauthoryear{{Lacki}, {Thompson}  \& {Quataert}}{{Lacki}
  et~al.}{2010}]{FIR_cooling}
{Lacki} B.~C.,  {Thompson} T.~A.,   {Quataert} E.,  2010, \mn@doi [\apj]
  {10.1088/0004-637X/717/1/1}, \href
  {https://ui.adsabs.harvard.edu/abs/2010ApJ...717....1L} {717, 1}

\bibitem[\protect\citeauthoryear{{Laki{\'c}evi{\'c}}
  et~al.,}{{Laki{\'c}evi{\'c}} et~al.}{2015}]{lacklack}
{Laki{\'c}evi{\'c}} M.,  et~al., 2015, \mn@doi [\apj]
  {10.1088/0004-637X/799/1/50}, \href
  {https://ui.adsabs.harvard.edu/abs/2015ApJ...799...50L} {799, 50}

\bibitem[\protect\citeauthoryear{{Lawton} et~al.,}{{Lawton}
  et~al.}{2010}]{70sfr}
{Lawton} B.,  et~al., 2010, \mn@doi [\apj] {10.1088/0004-637X/716/1/453}, \href
  {https://ui.adsabs.harvard.edu/abs/2010ApJ...716..453L} {716, 453}

\bibitem[\protect\citeauthoryear{{Leroy} et~al.,}{{Leroy}
  et~al.}{2011}]{Leroy_CO}
{Leroy} A.~K.,  et~al., 2011, \mn@doi [\apj] {10.1088/0004-637X/737/1/12},
  \href {https://ui.adsabs.harvard.edu/abs/2011ApJ...737...12L} {737, 12}

\bibitem[\protect\citeauthoryear{{Loiseau}, {Klein}, {Greybe}, {Wielebinski}
  \& {Haynes}}{{Loiseau} et~al.}{1987}]{smc_l}
{Loiseau} N.,  {Klein} U.,  {Greybe} A.,  {Wielebinski} R.,   {Haynes} R.~F.,
  1987, \aap, \href {https://ui.adsabs.harvard.edu/abs/1987A&A...178...62L}
  {178, 62}

\bibitem[\protect\citeauthoryear{{Longair}}{{Longair}}{2011}]{Longair}
{Longair} M.~S.,  2011, {High Energy Astrophysics}

\bibitem[\protect\citeauthoryear{{Lonsdale} et~al.,}{{Lonsdale}
  et~al.}{2009}]{mwa_design}
{Lonsdale} C.~J.,  et~al., 2009, \mn@doi [Proceedings of the IEEE]
  {10.1109/JPROC.2009.2017564}, 97, 1497

\bibitem[\protect\citeauthoryear{{Maggi} et~al.,}{{Maggi}
  et~al.}{2019}]{smc_snr}
{Maggi} P.,  et~al., 2019, \mn@doi [\aap] {10.1051/0004-6361/201936583}, \href
  {https://ui.adsabs.harvard.edu/abs/2019A&A...631A.127M} {631, A127}

\bibitem[\protect\citeauthoryear{Mao, Gaensler, Stanimirović, Haverkorn,
  McClure‐Griffiths, Staveley‐Smith  \& Dickey}{Mao
  et~al.}{2008}]{Mao_2008}
Mao S.~A.,  Gaensler B.~M.,  Stanimirović S.,  Haverkorn M.,
  McClure‐Griffiths N.~M.,  Staveley‐Smith L.,   Dickey J.~M.,  2008,
  \mn@doi [The Astrophysical Journal] {10.1086/590546}, 688, 1029–1049

\bibitem[\protect\citeauthoryear{{Mao} et~al.,}{{Mao} et~al.}{2012}]{Mao2012}
{Mao} S.~A.,  et~al., 2012, \mn@doi [\apj] {10.1088/0004-637X/759/1/25}, \href
  {https://ui.adsabs.harvard.edu/abs/2012ApJ...759...25M} {759, 25}

\bibitem[\protect\citeauthoryear{McLeod, Dale, Evans, Ginsburg, Kruijssen,
  Pellegrini, Ramsay  \& Testi}{McLeod et~al.}{2018}]{McLeod_2018}
McLeod A.~F.,  Dale J.~E.,  Evans C.~J.,  Ginsburg A.,  Kruijssen J. M.~D.,
  Pellegrini E.~W.,  Ramsay S.~K.,   Testi L.,  2018, \mn@doi [Monthly Notices
  of the Royal Astronomical Society] {10.1093/mnras/sty2696}, 486, 5263–5288

\bibitem[\protect\citeauthoryear{{Meixner} et~al.,}{{Meixner}
  et~al.}{2006}]{sage_lmc}
{Meixner} M.,  et~al., 2006, \mn@doi [\aj] {10.1086/508185}, \href
  {https://ui.adsabs.harvard.edu/abs/2006AJ....132.2268M} {132, 2268}

\bibitem[\protect\citeauthoryear{{Meixner} et~al.,}{{Meixner}
  et~al.}{2010}]{mx2010}
{Meixner} M.,  et~al., 2010, \mn@doi [\aap] {10.1051/0004-6361/201014662},
  \href {https://ui.adsabs.harvard.edu/abs/2010A&A...518L..71M} {518, L71}

\bibitem[\protect\citeauthoryear{{Meixner} et~al.,}{{Meixner}
  et~al.}{2013}]{meixner2013}
{Meixner} M.,  et~al., 2013, \mn@doi [\aj] {10.1088/0004-6256/146/3/62}, \href
  {https://ui.adsabs.harvard.edu/abs/2013AJ....146...62M} {146, 62}

\bibitem[\protect\citeauthoryear{{Meixner} et~al.,}{{Meixner}
  et~al.}{2015}]{meixner2015}
{Meixner} M.,  et~al., 2015, \mn@doi [\aj] {10.1088/0004-6256/149/2/88}, \href
  {https://ui.adsabs.harvard.edu/abs/2015AJ....149...88M} {149, 88}

\bibitem[\protect\citeauthoryear{{Mills}}{{Mills}}{1959}]{millss}
{Mills} B.~Y.,  1959, \mn@doi [Handbuch der Physik]
  {10.1007/978-3-642-45932-0\_6}, \href
  {https://ui.adsabs.harvard.edu/abs/1959HDP....53..239M} {53, 239}

\bibitem[\protect\citeauthoryear{{Murphy}, {Porter}, {Moskalenko}, {Helou}  \&
  {Strong}}{{Murphy} et~al.}{2012}]{CREs_30dor}
{Murphy} E.~J.,  {Porter} T.~A.,  {Moskalenko} I.~V.,  {Helou} G.,   {Strong}
  A.~W.,  2012, \mn@doi [\apj] {10.1088/0004-637X/750/2/126}, \href
  {https://ui.adsabs.harvard.edu/abs/2012ApJ...750..126M} {750, 126}

\bibitem[\protect\citeauthoryear{{Oey} \& {Kennicutt}}{{Oey} \&
  {Kennicutt}}{1997}]{Oey1997}
{Oey} M.~S.,  {Kennicutt} R.~C. J.,  1997, \mn@doi [\mnras]
  {10.1093/mnras/291.4.827}, \href
  {https://ui.adsabs.harvard.edu/abs/1997MNRAS.291..827O} {291, 827}

\bibitem[\protect\citeauthoryear{Oster}{Oster}{1961}]{oster}
Oster L.,  1961, \mn@doi [Rev. Mod. Phys.] {10.1103/RevModPhys.33.525}, 33, 525

\bibitem[\protect\citeauthoryear{{Osterbrock}}{{Osterbrock}}{1989}]{Osterbrock}
{Osterbrock} D.~E.,  1989, {Astrophysics of gaseous nebulae and active galactic
  nuclei}

\bibitem[\protect\citeauthoryear{{Paredes}, {Points}, {Smith}, {Rest}, {Damke},
  {Zenteno}  \& {MCELS Team}}{{Paredes} et~al.}{2015}]{paredes15}
{Paredes} L.,  {Points} S.~D.,  {Smith} R.~C.,  {Rest} A.,  {Damke} G.,
  {Zenteno} A.,   {MCELS Team} 2015, in {Points} S.,  {Kunder} A.,  eds,
  Astronomical Society of the Pacific Conference Series Vol. 491, Fifty Years
  of Wide Field Studies in the Southern Hemisphere: Resolved Stellar
  Populations of the Galactic Bulge and Magellanic Clouds. pp 366--369

\bibitem[\protect\citeauthoryear{{Payne}, {Filipovi{\'c}}, {Reid}, {Jones},
  {Staveley-Smith}  \& {White}}{{Payne} et~al.}{2004}]{smc_bkg}
{Payne} J.~L.,  {Filipovi{\'c}} M.~D.,  {Reid} W.,  {Jones} P.~A.,
  {Staveley-Smith} L.,   {White} G.~L.,  2004, \mn@doi [\mnras]
  {10.1111/j.1365-2966.2004.08287.x}, \href
  {https://ui.adsabs.harvard.edu/abs/2004MNRAS.355...44P} {355, 44}

\bibitem[\protect\citeauthoryear{{Peck}, {Goss}, {Dickel}, {Roelfsema},
  {Kesteven}, {Dickel}, {Milne}  \& {Points}}{{Peck} et~al.}{1997}]{Peck97}
{Peck} A.~B.,  {Goss} W.~M.,  {Dickel} H.~R.,  {Roelfsema} P.~R.,  {Kesteven}
  M.~J.,  {Dickel} J.~R.,  {Milne} D.~K.,   {Points} S.~D.,  1997, \mn@doi
  [\apj] {10.1086/304508}, \href
  {https://ui.adsabs.harvard.edu/abs/1997ApJ...486..329P} {486, 329}

\bibitem[\protect\citeauthoryear{{Pellegrini}, {Oey}, {Winkler}, {Points},
  {Smith}, {Jaskot}  \& {Zastrow}}{{Pellegrini} et~al.}{2012}]{Pellegrini_2012}
{Pellegrini} E.~W.,  {Oey} M.~S.,  {Winkler} P.~F.,  {Points} S.~D.,  {Smith}
  R.~C.,  {Jaskot} A.~E.,   {Zastrow} J.,  2012, \mn@doi [\apj]
  {10.1088/0004-637X/755/1/40}, \href
  {https://ui.adsabs.harvard.edu/abs/2012ApJ...755...40P} {755, 40}

\bibitem[\protect\citeauthoryear{{Pennock} et~al.,}{{Pennock}
  et~al.}{2021}]{askap_lmc}
{Pennock} C.~M.,  et~al., 2021, \mn@doi [\mnras] {10.1093/mnras/stab1858},
  \href {https://ui.adsabs.harvard.edu/abs/2021MNRAS.506.3540P} {506, 3540}

\bibitem[\protect\citeauthoryear{Pietrzyński et~al.,}{Pietrzyński
  et~al.}{2019}]{Pietrzyski2019}
Pietrzyński G.,  et~al., 2019, Nature, 567, 200

\bibitem[\protect\citeauthoryear{{Points}, {Smith}  \& {Chu}}{{Points}
  et~al.}{2005}]{Points2005}
{Points} S.~D.,  {Smith} R.~C.,   {Chu} Y.~H.,  2005, in American Astronomical
  Society Meeting Abstracts. p. 132.11

\bibitem[\protect\citeauthoryear{{Schleicher} \& {Beck}}{{Schleicher} \&
  {Beck}}{2013}]{bsfr}
{Schleicher} D. R.~G.,  {Beck} R.,  2013, \mn@doi [\aap]
  {10.1051/0004-6361/201321707}, \href
  {https://ui.adsabs.harvard.edu/abs/2013A&A...556A.142S} {556, A142}

\bibitem[\protect\citeauthoryear{{Shain}}{{Shain}}{1959}]{lowfreq}
{Shain} C.~A.,  1959, in {Bracewell} R.~N.,  ed., ~ Vol. 9, URSI Symp. 1: Paris
  Symposium on Radio Astronomy. p.~328

\bibitem[\protect\citeauthoryear{{Siejkowski}, {Soida}  \&
  {Chy{\.z}y}}{{Siejkowski} et~al.}{2018}]{CR_driven}
{Siejkowski} H.,  {Soida} M.,   {Chy{\.z}y} K.~T.,  2018, \mn@doi [\aap]
  {10.1051/0004-6361/201730566}, \href
  {https://ui.adsabs.harvard.edu/abs/2018A&A...611A...7S} {611, A7}

\bibitem[\protect\citeauthoryear{{Smith} \& {MCELS Team}}{{Smith} \& {MCELS
  Team}}{1999}]{smith99}
{Smith} R.~C.,  {MCELS Team} 1999, in {Chu} Y.-H.,  {Suntzeff} N.,  {Hesser}
  J.,   {Bohlender} D.,  eds,  IAU Symposium Vol. 190, New Views of the
  Magellanic Clouds. p.~28

\bibitem[\protect\citeauthoryear{{Smith}, {Points}, {Chu}, {Winkler},
  {Aguilera}, {Leiton}  \& {MCELS Team}}{{Smith} et~al.}{2005}]{smith2005}
{Smith} R.~C.,  {Points} S.~D.,  {Chu} Y.~H.,  {Winkler} P.~F.,  {Aguilera} C.,
   {Leiton} R.,   {MCELS Team} 2005, in American Astronomical Society Meeting
  Abstracts. p. 25.07

\bibitem[\protect\citeauthoryear{Stanimirović, Staveley‐Smith  \&
  Jones}{Stanimirović et~al.}{2004}]{Stanimirovi2004}
Stanimirović S.,  Staveley‐Smith L.,   Jones P.~A.,  2004, \mn@doi [The
  Astrophysical Journal] {10.1086/381869}, 604, 176–186

\bibitem[\protect\citeauthoryear{{Strong} \& {Mattox}}{{Strong} \&
  {Mattox}}{1996}]{x_co}
{Strong} A.~W.,  {Mattox} J.~R.,  1996, \aap, \href
  {https://ui.adsabs.harvard.edu/abs/1996A&A...308L..21S} {308, L21}

\bibitem[\protect\citeauthoryear{{Subramanian} \& {Subramaniam}}{{Subramanian}
  \& {Subramaniam}}{2015}]{smc_i}
{Subramanian} S.,  {Subramaniam} A.,  2015, \mn@doi [\aap]
  {10.1051/0004-6361/201424248}, \href
  {https://ui.adsabs.harvard.edu/abs/2015A&A...573A.135S} {573, A135}

\bibitem[\protect\citeauthoryear{Tabatabaei, Beck, Krügel, Krause,
  Berkhuijsen, Gordon  \& Menten}{Tabatabaei et~al.}{2007}]{tab_2007}
Tabatabaei F.~S.,  Beck R.,  Krügel E.,  Krause M.,  Berkhuijsen E.~M.,
  Gordon K.~D.,   Menten K.~M.,  2007, \mn@doi [Astronomy & Astrophysics]
  {10.1051/0004-6361:20078174}, 475, 133–143

\bibitem[\protect\citeauthoryear{{Tabatabaei}, {Krause}, {Fletcher}  \&
  {Beck}}{{Tabatabaei} et~al.}{2008}]{trt2008}
{Tabatabaei} F.~S.,  {Krause} M.,  {Fletcher} A.,   {Beck} R.,  2008, \mn@doi
  [\aap] {10.1051/0004-6361:200810590}, \href
  {https://ui.adsabs.harvard.edu/abs/2008A&A...490.1005T} {490, 1005}

\bibitem[\protect\citeauthoryear{Tabatabaei et~al.,}{Tabatabaei
  et~al.}{2013a}]{tab_2013}
Tabatabaei F.~S.,  et~al., 2013a, \mn@doi [Astronomy & Astrophysics]
  {10.1051/0004-6361/201220249}, 552, A19

\bibitem[\protect\citeauthoryear{{Tabatabaei}, {Berkhuijsen}, {Frick}, {Beck}
  \& {Schinnerer}}{{Tabatabaei} et~al.}{2013b}]{tab_2013b}
{Tabatabaei} F.~S.,  {Berkhuijsen} E.~M.,  {Frick} P.,  {Beck} R.,
  {Schinnerer} E.,  2013b, \mn@doi [\aap] {10.1051/0004-6361/201218909}, \href
  {https://ui.adsabs.harvard.edu/abs/2013A&A...557A.129T} {557, A129}

\bibitem[\protect\citeauthoryear{Tabatabaei et~al.,}{Tabatabaei
  et~al.}{2017}]{Tabatabaei_2017}
Tabatabaei F.~S.,  et~al., 2017, \mn@doi [The Astrophysical Journal]
  {10.3847/1538-4357/836/2/185}, 836, 185

\bibitem[\protect\citeauthoryear{Tabatabaei, M{\'i}nguez, Prieto  \&
  Fern'andez-Ontiveros}{Tabatabaei et~al.}{2018}]{tab_2018}
Tabatabaei F.~S.,  M{\'i}nguez P.,  Prieto M. A.~A.,   Fern'andez-Ontiveros
  J.~A.,  2018, Nature Astronomy, 2, 83

\bibitem[\protect\citeauthoryear{{Tingay} et~al.,}{{Tingay}
  et~al.}{2013}]{Tingay2013}
{Tingay} S.~J.,  et~al., 2013, \mn@doi [\pasa] {10.1017/pasa.2012.007}, \href
  {https://ui.adsabs.harvard.edu/abs/2013PASA...30....7T} {30, e007}

\bibitem[\protect\citeauthoryear{{Valls-Gabaud}}{{Valls-Gabaud}}{1998}]{vallas}
{Valls-Gabaud} D.,  1998, \mn@doi [\pasa] {10.1071/AS98111}, \href
  {https://ui.adsabs.harvard.edu/abs/1998PASA...15..111V} {15, 111}

\bibitem[\protect\citeauthoryear{{Vermeij} \& {van der Hulst}}{{Vermeij} \&
  {van der Hulst}}{2002}]{Vermeij}
{Vermeij} R.,  {van der Hulst} J.~M.,  2002, \mn@doi [\aap]
  {10.1051/0004-6361:20020864}, \href
  {https://ui.adsabs.harvard.edu/abs/2002A&A...391.1081V} {391, 1081}

\bibitem[\protect\citeauthoryear{{Vogt}, {Seitenzahl}, {Dopita}  \&
  {Ghavamian}}{{Vogt} et~al.}{2017}]{snr102}
{Vogt} F. P.~A.,  {Seitenzahl} I.~R.,  {Dopita} M.~A.,   {Ghavamian} P.,  2017,
  \mn@doi [\aap] {10.1051/0004-6361/201730756}, \href
  {https://ui.adsabs.harvard.edu/abs/2017A&A...602L...4V} {602, L4}

\bibitem[\protect\citeauthoryear{{Wayth} et~al.,}{{Wayth}
  et~al.}{2015}]{Wayth2015}
{Wayth} R.~B.,  et~al., 2015, \mn@doi [\pasa] {10.1017/pasa.2015.26}, \href
  {https://ui.adsabs.harvard.edu/abs/2015PASA...32...25W} {32, e025}

\bibitem[\protect\citeauthoryear{{Weilbacher}, {Streicher}  \&
  {Palsa}}{{Weilbacher} et~al.}{2016}]{muse_pipeline}
{Weilbacher} P.~M.,  {Streicher} O.,   {Palsa} R.,  2016, {MUSE-DRP: MUSE Data
  Reduction Pipeline} (\mn@eprint {ascl} {1610.004})

\bibitem[\protect\citeauthoryear{{Weingartner} \& {Draine}}{{Weingartner} \&
  {Draine}}{2001}]{Weingartner}
{Weingartner} J.~C.,  {Draine} B.~T.,  2001, \mn@doi [\apj] {10.1086/318651},
  \href {https://ui.adsabs.harvard.edu/abs/2001ApJ...548..296W} {548, 296}

\bibitem[\protect\citeauthoryear{{Winkler}, {Smith}, {Points}  \& {MCELS
  Team}}{{Winkler} et~al.}{2015}]{winkler15}
{Winkler} P.~F.,  {Smith} R.~C.,  {Points} S.~D.,   {MCELS Team} 2015, in
  {Points} S.,  {Kunder} A.,  eds,  Astronomical Society of the Pacific
  Conference Series Vol. 491, Fifty Years of Wide Field Studies in the Southern
  Hemisphere: Resolved Stellar Populations of the Galactic Bulge and Magellanic
  Clouds. p.~343

\bibitem[\protect\citeauthoryear{{Wong}, {Filipovic}, {Crawford}, {de Horta},
  {Galvin}, {Draskovic}  \& {Payne}}{{Wong} et~al.}{2011a}]{wong2011}
{Wong} G.~F.,  {Filipovic} M.~D.,  {Crawford} E.~J.,  {de Horta} A.~Y.,
  {Galvin} T.,  {Draskovic} D.,   {Payne} J.~L.,  2011a, \mn@doi [Serbian
  Astronomical Journal] {10.2298/SAJ1182043W}, \href
  {https://ui.adsabs.harvard.edu/abs/2011SerAJ.182...43W} {182, 43}

\bibitem[\protect\citeauthoryear{{Wong} et~al.,}{{Wong}
  et~al.}{2011b}]{co_2011}
{Wong} T.,  et~al., 2011b, \mn@doi [\apjs] {10.1088/0067-0049/197/2/16}, \href
  {https://ui.adsabs.harvard.edu/abs/2011ApJS..197...16W} {197, 16}

\bibitem[\protect\citeauthoryear{{Wong} et~al.,}{{Wong} et~al.}{2017}]{co_2017}
{Wong} T.,  et~al., 2017, \mn@doi [\apj] {10.3847/1538-4357/aa9333}, \href
  {https://ui.adsabs.harvard.edu/abs/2017ApJ...850..139W} {850, 139}

\bibitem[\protect\citeauthoryear{{Yao}, {Manchester}  \& {Wang}}{{Yao}
  et~al.}{2017}]{yao}
{Yao} J.~M.,  {Manchester} R.~N.,   {Wang} N.,  2017, \mn@doi [\apj]
  {10.3847/1538-4357/835/1/29}, \href
  {https://ui.adsabs.harvard.edu/abs/2017ApJ...835...29Y} {835, 29}

\bibitem[\protect\citeauthoryear{{de Jong}, {Klein}, {Wielebinski}  \&
  {Wunderlich}}{{de Jong} et~al.}{1985}]{dejong}
{de Jong} T.,  {Klein} U.,  {Wielebinski} R.,   {Wunderlich} E.,  1985, \aap,
  \href {https://ui.adsabs.harvard.edu/abs/1985A&A...147L...6D} {147, L6}

\bibitem[\protect\citeauthoryear{{van der Kruit}}{{van der
  Kruit}}{1971}]{vanderKruit}
{van der Kruit} P.~C.,  1971, \aap, \href
  {https://ui.adsabs.harvard.edu/abs/1971A&A....15..110V} {15, 110}

\bibitem[\protect\citeauthoryear{{van der Marel} \& {Kallivayalil}}{{van der
  Marel} \& {Kallivayalil}}{2014}]{lmc_ii}
{van der Marel} R.~P.,  {Kallivayalil} N.,  2014, \mn@doi [\apj]
  {10.1088/0004-637X/781/2/121}, \href
  {https://ui.adsabs.harvard.edu/abs/2014ApJ...781..121V} {781, 121}

\makeatother
\end{thebibliography}




\bsp	
\label{lastpage}

\end{document}